\documentclass[twocolumn,aps,prx,superscriptaddress,preprintnumbers,bibnotes]{revtex4-2}

\usepackage[T1]{fontenc}

\usepackage{natbib}
\usepackage{graphicx}
\usepackage{bm}
\usepackage{color}
\usepackage{amsmath}
\usepackage{gensymb} 
\usepackage{physics}
\usepackage{tabularx}
\usepackage{ragged2e}
\usepackage{multirow}
\usepackage{bigdelim}
\usepackage{amssymb}

\renewcommand{\vec}[1]{\ensuremath{\bm{#1}}}

\newcommand\unit[2]{\ensuremath{#1~\mathrm{{#2}}}}
\newcommand\unitonly[1]{\ensuremath{\mathrm{{#1}}}}

\newcommand\Ket[1]{\ensuremath{|{#1}\rangle}}
\newcommand\Bra[1]{\ensuremath{\langle{#1}|}}

\begin{document}

\title{The $^{1}\mathrm{S}_0$-$^{3}\mathrm{P}_2$ magnetic quadrupole transition in neutral strontium}

\author{J. Trautmann}
\author{D. Yankelev}
\author{V. Kl\"{u}sener}
\author{A. J. Park}
\affiliation{
  Max-Planck-Institut f{\"u}r Quantenoptik,
  Hans-Kopfermann-Stra{\ss}e 1,
  85748 Garching, Germany}
\affiliation{
  Munich Center for Quantum Science and Technology,
  80799 M{\"u}nchen, Germany}
\author{I. Bloch}
\affiliation{
  Max-Planck-Institut f{\"u}r Quantenoptik,
  Hans-Kopfermann-Stra{\ss}e 1,
  85748 Garching, Germany}
\affiliation{
  Fakult{\"a}t f{\"u}r Physik,
  Ludwig-Maximilians-Universit{\"a}t M{\"u}nchen,
  80799 M{\"u}nchen, Germany}
\affiliation{
  Munich Center for Quantum Science and Technology,
  80799 M{\"u}nchen, Germany}
\author{S. Blatt}
\email{sebastian.blatt@mpq.mpg.de}
\affiliation{
  Max-Planck-Institut f{\"u}r Quantenoptik,
  Hans-Kopfermann-Stra{\ss}e 1,
  85748 Garching, Germany}
\affiliation{
	  Fakult{\"a}t f{\"u}r Physik,
  Ludwig-Maximilians-Universit{\"a}t M{\"u}nchen,
  80799 M{\"u}nchen, Germany}
\affiliation{
  Munich Center for Quantum Science and Technology,
  80799 M{\"u}nchen, Germany}

\date{\today}

\begin{abstract}
We present a detailed investigation of the ultranarrow magnetic-quadrupole $^{1}\mathrm{S}_0$-$^{3}\mathrm{P}_2$ transition in neutral strontium and show how it can be made accessible for quantum simulation and quantum computation. By engineering the light shift in a one-dimensional optical lattice, we perform high-resolution spectroscopy and observe the characteristic absorption patterns for a magnetic quadrupole transition. We measure an absolute transition frequency of \unit{446,647,242,704(2)}{kHz} in $^{88}\mathrm{Sr}$ and an $^{88}\mathrm{Sr}$-$^{87}\mathrm{Sr}$ isotope shift of \unit{+62.91(4)}{MHz}. In a proof-of-principle experiment, we use this transition to demonstrate local addressing in an optical lattice with \unit{532}{nm} spacing with a Rayleigh-criterion resolution of \unit{494(45)}{nm}. Our results pave the way for applications of the magnetic quadrupole transition as an optical qubit and for single-site addressing in optical lattices.
\end{abstract}

\maketitle

\section{Introduction}
\label{sec:intro}

\begin{figure}
  \includegraphics{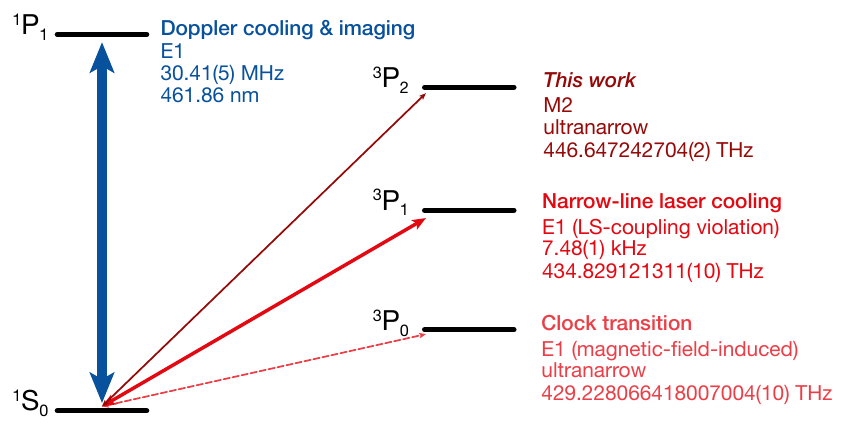}
  \caption{Level diagram of $^{88}\mathrm{Sr}$ including the important optical transitions from the $^{1}\mathrm{S}_0$ ground state. The broad transition to $^{1}\mathrm{P}_1$ is electric-dipole (E1) allowed and is used for Doppler cooling and imaging. The intercombination transitions to the lowest $^{3}\mathrm{P}_J$ states should be electric-dipole forbidden. However, LS-coupling~\cite{sobelman12} is violated, which leads to a narrow E1 transition to $^{3}\mathrm{P}_1$~\cite{boyd07}, used for narrow-line laser cooling. The ultranarrow clock transition to the $^{3}\mathrm{P}_0$ state is doubly-forbidden, and only becomes E1-allowed by perturbatively mixing $^{3}\mathrm{P}_0$ with $^{3}\mathrm{P}_1$ via a bias magnetic field~\cite{taichenachev06}. In contrast, the $^{1}\mathrm{S}_0$-$^{3}\mathrm{P}_2$ transition is magnetic-quadrupole (M2) allowed even at zero bias field~\cite{garstang67}. The values for the transitions from the $^{1}\mathrm{S}_0$ ground state to the $^{1}\mathrm{P}_1$, $^{3}\mathrm{P}_0$, and $^{3}\mathrm{P}_1$ states were obtained from Refs.~\cite{heinz20}, \cite{takano17,riehle18}, and \cite{ferrari03,nicholson15}, respectively.}
  \label{fig:level-scheme}
\end{figure}

Ultracold neutral two-electron atoms have emerged as a promising platform for metrology~\cite{takamoto05,ludlow06,ludlow15,bothwell22,zheng22}, quantum simulation~\cite{sonderhouse20,ozawa18,schafer20,young22}, and quantum computation~\cite{schine22,madjarov20,ma22,jenkins22}.
One pillar of their success are their ultranarrow optical transitions between the singlet $^{1}\mathrm{S}_0$ ground state and the long-lived metastable triplet $^{3}\mathrm{P}_J$ states, as shown in Fig.~\ref{fig:level-scheme} for $^{88}\mathrm{Sr}$.

The majority of the research in metrology, quantum simulation, and quantum computation with two-electron atoms relies on the $^{1}\mathrm{S}_0$-$^{3}\mathrm{P}_0$ clock transition, which is insensitive to most environmental effects.
This transition is forbidden by spin and angular momentum selection rules.
However, an electric dipole (E1) transition can be enabled by hyperfine mixing~\cite{boyd07} in fermionic isotopes with nuclear spin or by magnetic-field-induced mixing~\cite{taichenachev06} in bosonic isotopes without nuclear spin.
On the clock transition, coherence times of tens of seconds have been experimentally demonstrated~\cite{campbell17,young20}.
This transition forms the basis of state-of-the-art optical lattice clocks~\cite{ludlow15}, and it is a promising qubit candidate for quantum computation schemes~\cite{madjarov20,schine22}.
Accordingly, the clock transition has been thoroughly investigated in strontium~\cite{takamoto05,brusch06,ludlow06,bishof11,nicholson15,riehle18,young20,muniz21,bothwell22,zheng22}, ytterbium~\cite{riehle18,mcgrew18,pizzocaro20,kobayashi22}, and mercury~\cite{tyumenev16,riehle18,ohmae20}.

In contrast to the clock transition, the $^{1}\mathrm{S}_0$-$^{3}\mathrm{P}_2$ transition in bosonic isotopes without nuclear spin can be driven even at zero magnetic field, because it is magnetic-quadrupole (M2) allowed~\cite{mizushima64,mizushima66,garstang67}.
This transition offers several attractive features for quantum science and technology.

Because the $^{1}\mathrm{S}_0$ ground state has zero angular momentum, it has no internal structure in bosonic isotopes with zero nuclear spin and the Clebsch-Gordan coefficients of the transitions out of $^{1}\mathrm{S}_0$ to all Zeeman sublevels of any excited state are equal.
Compared to atoms with more complex ground-state structure, this allows for simpler interpretation of spectroscopic measurements, because of the absence of dark states.
This property makes such isotopes ideal for polarization-insensitive, high-quality spectroscopy and the study of fundamental light-matter interactions, such as collective emission phenomena~\cite{bromley16}.

In addition, the non-vanishing electronic angular momentum of $^{3}\mathrm{P}_2$ allows high flexibility in engineering the atomic polarizability~\cite{ido03,cooper18,norcia18}.
The $^{3}\mathrm{P}_2$ state also possesses a large magnetic moment, allowing control over the excited state's energy with external magnetic fields.
In combination with a natural lifetime of hundreds of seconds~\cite{yasuda04} this tunability provides new opportunities for quantum computing and quantum simulation~\cite{daley08,shibata09}.

For example, the $^{1}\mathrm{S}_0$ state and the $^{3}\mathrm{P}_2$ Zeeman sublevels with $m_J\ne0$ support magnetically-sensitive transitions that can be used for single-site addressing in an optical lattice within a magnetic-field gradient~\cite{daley08,shibata09,kato12,shibata14,yamamoto16}.
Single-site addressing allows the manipulation and readout of quantum states on the level of single atoms and the preparation of a single lattice layer for a quantum gas microscope~\cite{bakr09,sherson10,yamamoto16}.

So far, the $^{1}\mathrm{S}_0$-$^{3}\mathrm{P}_2$ transition has been studied and used in quantum simulation experiments with Yb~\cite{kato12,shibata14,kato13,yamaguchi08,okuno22,yamamoto16}.
Only very recently, a pioneering experiment measured the transition frequency in fermionic $^{87}\mathrm{Sr}$ with an uncertainty of \unit{30}{MHz}~\cite{onishchenko19}.

In this article, we demonstrate the advantages of the ultranarrow $^{1}\mathrm{S}_0$-$^{3}\mathrm{P}_2$ M2 transition for quantum simulation and quantum computing with $^{88}\mathrm{Sr}$.
First, we present a theoretical framework of the selection rules and the transition amplitude's dependence on the light polarization and propagation direction for M2 transitions and compare these results to the well-known E1 transitions.
We experimentally demonstrate this dependence for the $^{1}\mathrm{S}_0$-$^{3}\mathrm{P}_2$ transition in $^{88}\mathrm{Sr}$.
Second, we study the specific properties and applications of the $^{1}\mathrm{S}_0$-$^{3}\mathrm{P}_2$ transition in $^{88}\mathrm{Sr}$.
By engineering the polarizability of the $^{3}\mathrm{P}_2$ state via the trap polarization and the bias magnetic field, we perform Doppler- and Stark-shift-free spectroscopy for magnetic-field-insensitive and magnetic-field-sensitive transitions.
We measure the absolute frequency of the $^{1}\mathrm{S}_0$-$^{3}\mathrm{P}_2$ transition in $^{88}\mathrm{Sr}$ and $^{87}\mathrm{Sr}$, allowing us to extract the $^{87}\mathrm{Sr}$-$^{88}\mathrm{Sr}$ isotope shift.
In a final proof-of-principle experiment, we demonstrate local addressing in an optical lattice with single-site resolution.

\section{Multipole transitions}
\label{sec:hamiltonian}

We are interested in describing the transition amplitude for a general multipole transition between an atomic ground state $|J,m_J\rangle$ and an excited state $|J^\prime,m^\prime_J\rangle$.
For the well-known E1 transitions, several characteristics allow for a simplified description.
Assuming a plane wave, only the photon's \emph{spin} -- its polarization -- determines the change in the atom's orbital angular momentum.
For example, when driving a transition that does not change the atom's magnetic sublevel ($\Delta m_J \equiv m^\prime_J - m_J = 0$), only the $\pi$ polarization component will determine the transition amplitude.
In addition, the amplitude of an E1 transition does not depend on the driving field's propagation direction $\hat{\mathbf{k}}$, beyond the constraints that it imposes on the polarization vector $\hat{\vec{\epsilon}}$.

These simplifications do not generally hold for higher-order multipole transitions, which also allow the transfer of additional \emph{orbital} angular momentum besides the photon spin, even for a plane wave.
Hence, all three polarization components can contribute to driving a certain transition, while the orbital angular momentum transfer satisfies the conservation of total angular momentum.
Since the orbital angular momentum depends on the propagation direction, this direction now explicitly affects the transition amplitude resulting in an \emph{angular dependence}~\cite{hertel14} of the amplitude.

While the theory of atomic multipole transitions is well understood~\cite{johnson07}, the nuances of the contribution of the different polarization components and of the transition amplitude's angular dependence are rarely discussed.
In this section, we present an overview of these dependencies to provide a convenient framework for our experimental work on the $^{1}\mathrm{S}_0$-$^{3}\mathrm{P}_2$ M2 transition.

We begin by reviewing the main results obtained from expanding the light-matter interaction Hamiltonian into multipole orders under simplifying assumptions, described in detail in Appendix~\ref{sec:transition_amplitude}.
Based on these results, we discuss the effects of the polarization and the angular dependence for a general multipole transition.
Next, we focus on two specific multipole orders: the electric dipole, where we show that the transfer of orbital angular momentum and the angular dependence disappear as expected; and the magnetic quadrupole, the scenario of interest for the $^{1}\mathrm{S}_0$-$^{3}\mathrm{P}_2$ transition in two-electron atoms.

\subsection{General theory of multipole transitions}
\label{subsec:hamiltonian1}

Optical atomic transitions result from the coupling of the electromagnetic vector potential $\mathbf{A}$ with the atom's valence electrons and can be described by the minimal-coupling interaction Hamiltonian
\begin{equation}
  \label{eq:light_matter_hamiltonian}
  \begin{aligned}
    H_{\mathrm{int}} &= \sum_{i=1}^N \frac{e}{m_\mathrm{e}}\mathbf{p}_i \cdot
    \mathbf{A}(\mathbf{r}_i)
    + \frac{e g_{\mathrm{s}}}{2 m_\mathrm{e}} \mathbf{s}_i \cdot \bigl[\vec{\nabla}_i \times
    \mathbf{A}(\mathbf{r}_i)\bigr],\\
  \end{aligned}
  \end{equation}
in the Coulomb gauge~\cite{mizushima64,raab75,lamb87,sobelman12}, where $N$ is the number of valence electrons.
Here, $e$ is the elementary charge, $m_{\mathrm{e}}$ is the electron mass, and $g_{\mathrm{s}}\simeq + 2.002$ is the electron spin $g$-factor.
The operators $\mathbf{p}_i$, $\mathbf{s}_i$, $\mathbf{r}_i$, and $\vec{\nabla}_i\times$ are the momentum, spin, position, and curl with respect to the position of the $i$-th valence electron, respectively.
Further notes on Eq.~\eqref{eq:light_matter_hamiltonian} are summarized in Appendix~\ref{sec:transition_amplitude}.

The vector potential at the position of the atom can be expanded in different multipole orders $K$~\cite{johnson07}, where $K=1$ corresponds to the dipole terms, $K=2$ corresponds to the quadrupole terms, and so forth.
In the multipole expansion, the interaction Hamiltonian separates into electric and magnetic parts $H^\mathrm{(el)}_{K,q}$ and $H^\mathrm{(mg)}_{K,q}$, respectively.
These terms contain spherical tensor operators~\cite{edmonds60,varshalovich88} of rank $K$ with components $q=-K,\ldots ,+K$.
The natural frame to describe these tensors is in the basis consisting of the vectors $\hat{\mathbf{e}}_0 = \hat{\mathbf{z}}$ and $\hat{\mathbf{e}}_\pm = \mp(\hat{\mathbf{x}} \pm i \hat{\mathbf{y}})/\sqrt{2}$, corresponding to the atomic-frame polarization components of $\pi$- and $\sigma^\pm$-polarization, respectively.

To obtain the specific form of the interaction Hamiltonian relevant for the experiments described in this work, we assume that the vector potential can be approximated by a plane wave with wave-vector $\mathbf{k}$ and polarization $\hat{\vec{\epsilon}}$, such that $\mathbf{k}\cdot\hat{\vec{\epsilon}} = 0$.
As shown in Appendix~\ref{sec:transition_amplitude}, the multipole decomposition of a plane wave leads to the decomposition of the interaction Hamiltonian
\begin{equation}
  \begin{aligned}
		H_{\mathrm{int}} = 4\pi \sum_{K=1}^\infty  \sum_{q=-K}^K i^K \biggl[ & -i \left( \mathbf{Y}_{K,q}^{\mathrm{(el)}}(\hat{\mathbf{k}}) \cdot \hat{\vec{\epsilon}}\right) H_{K,q}^{\mathrm{(el)}} \\
		& +  \left( \mathbf{Y}_{K,q}^{\mathrm{(mg)}}(\hat{\mathbf{k}}) \cdot \hat{\vec{\epsilon}}\right) H_{K,q}^{\mathrm{(mg)}} \biggr] ,
	\end{aligned}
	\label{eq:hamiltonian_multipole_decomposition}
  \end{equation}
into the multipole terms
\begin{equation}
	\begin{aligned}
        H_{K,q}^{(\lambda)} = \sum_{i=1}^N & \frac{e}{m_{\mathrm{e}}} \mathbf{p}_i \cdot \mathbf{a}_{K,q}^{(\lambda)}(k, \mathbf{r}_i) \\
        & + \frac{e g_s}{2 m_\mathrm{e}} \mathbf{s}_i \cdot \left[\vec{\nabla}_i \times \mathbf{a}_{K,q}^{(\lambda)}(k,\mathbf{r}_i) \right], \\
	\end{aligned}
	\label{eq:individual_hamiltonian_term}
  \end{equation}
corresponding to the electric ($\lambda\to\mathrm{el}$) and magnetic ($\lambda\to\mathrm{mg}$) interactions, respectively.
Here, $\mathbf{a}_{K,q}^{(\lambda)}$ is the coefficient corresponding to component $q$ of the order-$K$ multipole expansion of $\mathbf{A}$, as described in Appendix~\ref{sec:transition_amplitude}.

The multipole operator $H_{K,q}^{(\lambda)}$ describes the light-atom interaction of a multipole transition, including all atomic selection rules and Clebsch-Gordan coefficients.
The Clebsch-Gordan coefficients for a transition between two atomic states $|J,m_J\rangle$ and $|J^\prime,m^\prime_J\rangle$ can be derived from the matrix elements $\langle J^\prime,m^\prime_J|H_{K,q}^{(\lambda)}|J,m_J\rangle$.
Since the multipole operators are irreducible tensor operators~\cite{mizushima66,auzinsh10}, the matrix element can be calculated using the Wigner-Eckart theorem~\cite{edmonds60,sobelman12,hertel14}, leading to the selection rules $|J^\prime-J| \le K \le J^\prime+J$, and $q=\Delta m_J$.

The term $\mathbf{Y}_{K,q}^{(\lambda)}(\hat{\mathbf{k}})\cdot\hat{\vec{\epsilon}}$ describes the transition amplitude's dependence on the light field propagation direction and its polarization, where $\mathbf{Y}_{K,q}^{(\lambda)}(\hat{\mathbf{k}})$ are the vector spherical harmonics (see Appendix~\ref{sec:transition_amplitude})~\cite{edmonds60,varshalovich88,johnson07}. For the remainder of this section, we will focus on this dependence.

By decomposing the term describing the angular dependence into the atomic polarization basis $\hat{\mathbf{e}}_{s}$, we find

\begin{align}
	\mathbf{Y}_{K,q}^{\mathrm{(el)}}(\hat{\mathbf{k}})\cdot\hat{\vec{\epsilon}} &= \sum_{s=-1}^{1}  \Bigl[ c_{K,q,s}^{(-1)} Y_{K-1}^{q-s}(\hat{\mathbf{k}}) \nonumber \\
	& \qquad\quad\enskip + c_{K,q,s}^{(+1)} Y_{K+1}^{q-s}(\hat{\mathbf{k}}) \Bigr] (\hat{\mathbf{e}}_{s}\cdot\hat{\vec{\epsilon}}),
	\label{eq:electric_Y_decomp} \\
	\mathbf{Y}_{K,q}^{\mathrm{(mg)}}(\hat{\mathbf{k}})\cdot\hat{\vec{\epsilon}} &= \sum_{s=-1}^{1} c_{K,q,s}^{(0)} Y_{K}^{q-s}(\hat{\mathbf{k}}) (\hat{\mathbf{e}}_{s}\cdot\hat{\vec{\epsilon}}),
	\label{eq:magnetic_Y_decomp}
\end{align}
where $Y_K^q(\hat{\mathbf{k}})$ are the scalar spherical harmonics as a function of the unit vector $\hat{\mathbf{k}}$. The explicit values of the $c_{K,q,s}^{(j)}$ coefficients for the dipole and quadrupole transitions are given in Appendix~\ref{sec:transition_amplitude}.

From Eqs.~\eqref{eq:electric_Y_decomp} and \eqref{eq:magnetic_Y_decomp}, we observe that, in general, all three polarization components can contribute to the amplitude of a given transition, even if their polarization-determined spin $s$ is different from the change in the electronic angular momentum $q$.
To conserve angular momentum in such cases, the light field transfers $q-s$ quanta of orbital angular momentum along the quantization axis.
We further note the dependence of the transition strength on the direction of $\hat{\mathbf{k}}$ of the light field given by $Y_K^{q-s}(\hat{\mathbf{k}})$.

\subsection{Angular dependence of E1 and M2 transitions}
\label{subsec:hamiltonian2}

\begin{figure}
	\centering
	\includegraphics{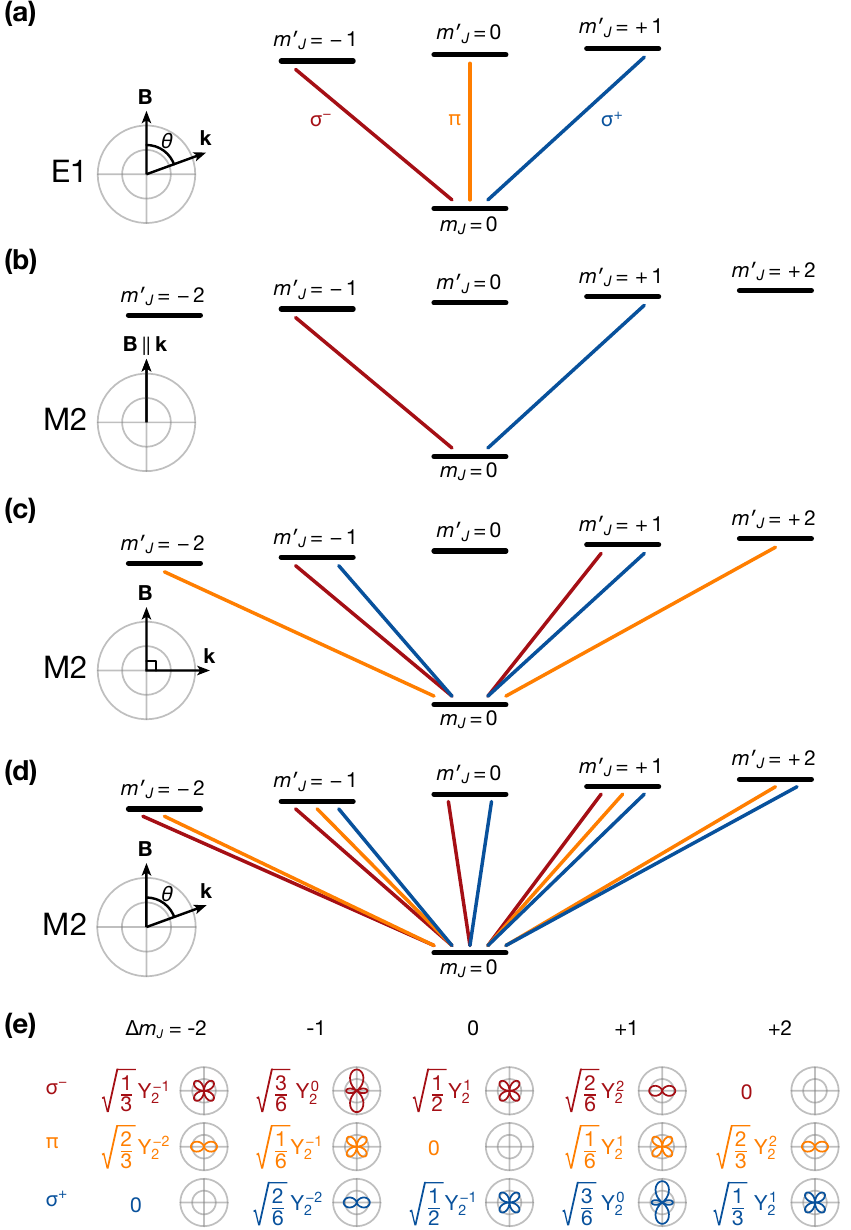}
	\caption{The contributions of the atomic-frame light polarization components of the driving beam to the different $\Delta m_J$ transitions, where the magnetic field $\mathbf{B}$ defines the quantization axis. The full transition amplitude is obtained by decomposing the physical polarization and coherently summing the contributions from all components (phase information not depicted in figure). (a) The case of an E1 transition for $J=0$ and $J^\prime=1$. The $\pi$ and $\sigma^\pm$ polarizations drive only the $\Delta m_J=0$ and $\pm1$ transitions, respectively. The transition amplitude does not depend on the propagation direction beyond the constraints it imposes on the polarization. (b) An M2 transition driven by a probe beam propagating along the quantization axis for $J=0$ and $J^\prime=2$. As for the E1 transition, $\sigma^\pm$ polarizations drive only $\Delta m_J=\pm1$, respectively. (c) As in panel (b), but for propagation perpendicular to the quantization axis. Here both $\sigma^\pm$ contribute to each of the $\Delta m_J=\pm1$ transitions, while the $\pi$ polarization drives only the $\Delta m_J=\pm2$ transitions. (d) As in panel (c), but for an arbitrary relative angle $\theta$ between propagation and quantization axis. Nearly all polarization components can contribute to all $\Delta m_J$ transitions, with the transition amplitudes depending on $\theta$. (e) The angular dependence of the amplitudes of M2 transitions for each polarization and $\Delta m_J$ as a function of $\theta$, as defined in panel (d).} \label{fig:hamiltonian}
\end{figure}

In the following, we specialize the general formalism to two scenarios of interest, the well-known E1 transition and the M2 transition relevant for our experimental studies.
For an E1 transition, we show in Appendix~\ref{sec:transition_amplitude} that the angular dedependence in Eq.~\eqref{eq:electric_Y_decomp} simplifies to
\begin{equation}
\mathbf{Y}_{1,q}^{\mathrm{(el)}}(\hat{\mathbf{k}})\cdot\hat{\vec{\epsilon}} = \sqrt{\frac{3}{8\pi}}(\hat{\mathbf{e}}_{q}\cdot\hat{\vec{\epsilon}}).
\end{equation}
The contributions of the polarization components with $s \ne q$ vanish, such that, \emph{e.g.}, for $q=0$ only the $\pi$ polarization component contributes to the transition amplitude.
The explicit dependence on $\hat{\mathbf{k}}$ also disappears.
Hence, we recover the expected results from the standard dipole interaction Hamiltonian, visualized in Fig.~\ref{fig:hamiltonian}(a) for a $J=0$ to $J^\prime=1$ transition.

Next, we discuss in detail the M2 transition, the scenario of interest for the $^{1}\mathrm{S}_0$-$^{3}\mathrm{P}_2$ transition in $^{88}\mathrm{Sr}$.
For the relevant case of $J=0$ to $J^\prime=2$, all Clebsch-Gordan coefficients are equal, and only the angular dependence determines the relative transition strengths.

First, we can consider the limiting cases of the light field propagating parallel to the quantization axis or perpendicular to the quantization axis.
For parallel propagation, the allowed transitions in Fig.~\ref{fig:hamiltonian}(b) are similar to the ones of an E1 transition.
However, the allowed transitions for perpendicular propagation shown in Fig.~\ref{fig:hamiltonian}(c) are very different.
In this case, $\pi$ polarization only drives the $\Delta m_J=\pm2$ transitions, while the $\Delta m_J=\pm1$ transitions are each driven by both $\sigma^\pm$-polarization components.

The case of an arbitrary propagation direction is visualized in Fig.~\ref{fig:hamiltonian}(d).
For a given $\Delta m_J$, the transition amplitude will generally have contributions from all polarization components such that $|s - \Delta m_J| \le 2$.
This is as expected from a magnetic transition with $K=2$, which can deliver up to $2$ quanta of angular momentum.
A notable exception is the $\Delta m=0$ transition, which cannot be driven with $\pi$-polarized light at all, despite not being forbidden by angular momentum considerations.

The relative magnitude of each polarization component's contribution to the transition amplitude into a specific Zeeman sublevel as a function of the propagation direction is plotted in Fig.~\ref{fig:hamiltonian}(e).
This visualization is meant to be used as follows.
First, decompose the polarization vector into the spherical tensor basis $\hat{\mathbf{e}}_q$~\cite{varshalovich88}.
Next, weight each row of Fig.~\ref{fig:hamiltonian}(e) accordingly, and finally coherently sum over the rows.
The angular dependencies apply to all M2 transitions, regardless of the values of $J$ and $J^\prime$.
For $J\neq 0$, the calculation of the relative transition amplitudes only needs to be extended by taking into account the different Clebsch-Gordan coefficients.
A detailed analysis is presented in Appendix~\ref{sec:transition_amplitude}.

Note that the above derivations are based on plane waves under the assumption of transverse fields ($\mathbf{k}\cdot\hat{\vec{\epsilon}}=0$).
Our derivations can be extended to more general scenarios, such as strongly focused laser beams~\cite{lodahl17} and laser beams carrying orbital angular momentum~\cite{schulz20,schmiegelow12,schmiegelow16}.

\section{Experimental setup}
\label{sec:setup}

High-resolution spectroscopy of an ultranarrow optical transition requires a long interrogation time and the suppression of systematic effects that lead to frequency shifts and line broadening, most notably the Doppler and ac Stark effects.
For this reason, we perform the $^{1}\mathrm{S}_0$-$^{3}\mathrm{P}_2$ spectroscopy using a sample of ultracold strontium atoms trapped in an optical lattice.

For the measurements presented in this work, we prepare a sample of strontium $^{88}\mathrm{Sr}$ in a fast and dense magneto-optical trap~\cite{snigirev19} and then optically transport it into the interrogation region~\cite{park22} using a moving optical lattice~\cite{schmid06}.
In the interrogation region, we adiabatically load the atoms into a vertical, retro-reflected, one-dimensional (1D) optical lattice at \unit{1064}{nm}, sketched in Fig.~\ref{fig:mstates}(a).
We operate at sufficient lattice depth to suppress Doppler broadening in the Lamb-Dicke and resolved-sideband regimes~\cite{blatt09}.
The lattice polarization $\hat{\vec{\epsilon}}_{\mathrm{l}}$ can be dynamically controlled by motorized waveplates, allowing control over the excited-state polarizability, as will be explained in detail in the following sections.
After loading, we cool the atoms to the axial vibrational ground state using direct sideband cooling on the $^{1}\mathrm{S}_0$-$^{3}\mathrm{P}_1$ $\Delta m_{J}=\pm1$ transitions~\cite{park22}.
Typically, we achieve axial temperatures of $\sim$\unit{1.5}{\mu K}, measured with time-of-flight expansion.

To probe the $^{1}\mathrm{S}_0$-$^{3}\mathrm{P}_2$ transition, we interrogate the atoms with the spectroscopy laser beam.
The beam is derived from a diode laser stabilized to a reference cavity with a finesse of ${\sim}25,000$, resulting in a laser linewidth on the order of \unit{1}{kHz}.
The spectroscopy beam propagation direction $\hat{\mathbf{k}}$ is collinear with the vertical lattice $\hat{\mathbf{k}}_\mathrm{l}$ and is focused to a $1/e^{2}$ waist of \unit{{\sim}450}{\mu m} at the position of the atoms.
The spectroscopy beam polarization $\hat{\vec{\epsilon}}$ is linear and fixed throughout all the measurements reported in this work.
Typically, the spectroscopy is performed with a beam power of \unit{25.8}{mW}, and interrogation times range from tens to hundreds of milliseconds.

After applying the spectroscopy pulse, we measure the number $N_g$ of $^{1}\mathrm{S}_0$ ground state atoms with absorption imaging.
Since the $^{3}\mathrm{P}_2$ state has a long natural lifetime~\cite{yasuda04}, atoms excited to $^{3}\mathrm{P}_2$ do not spontaneously decay back to the ground state over the duration of the experiment.
Hence, the spectroscopic signal manifests as a reduction of the $^{1}\mathrm{S}_0$ atom number.
The excited atoms are lost from the trap through inelastic collisions between metastable triplet atoms~\cite{traverso09,yamaguchi08,lisdat09,bishof11,park22}.
To compensate for possible drifts in atom number, we interleave identical experiments, but with an off-resonant spectroscopy pulse.
The atom number $N_0$ measured in these reference measurements is used to normalize the baseline of the measured spectra.

Further details about the experimental setup and sequence can be found in Appendix~\ref{sec:si_setup}.

\section{Quadrupole transition angular dependence}
\label{sec:bfield_angle}

\begin{figure}
  \centering
  \includegraphics{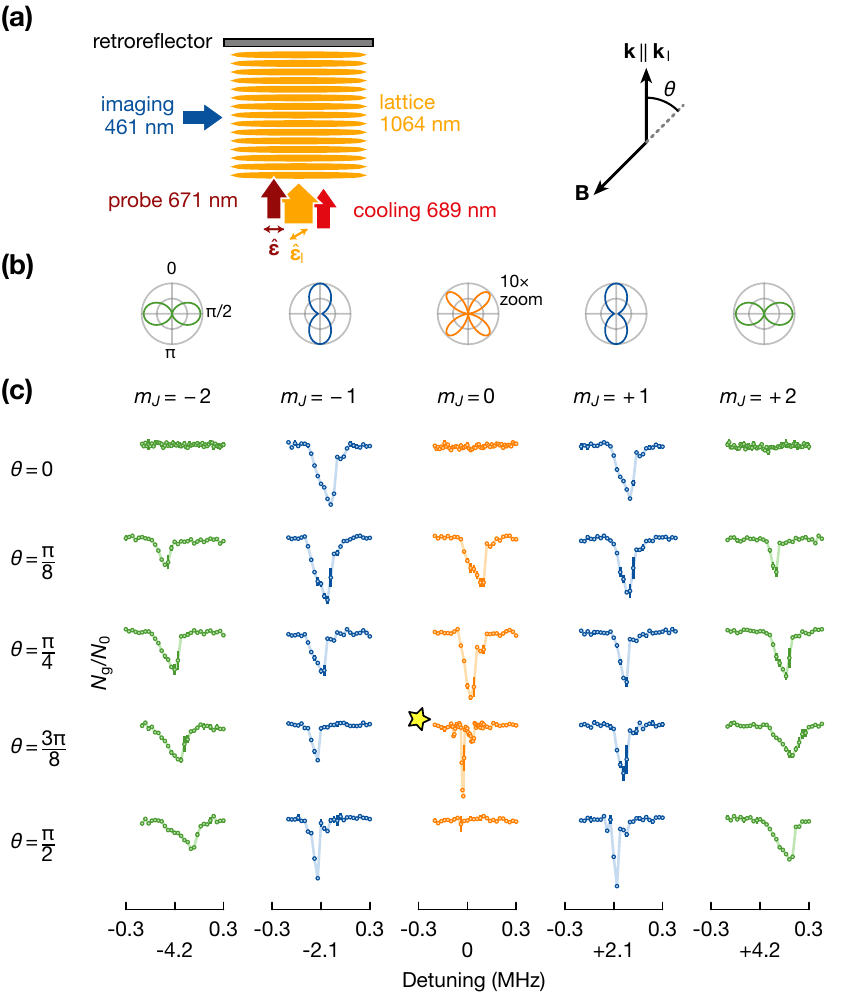}
  \caption{Magnetic quadrupole transition in $^{88}\mathrm{Sr}$. (a) Schematic of the experimental setup, including lattice (probe) wave vector $\hat{\mathbf{k}}_\mathrm{l}$ ($\hat{\mathbf{k}}$) and magnetic field $\mathbf{B}$ tilted at an angle $\theta$. The probe polarization $\hat{\vec{\epsilon}}$ is linear with the major component in the plane spanned by $\hat{\mathbf{k}}_\mathrm{l}$ and $\mathbf{B}$; the lattice polarization  $\hat{\vec{\epsilon}}_\mathrm{l}$ is also linear. (b) Calculated angular dependence of the transition amplitude as a function of $\theta$ for the probe polarization $\hat{\vec{\epsilon}}$ discussed in the main text. (c) Experimental study of the $^{1}\mathrm{S}_0$-$^{3}\mathrm{P}_2$ transition for all $m_{J}$-states as a function of $\theta$. The line strengths of the measured spectra follow the calculated patterns of panel (b), but the spectra are broadened due to the differential ac Stark shift of the $^{1}\mathrm{S}_0$ and the $^{3}\mathrm{P}_2$ $m_{J}$ states. The markers represent an average of 3 measurements, the error bars represent the standard error of the mean, and the solid lines are a  guide to the eye.}
  \label{fig:mstates}
\end{figure}

We experimentally investigate the angular dependence of the $^{1}\mathrm{S}_0$-$^{3}\mathrm{P}_2$ M2 transition by varying the angle $\theta$ between the quantization axis, defined by a bias magnetic field  $\mathbf{B}$, and the probe beam propagation direction $\hat{\mathbf{k}}$.

In practice, we vary $\theta$ by changing the direction of $\mathbf{B}$, while keeping all beam propagation directions and polarizations fixed, as sketched in Fig.~\ref{fig:mstates}(a).
We use a bias field magnitude of \unit{1}{G}, which splits the transition to each magnetic sublevel of the $^{3}\mathrm{P}_2$ state by \unit{2.1}{MHz}, and use a linear lattice polarization $\hat{\vec{\epsilon}}_{\mathrm{l}}$.
The bias field magnitude is a compromise between having a well-defined quantization axis and the requirement to minimize contributions of possible E1 transitions due to state mixing~\cite{taichenachev06}.

We label the plane in which we vary $\mathbf{B}$ in this measurement as the $xz$-plane, where the $z$ is vertical.
In this coordinate system, the probe polarization is oriented $5\degree$ from the $x$ axis in the $xy$-plane.
For this configuration, the $\Delta m_J=0$ transition is expected to be much weaker than the other transitions, since only the small out-of-plane $y$ component contributes to driving the transition (see Appendix~\ref{sec:transition_amplitude}).
To ensure that even weak spectral features, such as the $\Delta m_J=0$ transition for all angles and the $\Delta m_J=\pm1$ transition for $\theta=\pi/2$, will be made visible, we choose a long interrogation time of \unit{500}{ms}.
We note that for an E1 transition, such a geometry would result in a significant transition amplitude for $\Delta m_J=0$ as $\theta$ approaches $\pi/2$, since in the atomic frame the probe beam becomes predominantly $\pi$-polarized.
Therefore, this choice of orientations allows us to discern the existence of possible residual E1 contributions to the transition amplitude.

In Fig.~\ref{fig:mstates}(b), we plot the expected absorption patterns as a function of $\theta$.
The transition amplitude depends on $\theta$ for each polarization, both explicitly as described by Eq.~\eqref{eq:magnetic_Y_decomp} and visualized in Fig.~\ref{fig:hamiltonian}(e), and implicitly through the change of the probe beam's polarization decomposition into the atomic frame.
The experimental spectra of the individual $^{1}\mathrm{S}_0$-$^{3}\mathrm{P}_2$ $\Delta m_{J}$ transitions are shown in Fig.~\ref{fig:mstates}(c).

We observe spectra that are consistent with the expectations based on Fig.~\ref{fig:mstates}(b).
First, for $\theta=0$, we can only drive the transitions to the $\Delta m_{J}=\pm1$ states, as for an E1 transition with the same polarization.
Next, for $0<\theta<\pi/2$, we observe transitions to all excited states, including $\Delta m_{J}=\pm 2$, which are forbidden for an E1 transition.
Finally, for $\theta=\pi/2$, we once again see a complete suppression of the $\Delta m_{J}=0$ transition, while still being able to drive $\Delta m_J=\pm1$ and $\Delta m_J=\pm2$.

We also observe that each of the lines is broadened and shifted by the differential ac Stark shift of the $^{1}\mathrm{S}_0$ and $^{3}\mathrm{P}_2$ states from the trapping lattice, which depends on the $^{3}\mathrm{P}_2$ Zeeman sublevel and $\theta$.
These light shifts complicate a quantitative analysis of the transition strengths in the general case.
Nonetheless, the observed transition strengths qualitatively agree with the calculated angular dependence of the M2 transition amplitude, allowing us to clearly see the unique properties of the M2 transition and how it differs from an E1 transition.

Considering the strong suppression of the $\Delta m_{J}=0$ transition at $\theta=\pi/2$, we conclude that at a magnetic field of \unit{1}{G}, the transition is entirely due to M2 coupling, with a negligible E1 contribution (\textit{e.g.} via state mixing).
In separate measurements, we tested the amplitude of the $\Delta m_{J}=1$ as a function of magnetic field magnitude up to \unit{200}{G} at fixed $\theta=0$, and observed no visible effect.
From these experiments we conclude that the contribution of the E1 coupling due to state mixing is negligible compared to the M2 coupling, for a magnetic field of up to \unit{200}{G}.

In addition, we verified that a magnetic field magnitude of \unit{1}{G} defined the quantization axis sufficiently well for this measurement.
Comparing measurements at fields of \unit{1}{G} and \unit{5}{G} under otherwise identical conditions, we observe no change to the spectral line's properties.

Finally, we note that the spectrum of the $\Delta m_{J}=0$ transition for $\theta=3\pi/8$, marked with a star, is significantly less broadened than the other spectra presented in Fig.~\ref{fig:mstates}(c).
It is sufficiently narrow to observe motional side bands at the trap frequency.
Based on this observation, we engineer in the following sections Stark-shift-free trapping conditions for specific applications.

\section{Magnetic-field-insensitive transition}
\label{sec:magnetic-field-insensitive}

As demonstrated in the previous section, high-precision spectroscopy in optical lattices requires understanding and control of the differential ac Stark shift due to the optical trap.
These differential Stark shifts can be suppressed by engineering the response of the atomic states involved in the transition.
For the $^{1}\mathrm{S}_0$-$^{3}\mathrm{P}_0$ clock transition in Sr, this is typically done by working at the so-called ``magic wavelength'' around \unit{813.4}{nm}~\cite{safronova13,safronova15}.
As we have observed in Fig.~\ref{fig:mstates}, such a magic condition can also be found for the magnetically insensitive $^{1}\mathrm{S}_0$-$^{3}\mathrm{P}_2$ $\Delta m_J = 0$ transition by tilting the magnetic field angle $\theta$.

In this section, we experimentally determine the magic field angle which allows us to measure the absolute frequency of the $^{1}\mathrm{S}_0$-$^{3}\mathrm{P}_2$ transition in bosonic $^{88}\mathrm{Sr}$.
In a complementary measurement, we also find the absolute frequency of the $^{1}\mathrm{S}_0$-$^{3}\mathrm{P}_2$ in  fermionic $^{87}\mathrm{Sr}$ and extract the isotope shift.

\subsection{Tensor polarizability tuning}
\label{subsec:tensor-polarizability}

Stark shifts in high-resolution Doppler-free spectroscopy are well-understood~\cite{lekien13,safronova15,heinz20}.
Here, we briefly summarize the relevant information.

The Stark shift of an atomic level $\Ket{i}$ is proportional to the light intensity and the dynamic dipole polarizability $\alpha_{i}$.
The polarizability can be decomposed into scalar, vector and tensor components~\cite{lekien13, heinz20, cooper18}, with different contributions depending on the Zeeman sublevel and the optical lattice polarization
\begin{equation}
	\label{eq:polarizability}
  \begin{aligned}
    \alpha_{i}(\lambda_{\mathrm{l}}, \beta, \gamma)& =  \alpha_{i}^{\mathrm{s}}(\lambda_{\mathrm{l}}) + \alpha_{i}^{\mathrm{v}}(\lambda_{\mathrm{l}}) \sin(2 \gamma) \frac{m_{J}}{2 J}\\
    & + \alpha_{i}^{\mathrm{t}}(\lambda_{\mathrm{l}})\frac{3 \cos^{2}\beta -1 }{2} \frac{3m_{J}^{2} - J (J+1)}{J (2J -1)}.\\
      \end{aligned}
\end{equation}
The coefficients $\alpha_{i}^{\mathrm{s}}$, $\alpha_{i}^{\mathrm{v}}$ and $\alpha_{i}^{\mathrm{t}}$ of the scalar, vector, and tensor terms, respectively, depend only on the atomic state and the wavelength $\lambda_{\mathrm{l}} = 2\pi/k_\mathrm{l}$ of the optical lattice.
The lattice polarization $\hat{\vec{\epsilon}}_{\mathrm{l}}$ is characterized by the angles $\beta$ and $\gamma$, where $\cos^2\beta=|\hat{\vec{\epsilon}}_{\mathrm{l}}\cdot\hat{\mathbf{B}}|^2$, and $\gamma$ is the ellipticity angle~\cite{rosenbusch09, cooper18} .
We can write any elliptical lattice polarization as $\hat{\vec{\epsilon}}_{\mathrm{l}}=\cos\gamma \hat{\mathbf{e}}_\mathrm{l,1}+i\sin\gamma \hat{\mathbf{e}}_\mathrm{l,2}$, where $\hat{\mathbf{e}}_\mathrm{l,1}$ and $\hat{\mathbf{e}}_\mathrm{l,2}$ are orthogonal basis vectors of the lattice polarization plane perpendicular to $\mathbf{k}_\mathrm{l}$.
For a linear lattice polarization, $\beta$ is the angle between the polarization vector and the quantization axis, and $\gamma=0$.
A more detailed discussion can be found in Appendix~\ref{sec:polarizability}.

The vector and tensor polarizabilities of the $^{1}\mathrm{S}_0$ ground state vanish because $J=0$ and $m_{J}=0$, and hence the ground state polarizability is independent of $\beta$ and $\gamma$.

The results presented in section~\ref{sec:bfield_angle} serve as a good demonstration of tuning the tensor polarizability.
In that measurement, we used a linearly polarized lattice, which resulted in a vanishing vector component also for the $^{3}\mathrm{P}_2$ state.
The tensor polarizability of the $^{3}\mathrm{P}_2$ state does depend on $\theta$, as the rotation of the magnetic field also changes $\beta$, leading to different light shifts for each value of $\theta$ and each $m_{J}$ state.

This variation is most strongly apparent when considering the spectra for the $\Delta m_{J}=0$ transition in Fig.~\ref{fig:mstates}(d).
As previously discussed, the width of this transition changes dramatically as we vary $\theta$, reaching a narrow linewidth at $\theta=3\pi/8$ (marked with a star).

Varying $\theta$ more finely around this value allows one to find the angle for which the differential ac Stark shift vanishes.
Taking into account the precise orientation of the lattice polarization $\hat{\vec{\epsilon}}_{\mathrm{l}}$, we find that the ``magic angle'' for the $^{1}\mathrm{S}_0$-$^{3}\mathrm{P}_2$ $\Delta m_{J}=0$ transition is $\beta_0=0.09(1)\,\pi$, in excellent agreement with the theoretically expected value of $0.089 \, \pi$ (see Appendix~\ref{sec:polarizability}).

\subsection{Spectroscopy of the $\Delta m_J = 0$ transition}
\label{subsec:m0_spectroscopy}

\begin{figure}
	\centering
	\includegraphics{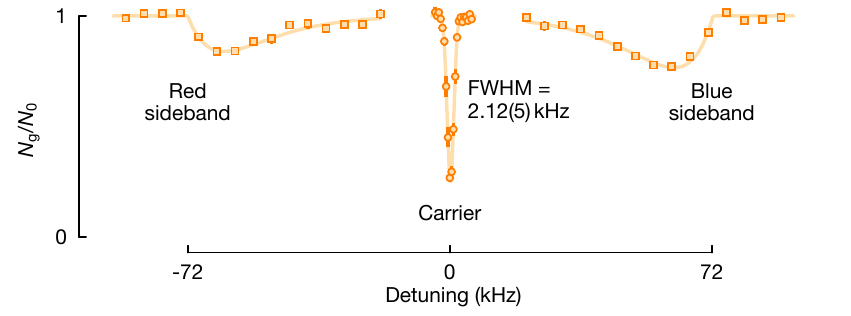}
	\caption{Carrier (circles) and motional sideband (squares) spectrum of the $^{1}\mathrm{S}_0$-$^{3}\mathrm{P}_2$ $\Delta m_J=0$ transition in a Stark-shift-free 1D optical lattice. The sidebands are probed for ten times longer than the carrier. The markers represent an average of 10 measurements, and the error bars are the standard error of the mean. The solid lines are fits, with the carrier fitted to a Gaussian and the sidebands to the sideband line shape discussed in Appendix~\ref{sec:spectroscopy}.}
	\label{fig:spectroscopy}
\end{figure}

To find the absolute transition frequency, we probe the magnetic-field-insensitive $^{1}\mathrm{S}_0$-$^{3}\mathrm{P}_2$ $\Delta m_{J}=0$ transition in a Stark-shift-free lattice.
We enhance the transition amplitude compared to the measurement performed in Sec.~\ref{sec:bfield_angle} by choosing a different orientation for the magnetic field.
  We use a lattice polarization $\hat{\vec{\epsilon}}_\mathrm{l} = \hat{\mathbf{y}}$, and rotate the magnetic field in the $yz$-plane.
In these measurements, we use a bias magnetic field of \unit{5}{G}.

At $\beta_0$, we probe the transition with an interrogation time of \unit{35}{ms} and find the spectrum shown in Fig.~\ref{fig:spectroscopy}.
By fitting the carrier to a Gaussian lineshape, we extract a full-width-at-half-maximum (FWHM) linewidth of \unit{2.12(5)}{kHz}, confirming that our spectroscopy laser allows resolving spectral features on the kHz scale.

Due to the narrow transition linewidth, we can spectroscopically resolve the motional sidebands.
At a detuning corresponding to the blue (red) sideband, atoms are transferred to higher (lower) vibrational states of the lattice.
The sidebands are asymmetrically broadened toward the carrier due to the radial variation of the trap frequency over the spatial extent of the cloud~\cite{blatt09}.
To compensate for the suppressed sideband amplitude in the Lamb-Dicke regime~\cite{leibfried03,ido03,blatt09}, we probe the sidebands for \unit{350}{ms}.
We further enhance the red sideband by working with a hot atomic sample.
The sideband spectrum for a cold sample is shown in Fig.~\ref{fig:si_spectrosocpy} in Appendix~\ref{sec:spectroscopy}.

Finally, we use an optical frequency comb to obtain the absolute frequency of the spectroscopy laser at the carrier resonance conditions.
We find the absolute transition frequency of the $^{1}\mathrm{S}_0$-$^{3}\mathrm{P}_2$ transition in $^{88}\mathrm{Sr}$ to be $446,647,242,704 \pm 0.04_\mathrm{stat} \pm 2_\mathrm{sys}\,\mathrm{kHz}$.

We additionally measure the absolute transition frequency for $^{87}\mathrm{Sr}$.
While the presented light-shift-engineering techniques also enable the realization of a Stark-shift-free lattice for the various states of $^{87}\mathrm{Sr}$, we leave the determination of suitable magic conditions for future work.
Nonetheless, we find the frequency of the $^{1}\mathrm{S}_0$-$^{3}\mathrm{P}_2$ ($F=9/2$) transition in $^{87}\mathrm{Sr}$ to be $446,647,798,443 \pm 5_\mathrm{stat} \pm 40_\mathrm{sys} \mathrm{kHz}$.
Taking into account the known hyperfine structure of $^{87}\mathrm{Sr}$~\cite{heider77}, our measurements lead to an isotope shift $\nu(^{88}\mathrm{Sr})-\nu(^{87}\mathrm{Sr})=+62.91(4)~\mathrm{MHz}$.

A detailed discussion of the spectroscopy and its uncertainties can be found in Appendix~\ref{sec:spectroscopy}.

\section{Magnetic-field-sensitive transition}
\label{sec:magnetic-field-sensitive}
\begin{figure}[t]
	\centering
	\includegraphics{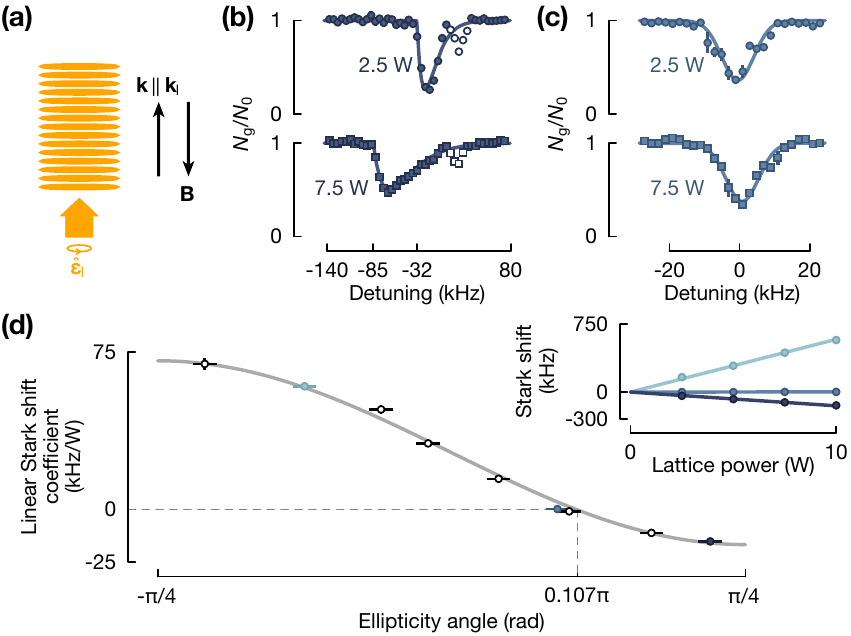}
	\caption{Characterization of the vector ac Stark shift of the $^{1}\mathrm{S}_0$-$^{3}\mathrm{P}_2$ $\Delta m_{J}=1$ transition. (a) Simplified schematic of the experimental setup, emphasizing the differences compared to Fig~\ref{fig:mstates}(a). Here, the magnetic field is oriented vertically, and the lattice polarization is elliptical. (b) Spectra in a lattice of elliptical polarization with an ellipticity angle of $\gamma = 0.17 \pi$ for a lattice power of \unit{2.5}{W} (dots) and \unit{7.5}{W} (squares). The sidebands (open markers) are excluded from the fit. The markers represent an average of 3 measurements, and the error bars are the standard error of the mean. The solid lines are fits to the expected ac-Stark-shift-broadened line shape as discussed in Appendix~\ref{sec:spectroscopy}. (c) As in panel (b), but for $\gamma = 0.09 \pi$. The solid lines are Gaussian fits. (d) Differential ac Stark shift as a function of the ellipticity angle. To extract the magic ellipticity angle $\gamma_{0}$, we fit the data with the function $a_{0} [\sin(2 \gamma) - \sin(2 \gamma_{0})]$ (solid line), see Appendix~\ref{sec:polarizability}. The inset shows an example ac Stark shift measurements for three different elliptical lattice polarizations. We fit the data with a linear function, where the slope is the linear Stark shift coefficient. Vertical error bars represent 1-$\sigma$ uncertainty on the plotted parameter. Horizontal error bars represent the experimental uncertainty on the lattice polarization, discussed in Appendix~\ref{sec:polarizability}.}
	\label{fig:stark_shift}
\end{figure}
Atoms trapped in optical lattices can be addressed and controlled by laser beams in combination with a magnetic field gradient that locally modifies the atomic resonance frequency~\cite{kato12, weitenberg11,yamamoto16}.
This so-called local addressing enables controlling individual atomic qubits in quantum computing schemes~\cite{shibata09, daley11}.
The achievable spatial resolution is only limited by the strength of the gradient and the effective transition linewidth -- taking into account all possible broadening mechanisms -- and thus can allow super-resolution addressing beyond the diffraction limit.

Here, we use this technique to demonstrate how to isolate a single layer of a 1D optical lattice in the focus of a quantum gas microscope~\cite{gross21, shibata14,yamamoto16}.
We aim to address only the atoms in a single lattice site using the magnetic-field-sensitive $^{1}\mathrm{S}_0$-$^{3}\mathrm{P}_2$ $\Delta m_{J}=1$ transition and a magnetic field gradient.
The gradient splits the resonance frequencies of neighboring lattice layers by tens of \unitonly{kHz}, much larger than the several kHz of broadening caused by mG-scale temporal and spatial variations in the magnetic field.
To make the magnetic gradient uniform across the transverse extent of the optical lattice, we combine the gradient with a $\sim$\unit{100}{G} bias field pointing along the lattice.

For this technique to work, we need to engineer the differential ac Stark shift for this transition such that the linewidth becomes smaller than the frequency splitting between neighboring layers.
However, since the bias magnetic field must point along the lattice, we cannot use it to tune the tensor polarizability as in Sec.~\ref{sec:magnetic-field-insensitive}.
Instead, here we use the vector light shift and adjust the ellipticity angle $\gamma$ of the lattice polarization in Eq.~\eqref{eq:polarizability}, until we find a ``magic ellipticity'' $\gamma_{0}$ where the differential ac Stark shift vanishes.

\subsection{Vector polarizability tuning}
\label{subsec:vector-polarizability}

As a first step, we investigate the differential light shift as a function of $\gamma$.
We define the quantization axis by applying a static magnetic field of \unit{20}{G} pointing along the $-z$-axis as indicated in Fig.~\ref{fig:stark_shift}(a).
The wave vectors of the lattice and the probe beam are aligned parallel to the quantization axis.
The lattice has an elliptical polarization described by $\gamma$ in the $xy$-plane.

For a fixed ellipticity, we probe the spectrum of the $^{1}\mathrm{S}_0$-$^{3}\mathrm{P}_2$ $\Delta m_{J}=1$ transition for several lattice powers, as shown in Fig.~\ref{fig:stark_shift}(b).
We observe a shift of the carrier frequency as a function of the lattice power and an asymmetric broadening of the line.
Spectra with an ellipticity close to the Stark-shift-free condition at $\gamma_{0}$ are shown in Fig.~\ref{fig:stark_shift}(c).
The residual linewidth is determined by magnetic field fluctuations on the $10^{-5}$ level.

To study the differential light shift, we extract the carrier frequencies by fitting the spectra with lineshapes discussed in Appendix~\ref{sec:spectroscopy}.
For each value of $\gamma$, we find that the ac Stark shift changes linearly with the lattice power, with some examples plotted in the inset of Fig.~\ref{fig:stark_shift}(d).
We plot the linear Stark shift coefficient as a function of the ellipticity angle $\gamma$ in Fig.~\ref{fig:stark_shift}(d).
By fitting these Stark-shift coefficients, we obtain the magic ellipticity angle $\gamma_{0}=\unit{0.106(3)}{\pi}$.
This value is in excellent agreement with the theoretical expectation of \unit{0.108}{\pi}.
In Appendix~\ref{sec:polarizability}, we show the corresponding polarizability calculations and provide further details on the experiments and their analysis.

\begin{figure}[t]
	\centering
	\includegraphics{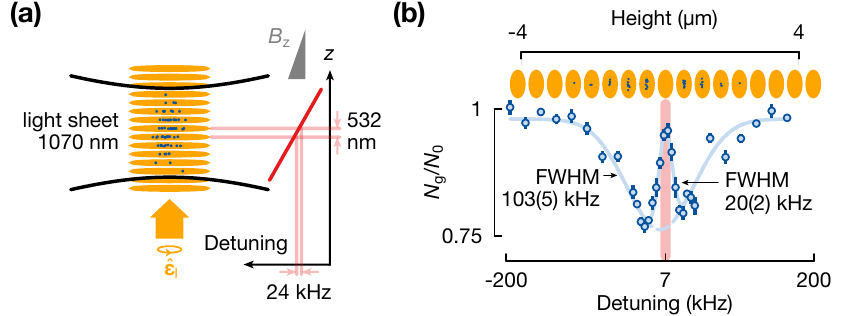}
	\caption{(a) Simplified schematic of the experimental setup, emphasizing the differences compared to Fig~\ref{fig:mstates}(a). We compress the atomic cloud using a light sheet such that the atoms populate only a few lattice sites (left). With a magnetic field gradient of \unit{215}{G/cm} we create a spatially-dependent Zeeman shift (right) that splits neighboring lattice sites by \unit{24}{kHz} on the $^{1}\mathrm{S}_0$-$^{3}\mathrm{P}_2$ $\Delta m_{J} = 1$ transition. (b) Local addressing in an 1D optical lattice using the $^{1}\mathrm{S}_0$-$^{3}\mathrm{P}_2$ transition in a magnetic field gradient. The spectrum is broadened according to the spatial extent of the atomic cloud and the magnetic field gradient. Atoms in the layer that is resonant with the detuning of preparation pulse (red) are depleted prior to the spectroscopy, resulting in a dip in the spectrum at the appropriate detuning. The markers represent an average of 10 measurements, and the error bars are the standard error of the mean. The data is fitted to a broad Gaussian with a narrow Gaussian dip, with the solid (dashed) line representing the fitted curve with (without) the contribution of the narrow dip.}
	\label{fig:hole_burning}
\end{figure}

\subsection{Local addressing}
\label{subsec:local addressing}

Finally, in a proof-of-principle experiment, we demonstrate local addressing on the $^{1}\mathrm{S}_0$-$^{3}\mathrm{P}_2$ $\Delta m_{J} = 1$ transition in a Stark-shift-free optical lattice at $\gamma_{0}$ within a magnetic field gradient.

To reduce the number of initially populated lattice layers and thereby enhance the signal-to-noise ratio of the spectroscopic data, we compress the atomic cloud vertically using a light sheet before loading the atoms into the vertical lattice (see Appendix~\ref{sec:si_setup}).
After this procedure, we obtain a significant atomic population over $\sim$8 lattice sites.
Then, we apply a magnetic field gradient of \unit{215}{G/cm} together with a bias magnetic field of \unit{80}{G} along the $-z$-axis as shown in Fig.~\ref{fig:hole_burning}(a).
This gradient splits the $^{1}\mathrm{S}_0$-$^{3}\mathrm{P}_2$ $\Delta m_{J} = 1$ resonance frequencies of neighboring layers by \unit{24}{kHz}.

Then, we perform a ``spectral hole burning'' measurement.
We address the atoms with a preparation pulse on the $^{1}\mathrm{S}_0$-$^{3}\mathrm{P}_2$ transition, followed by the spectroscopy sequence as in the previous sections.
The atoms that are excited to the $^{3}\mathrm{P}_2$ state by the preparation pulse are quickly lost from the trap due to inelastic collisions, as discussed in Sec.~\ref{sec:setup}.
Therefore, when setting the spectroscopy pulse to the same frequency, we expect not to find any remaining ground-state atoms to excite.
We use \unit{300}{ms}-long and \unit{200}{ms}-long pulses for preparation and spectroscopy, respectively.

Scanning the spectroscopy pulse detuning, we obtain the spectrum shown in Fig.~\ref{fig:hole_burning}(b).
It is inhomogeneously broadened by the magnetic field gradient across the atomic sample to a FWHM $>$\unit{100}{kHz}.
Around the detuning used for the preparation pulse, we observe a dip in the depletion curve.
From this dip we infer that the preparation pulse indeed addressed and depleted the atoms only in a specific layer, while atoms in other layers were not influenced.

We characterize the spatial resolution corresponding to the narrow depletion dip of the cloud with the one-dimensional Rayleigh criterion, which requires the dip between the peaks of two Gaussians drop to 81\% of the maximum~\cite{born13}.
If we apply this criterion to the observed depletion dip, we obtain a spatial resolution of \unit{494(45)}{nm} for the addressing pulse.
This resolution should be sufficient to clearly resolve single lattice layers in the future, for which a more stable mechanical mounting of the lattice retro-reflector is needed.
Our results pave the way to the preparation of a single lattice layer and to local addressing of qubits in quantum computing applications with neutral strontium atoms.

\section{Conclusion}
\label{sec:conclusion}

In this work, we demonstrate that the $^{1}\mathrm{S}_0$-$^{3}\mathrm{P}_2$ magnetic quadrupole transition in neutral strontium can be used as a precise tool for quantum simulation and quantum computation experiments.
We present the first Doppler- and Stark-shift-free optical spectroscopy of this transition in bosonic $^{88}\mathrm{Sr}$ with kilohertz precision.
We suppress differential light shifts by tuning the polarizability of the $^{3}\mathrm{P}_2$ state either by tilting the angle of the magnetic field with respect to the optical lattice or by adjusting the lattice polarization.
We measure the absolute transition frequency to an accuracy of better than \unit{10}{kHz} for $^{88}\mathrm{Sr}$ and to better than \unit{100}{kHz} for $^{87}\mathrm{Sr}$, orders-of-magnitude improvements over previous results~\cite{onishchenko19, sansonetti10}.
In addition, we present a theoretical framework and experimental demonstration of the polarization and propagation-direction dependence of the transition amplitude associated with a magnetic quadrupole transition, and high-order multipole transitions in general.
Finally, in a proof-of-principle experiment, we locally address atoms in an optical lattice with a magnetic field gradient and demonstrate a spatial resolution at the half-micrometer scale.

The tunability of the $^{3}\mathrm{P}_2$ state with external fields and its ultranarrow transition linewidth offer advantages in controlling and manipulating excited state atoms for quantum simulation.
In this work, we took advantage of the tunability of this state's ac Stark shift to generate differential-Stark-shift-free lattices for transitions between the ground state and different Zeeman sublevels of the $^{3}\mathrm{P}_2$ state at a wavelength of \unit{1064}{nm}, where high laser powers are readily available.
This tunability can be taken advantage of also for additional wavelengths and states, as discussed in detail in Appendix~\ref{sec:polarizability}.
Notably, we predict trapping conditions with the same ac Stark shift simultaneously for the $^{1}\mathrm{S}_0$,  $^{3}\mathrm{P}_0$, and $^{3}\mathrm{P}_2$ $m_J=0$ states.

The magnetic field sensitivity of the $^{3}\mathrm{P}_2$ state also allows single-site addressing in an optical lattice within a magnetic field gradient which is impractical with the insensitive $^{3}\mathrm{P}_0$ state.
This addressing enables the isolation of a single layer of an optical lattice required as a preparation step for quantum gas microscope experiments~\cite{sherson10,bakr09, haller15}.
Furthermore, the scattering properties between $^{1}\mathrm{S}_0$ and $^{3}\mathrm{P}_2$ atoms can be tuned with a magnetic field, enabling the search for magnetic Feshbach resonances in strontium, similar to those already observed in ytterbium~\cite{kato13}.

The properties of the magnetically insensitive $^{1}\mathrm{S}_0$-$^{3}\mathrm{P}_2$ $\Delta m_{J}=0$ transition create opportunities in quantum computation with alkaline earth atoms~\cite{daley08,shibata09,okuno22}.
Our experiments demonstrate the first steps towards the full control over this transition, which can also serve as an optical qubit.
One advantage of the M2 transition is that in bosonic isotopes the transition can be driven without applying a large magnetic field~\cite{taichenachev06}.
Recent theoretical proposals~\cite{pagano22} to use the $^{3}\mathrm{P}_0$ and $^{3}\mathrm{P}_2$ $m_{J}=0$ states as a fine-structure qubit for quantum computing highlight the need to investigate the $^{3}\mathrm{P}_2$ state in strontium experimentally.
Being able to excite atoms from $^{1}\mathrm{S}_0$ to both $^{3}\mathrm{P}_2$ and $^{3}\mathrm{P}_0$ states without differential Stark shifts in the same optical lattice paves the way to investigate this fine structure qubit.

\begin{acknowledgments}
  We thank M.\,Safronova for providing the matrix elements required for the polarizability calculation, and T.\,Udem, S.\,Stellmer, B.\,Ohayon, and D.\,DeMille for stimulating discussions.
  This work was supported by funding from the European Union (PASQuanS Grant No.
817482).
  A.\,J.\,P. was supported by a fellowship from the Natural Sciences and Engineering Research Council of Canada (NSERC), funding ref. no. 517029, and V.\,K. was supported by a Hector Fellow Academy fellowship.
\end{acknowledgments}

\appendix

\section{Multipole transition amplitudes}
\label{sec:transition_amplitude}

In this section, we present the derivation of the multipole decomposition of the light-matter interaction Hamiltonian in Eq.~\eqref{eq:light_matter_hamiltonian} that leads to  Eqs.~\eqref{eq:hamiltonian_multipole_decomposition} -- \eqref{eq:magnetic_Y_decomp}.

\subsection*{Light-matter interaction Hamiltonian}

Since the relevant literature on multipole transitions in two-electron atoms dates back to the 1960s, and is sometimes inconsistent in the sign conventions for $g$-factors, or sets $g_\mathrm{s}/2 \approx 1$ without mentioning the sign convention used, we show explicitly the steps in deriving Eq.~\eqref{eq:light_matter_hamiltonian} in the main text.

We start with the interaction between an electromagnetic field and a single electron with charge $q_\mathrm{e} = -e$ and mass $m_\mathrm{e}$, bound in the potential $V(\mathbf{r})$.
The electromagnetic field is described by a vector potential $\mathbf{A}$ and a scalar potential $\Phi$, and the interaction is modelled by the Hamiltonian~\cite{cohen-tannoudji04}
\begin{equation}
  \label{eq:1}
  H = \frac{1}{2 m_\mathrm{e}}[\mathbf{p} - q_\mathrm{e} \mathbf{A}(\mathbf{r})]^2 + q_\mathrm{e} \Phi(\mathbf{r}) + V(\mathbf{r}),
\end{equation}
in SI units.
In the Coulomb gauge, the interaction part of the Hamiltonian in Eq.~\eqref{eq:1} can be written as
the \emph{minimal-coupling} interaction Hamiltonian~\cite{lamb87,loudon00}
\begin{equation}
  \label{eq:3}
  H_\mathrm{int} = - \frac{q_\mathrm{e}}{m_\mathrm{e}} \mathbf{p}\cdot\mathbf{A}(\mathbf{r}) + \frac{q_\mathrm{e}^2}{2 m_\mathrm{e}} A(\mathbf{r})^2.
\end{equation}
Here, we neglect the second-order term in $\mathbf{A}$, but add the Zeeman interaction between the magnetic moment $\vec{\mu}$ of the electron and the magnetic field $\mathbf{B}(\mathbf{r})$ created by the vector potential~\cite{bransden03}
\begin{equation}
  \label{eq:4}
  H_Z = - \vec{\mu}\cdot\mathbf{B}(\mathbf{r}) = - \vec{\mu}\cdot[\vec{\nabla}\times\mathbf{A}(\mathbf{r})],
\end{equation}
where the negative sign ensures that it is energetically favorable for the magnetic moment to align parallel to the magnetic field.
The magnetic moment of the electron is antiparallel to its spin $\mathbf{s}$~\cite{bransden03}, which we write explicitly using a positive electron spin $g$-factor as
\begin{equation}
  \label{eq:5}
  \vec{\mu} = - g_\mathrm{s} \frac{\mu_B}{\hbar} \mathbf{s} = - g_\mathrm{s} \frac{e}{2 m_\mathrm{e}} \mathbf{s},
\end{equation}
where $\mu_B = e\hbar/2m_\mathrm{e}$ is the Bohr magneton, and $h = 2\pi\hbar$ is Planck's constant.
Note that we are free to model the antiparallel alignment of the spin with respect to the magnetic moment by including the minus sign in the $g$-factor or in the sign of the charge, but not both at the same time.
Here, we choose to work with positive charges and $g$-factors and write the signs explicitly.
Putting everything together, we arrive at
\begin{equation}
  \label{eq:6}
  H_\mathrm{int} \simeq + \frac{e}{m_\mathrm{e}} \mathbf{p}\cdot\mathbf{A}(\mathbf{r})
  + \frac{g_\mathrm{s} e}{2 m_\mathrm{e}} \mathbf{s}\cdot[\vec{\nabla}\times\mathbf{A}(\mathbf{r})],
\end{equation}
whose generalization to more than one valence electron is presented in Eq.~\eqref{eq:light_matter_hamiltonian}.

\subsection*{Full multipole Hamiltonian}
We begin the discussion by taking a closer look at the vector potential $\mathbf{A}$.
We approximate the potential as a plane wave propagating along $\mathbf{k}$, such that $\mathbf{A}(\mathbf{r}) = A_0\hat{\vec{\epsilon}} \exp(i \mathbf{k} \cdot \mathbf{r})$, where $\hat{\vec{\epsilon}}$ is the polarization vector, the hat marks a unit vector, and $A_0$ is the amplitude of the vector potential.

A natural approach would be to take advantage of the fact that the size of the atom is small compared to the wavelength of the laser beam, and expand the plane wave in terms of $\mathbf{k} \cdot \mathbf{r}\ll 1$.
However, this procedure turns out to be inconvenient for the separation of the individual electric and magnetic higher-order multipole contributions~\cite{sobelman12}.
Instead, we expand the vector potential in vector spherical harmonics $\mathbf{Y}_{Klq}$~\cite{johnson07}
\begin{equation}
  \begin{aligned}
    \mathbf{Y}_{Klq}(\theta, \phi) =& (-1)^{K-q}\sqrt{2K+1}\sum_{p=-1}^{+1} \\
    &\times \begin{pmatrix}
      K & 1 & l \\
      -q & p & q-p \\
    \end{pmatrix}
    Y_{l}^{q}(\theta, \phi) \hat{\vec{e}}_{p},\\
  \end{aligned}
\end{equation}
where the term in parentheses represents a Wigner-3$j$ symbol.
Then, the vector potential is given by
\begin{equation}
\mathbf{A}(\mathbf{r}) = 
\sum_{Klq} A_{Klq} \mathbf{Y}_{Klq}(\hat{\mathbf{r}}).
\end{equation}
In the literature~\cite{johnson07}, the vector spherical harmonics are typically labeled with $JLM$.
However, we choose the notation $Klq$ to avoid confusion about the labels with the angular momentum of the relevant atomic states.
To simplify the notation, we use the unit vector $\hat{\mathbf{r}}$ as the argument of the vector spherical harmonics to represent the polar angles $\theta_{r}$ and $\phi_{r}$ describing the orientation of $\mathbf{r}$.
Similarly, we use $\hat{\mathbf{k}}$ to represent $\theta$ and $\phi$ below as in the main text.
The expansion coefficients are given by
\begin{equation}
A_{Klq} = A_0\int_{0}^{\pi} \!\! d\theta \sin{\theta} \int_{0}^{2\pi} \!\! d\phi\,\,
\mathbf{Y}_{Klq}(\hat{\mathbf{r}}) \cdot \hat{\vec{\epsilon}}\enspace e^{i \mathbf{k} \cdot \mathbf{r}}.
  \label{eq:Aexpansion-coeff}
\end{equation}
Inserting the expansion of a plane wave in spherical harmonics $Y_{l}^{q}$ and spherical Bessel functions $j_{l}$~\cite{jackson99}
\begin{equation}
e^{i \mathbf{k} \cdot \mathbf{r}} = 4 \pi \sum_{l,q} i^{l} j_{l} (kr) Y_{l}^{q*}(\hat{\mathbf{k}}) Y_{l}^{q}(\hat{\mathbf{r}}),
\end{equation}
we solve the integral in Eqn.~\eqref{eq:Aexpansion-coeff} and obtain~\cite{johnson07}
\begin{equation}
\mathbf{A}(\mathbf{r})  = 4 \pi A_0\sum_{K,l,q} i^{l}(\mathbf{Y}_{Klq}(\hat{\mathbf{k}}) \cdot \hat{\vec{\epsilon}}) j_{l} (kr) \mathbf{Y}_{Klq}(\hat{\mathbf{r}}).
\end{equation}

\begin{widetext}
\noindent Before we continue, we define another, more convenient, set of vector spherical harmonics $\mathbf{Y}_{Kq}^{(\lambda)}$~\cite{johnson07}
  \begin{equation}
    \begin{aligned}
      \mathbf{Y}_{Kq}^{(-1)}(\hat{\mathbf{r}}) &= \sqrt{\frac{K}{2K+1}}\mathbf{Y}_{KK-1q}(\hat{\mathbf{r}}) - \sqrt{\frac{K+1}{2K+1}}\mathbf{Y}_{KK+1q}(\hat{\mathbf{r}})
      = \frac{\mathbf{r}}{r} Y_K^q(\hat{\mathbf{r}}),\\
      \mathbf{Y}_{Kq}^{(0)}(\hat{\mathbf{r}}) & = \mathbf{Y}_{KKq}(\hat{\mathbf{r}})
      = \frac{1}{\sqrt{K(K+1)}} (-i \mathbf{r}\times\vec{\nabla}) Y_K^q(\hat{\mathbf{r}}), \\
      \mathbf{Y}_{Kq}^{(1)}(\hat{\mathbf{r}}) & = \sqrt{\frac{K+1}{2K+1}}\mathbf{Y}_{KK-1q}(\hat{\mathbf{r}}) + \sqrt{\frac{K}{2K+1}}\mathbf{Y}_{KK+1q}(\hat{\mathbf{r}})
      = \frac{r}{\sqrt{K(K+1)}} \vec{\nabla} Y_K^q(\hat{\mathbf{r}}). \\
    \end{aligned}
  \end{equation}
  Using $\mathbf{Y}_{Kq}^{(\lambda)}$, the expansion of $\mathbf{A}(\mathbf{r})$ becomes~\cite{johnson07}
  \begin{equation}
    \mathbf{A}(\mathbf{r})  = 4 \pi \sum_{Kq\lambda} i^{K-\lambda}(\mathbf{Y}_{Kq}^{(\lambda)} (\hat{\mathbf{k}}) \cdot \hat{\vec{\epsilon}})\,\, \mathbf{a}_{Kq}^{(\lambda)}(\hat{\mathbf{r}}),
    \label{eq:multipole_exp}
  \end{equation}
  with
  \begin{equation}
    \begin{aligned}
      \mathbf{a}_{Kq}^{(0)}(\hat{\mathbf{r}}) =& A_0 j_{K}(kr)\mathbf{Y}_{Kq}^{(0)}(\hat{\mathbf{r}}),\\
      \mathbf{a}_{Kq}^{(1)}(\hat{\mathbf{r}}) =& A_0 \left[\sqrt{\frac{K+1}{2K+1}}j_{K-1}(kr) \mathbf{Y}_{KK-1q}(\hat{\mathbf{r}})
      - \sqrt{\frac{K}{2K+1}}j_{K+1}(kr) \mathbf{Y}_{KK+1q}(\hat{\mathbf{r}})\right]. \\
    \end{aligned}
  \end{equation}
  We identify the terms with $\lambda=0$ and $\lambda=1$ as the magnetic and electric multipole components, respectively~\cite{johnson07}.
  This allows us to write $\lambda=0$ (1) or $\lambda=\textrm{mg}$ ($\textrm{el}$), interchangeably.
  The term $\mathbf{Y}_{Kq}^{(-1)}$ is parallel to $\mathbf{k}$, and thus its contribution vanishes under the assumption of $\mathbf{k}\cdot\hat{\vec{\epsilon}}=0$.
  Using the expansion of the vector potential, the light-matter interaction Hamiltonian in the Coulomb gauge becomes
  \begin{equation}
    \begin{aligned}
      H_{\mathrm{int}} & = 4 \pi \sum_{\lambda\in\{0,1\}} \sum_{K=1}^\infty \sum_{q=-K}^{K}
      i^{K-\lambda}
      \left(\mathbf{Y}_{Kq}^{(\lambda)} (\hat{\mathbf{k}}) \cdot \hat{\vec{\epsilon}}\right)
      \sum_{i=1}^N \left\{ \frac{e}{m_{\mathrm{e}}}
        \mathbf{p}_i \cdot \mathbf{a}_{Kq}^{(\lambda)}(\hat{\mathbf{r}}_i)
        + \frac{e g_\mathrm{s}}{2 m_\mathrm{e}}
        \mathbf{s}_i \cdot \left[ \vec{\nabla}_i \times
          \mathbf{a}_{Kq}^{(\lambda)}(\hat{\mathbf{r}}_i) \right] \right\} \\
      & \equiv 4 \pi \sum_{\lambda\in\{0,1\}} \sum_{K=1}^\infty \sum_{q=-K}^{K} i^{K-\lambda}\left(\mathbf{Y}_{Kq}^{(\lambda)} (\hat{\mathbf{k}}) \cdot \hat{\vec{\epsilon}}\right) H_{K,q}^{(\lambda)}, \\
    \end{aligned}
  \end{equation}
  as presented in the main text.
\end{widetext}

\subsection*{Atomic transition terms}

\paragraph*{General expression. --- }
  We can write the explicit expressions for the electric and magnetic contributions as~\cite{mizushima64}
\begin{align}
  H_{K,q}^{\mathrm{(el)}} & = A_0 b_K k^K Q_{K,q}^{\mathrm{(el)}}(\vec{r}_1, \ldots, \vec{r}_N), \\
  H_{K,q}^{\mathrm{(mg)}} & = A_0 b_K k^K Q_{K,q}^{\mathrm{(mg)}}(\vec{r}_1, \ldots, \vec{r}_N),
\end{align}
where the proportionality factors are chosen to be consistent with the literature on multipole transitions~\cite{mizushima64,mizushima66,raab75}
\begin{equation}
  b_K = \sqrt{\frac{(2K+1)(K+1)}{4 \pi K}} \frac{1}{(2K+1)!!},
\end{equation}
and we emphasize that the multipole transition operators $Q_{K,q}^{(\lambda)}$ are functions of all valence electron positions $\vec{r}_i$.
We find
\begin{align}
	Q_{K,q}^{\mathrm{(el)}} =& e \sqrt{\frac{4 \pi}{(2K+1)}} \sum_{i=1}^N r_i^{K} Y_{K}^{q}(\hat{\mathbf{r}}_i), \label{eq:ele_mult_transition_operator}\\
	Q_{K,q}^{\mathrm{(mg)}} =& \frac{e}{m_\mathrm{e}} \sqrt{\frac{4 \pi}{(2K+1)}}\label{eq:mg_mult_transition_operator} \times \\
    &\sum_{i=1}^N\biggl[\vec{\nabla}_i \left[ r_i^{K} Y_{K}^{q} (\hat{\mathbf{r}}_i) \right]
	\left( \frac{1}{K+1} \mathbf{l}_i + \frac{g_{\mathrm{s}}}{2} \mathbf{s}_i \right)\biggr].
    \nonumber
\end{align}
Here, $\mathbf{l}_i$ is the orbital angular momentum operator of the $i$-th valence electron.
In contrast to older literature~\cite{blatt79}, we include $\hbar$ in both orbital angular momentum and spin operators.

\paragraph*{Selection rules and Clebsch-Gordan coefficients. --- }
Let us now consider an attempt to drive the atom from the state $| i\rangle=|\gamma,J,m_J\rangle$ to the state $| f\rangle=|\gamma^\prime,J^\prime,m_J^\prime\rangle$.
Here, $\gamma$ represents the state's radial wavefunction, $J$ is its total angular momentum, and $m_J$ is the projection of this angular momentum onto the quantization axis.
Since $Q_{K,q}^{\mathrm{(el)}}$ and $Q_{K,q}^{\mathrm{(mg)}}$ are irreducible tensor operators of rank $K$~\cite{mizushima66,auzinsh10}, we can use the Wigner-Eckart theorem~\cite{edmonds60,sobelman12,hertel14} to obtain
\begin{equation}
  \begin{aligned}
    \langle f | Q_{K,q}^{(\lambda)} | i\rangle =& (-1)^{ J^{\prime}-m^{\prime}}
    \begin{pmatrix}
      J^{\prime} & K & J \\
      -m^{\prime} & q & m \\
    \end{pmatrix} \\
	&\times \langle\gamma^{\prime}, J^{\prime}\| Q_{K}^{(\lambda)}\| \gamma, J\rangle, \\
  \end{aligned}
\end{equation}
where $\langle\gamma^{\prime}, J^{\prime}\| Q_{K}^{(\lambda)}\| \gamma, J\rangle$ is the reduced matrix element.
This procedure leads to the standard angular momentum selection rules tying $(K,q)$ to $(J,J^\prime,m_J,m_J^\prime)$ as described in the main text.

Taking into account the conservation of parity leads to additional selection rules.
Inverting the spatial coordinates of all electrons ($\mathbf{r}_i\rightarrow -\mathbf{r}_i$) in Eqs.~\eqref{eq:ele_mult_transition_operator} and \eqref{eq:mg_mult_transition_operator}, we find that the electric multipole operator $Q_{K,q}^{\mathrm{(el)}}$ has a parity of $(-1)^K$, while the magnetic multipole operator $Q_{K,q}^{\mathrm{(mg)}}$ has a parity of $(-1)^{K+1}$.
In other words, states with identical parity can be coupled by electric transitions with even $K$ and magnetic transitions with odd $K$; conversely, states with opposite parity can be coupled by electric transitions with odd $K$ and magnetic transitions with even $K$.

\paragraph*{Transitions in two-electron atoms. ---}
We now apply these rules to understand the important transitions from the ground-state, shown in Fig.~\ref{fig:level-scheme}, in two-electron atoms without nuclear spin, such as $^{88}\mathrm{Sr}$.

We note that the $^{1}\mathrm{S}_0$ state has $J=0$, and thus only the terms with $K=J'$ can contribute.
Since $K \ge 1$, the $^{1}\mathrm{S}_0$-$^{3}\mathrm{P}_0$ transition cannot be driven at all.
For $^{1}\mathrm{P}_1$ and $^{3}\mathrm{P}_1$, we can have only a dipole transition ($K=1$).
Because the parity of the ground and excited states is opposite, it has to be an E1 transition.
This is indeed the case for the $^{1}\mathrm{S}_0$-$^{1}\mathrm{P}_1$ transition, which is a strong E1-allowed transition.
In contrast, the $^{1}\mathrm{S}_0$-$^{3}\mathrm{P}_1$ transition is forbidden by another selection rule arising from Eq.~\eqref{eq:ele_mult_transition_operator}: the electric transition operators $Q_{K,q}^{\mathrm{(el)}}$ do not couple to the electronic spin and thus cannot change the spin quantum number.

The $^{1}\mathrm{S}_0$-$^{3}\mathrm{P}_1$ and $^{1}\mathrm{S}_0$-$^{3}\mathrm{P}_0$ transitions are only allowed due to state mixing in heavy atoms: the bare LS-coupling $^3\mathrm{P}_1^\circ$ state is mixed with the bare $^1\mathrm{P}_1^\circ$ state through LS-coupling violation~\cite{boyd07}, and the bare $^3\mathrm{P}_0^\circ$ state can be mixed with the $^{3}\mathrm{P}_1$ state by applying external magnetic fields~\cite{taichenachev06} or via hyperfine coupling in the fermionic isotope~\cite{boyd07}.
State-mixing effects relate to the Hamitonian of the unperturbed atom, rather than to the interaction Hamiltonian, and thus they are outside the scope of the derivations presented here.
In summary, the $^{1}\mathrm{S}_0$-$^{3}\mathrm{P}_1$ and $^{1}\mathrm{S}_0$-$^{3}\mathrm{P}_0$ transitions are fundamentally related to the $^{1}\mathrm{S}_0$-$^{1}\mathrm{P}_1$ transition, and thus have E1 characteristics.

Applying the same considerations to the $^{1}\mathrm{S}_0$-$^{3}\mathrm{P}_2$ transition, the transition must have a quadrupole character ($K=2$), and from parity considerations it must be a magnetic quadrupole transition.
Unlike the previous cases, this transition can be driven directly, even without considering state mixing.
However, as for the $^{1}\mathrm{S}_0$-$^{3}\mathrm{P}_0$ transition, it is possible to induce an E1 contribution by an external magnetic field.
As discussed in the main text, we find that for fields up to \unit{200}{G} this effect is negligible, and the transition can be considered to be entirely M2.

\subsection*{Angular dependence terms}
\begin{table}[h]
	\centering
	\caption{Angular dependence coefficients $c_{K,q,s}^{(j)}$ in Eqs.~\eqref{eq:electric_Y_decomp} and \eqref{eq:magnetic_Y_decomp} for electric dipole (E1), magnetic dipole (M1), electric quadrupole (E2), and magnetic quadrupole (M2) transitions.}
	\begin{tabularx}{\columnwidth}{
      @{}
      >{\RaggedLeft}X
      @{}
      >{\setlength\hsize{0.165\columnwidth}\RaggedLeft}X
      @{}
      >{\setlength\hsize{0.165\columnwidth}\RaggedLeft}X
      @{}
      >{\setlength\hsize{0.165\columnwidth}\RaggedLeft}X
      @{}
      >{\setlength\hsize{0.165\columnwidth}\RaggedLeft}X
      @{}
      >{\setlength\hsize{0.165\columnwidth}\RaggedLeft}X
      @{}}
		\hline \hline \\
		& &  & $s=-1$ & $s=0$ & $s=+1$\\
		& &  & $\sigma^{-}$ & $\pi$ & $\sigma^{+}$\\
		\hline \\
		\multicolumn{6}{c}{E1}\\
		\ldelim\{{4}{3mm} & $q=-1$ & $c_{1,-1,s}^{(-1)}$ & $\sqrt{\frac{20}{30}}$ & $0$ & $0$\\
		$Y_{0}^{q-s}$\hspace*{5mm} & $q=0$ & $c_{1,0,s}^{(-1)}$ & $0$ & $\sqrt{\frac{20}{30}}$ & $0$\\
		& $q=1$ & $c_{1,1,s}^{(-1)}$ & $0$ & $0$ & $\sqrt{\frac{20}{30}}$\\[1ex]
		\ldelim\{{4}{3mm} & $q=-1$ & $c_{1,-1,s}^{(+1)}$ & $\sqrt{\frac{1}{30}}$ & $-\sqrt{\frac{3}{30}}$ & $\sqrt{\frac{6}{30}}$\\
		$Y_{2}^{q-s}$\hspace*{5mm} & $q=0$ & $c_{1,0,s}^{(+1)}$ & $-\sqrt{\frac{3}{30}}$ & $-\sqrt{\frac{4}{30}}$ & $\sqrt{\frac{3}{30}}$\\
		& $q=1$ & $c_{1,1,s}^{(+1)}$ & $\sqrt{\frac{6}{30}}$ & $-\sqrt{\frac{3}{30}}$ & $\sqrt{\frac{1}{30}}$\\[3ex]
		\multicolumn{6}{c}{M1}\\
		\ldelim\{{4}{3mm} & $q=-1$ & $c_{1,-1,s}^{(0)}$ & $\sqrt{\frac{1}{2}}$ & $-\sqrt{\frac{1}{2}}$ & $0$\\
		$Y_{1}^{q-s}$\hspace*{5mm} & $q=0$ & $c_{1,0,s}^{(0)}$ & $\sqrt{\frac{1}{2}}$ & $0$ & $-\sqrt{\frac{1}{2}}$\\
		& $q=1$ & $c_{1,1,s}^{(0)}$ & $0$ & $\sqrt{\frac{1}{2}}$ & $-\sqrt{\frac{1}{2}}$\\[4ex]
		\multicolumn{6}{c}{E2}\\
		\ldelim\{{7.5}{3mm} & $q=-2$ & $c_{2,-2,s}^{(-1)}$ & $\sqrt{\frac{3}{5}}$ & $0$ & $0$\\
		& $q=-1$ & $c_{2,-1,s}^{(-1)}$ & $\sqrt{\frac{3}{10}}$ & $\sqrt{\frac{3}{10}}$ & $0$\\
		$Y_{1}^{q-s}$\hspace*{5mm} & $q=0$ & $c_{2,0,s}^{(-1)}$ & $\sqrt{\frac{1}{10}}$ & $\sqrt{\frac{2}{5}}$ & $\sqrt{\frac{1}{10}}$\\
		& $q=1$ & $c_{2,1,s}^{(-1)}$ & $0$ & $\sqrt{\frac{3}{10}}$ & $\sqrt{\frac{3}{10}}$\\
		& $q=2$ & $c_{2,2,s}^{(-1)}$ & $0$ & $0$ & $\sqrt{\frac{3}{5}}$\\[1ex]
		\ldelim\{{7.5}{3mm} & $q=-2$ & $c_{2,-2,s}^{(+1)}$ & $\sqrt{\frac{2}{105}}$ & $-\sqrt{\frac{2}{21}}$ & $\sqrt{\frac{2}{7}}$\\
		& $q=-1$ & $c_{2,-1,s}^{(+1)}$ & $\sqrt{\frac{2}{35}}$ & $-\sqrt{\frac{16}{105}}$ & $\sqrt{\frac{4}{21}}$\\
		$Y_{3}^{q-s}$\hspace*{5mm} & $q=0$ & $c_{2,0,s}^{(+1)}$ & $\sqrt{\frac{4}{35}}$ & $-\sqrt{\frac{6}{35}}$ & $\sqrt{\frac{4}{35}}$\\
		& $q=1$ & $c_{2,1,s}^{(+1)}$ & $\sqrt{\frac{4}{21}}$ & $-\sqrt{\frac{16}{105}}$ & $\sqrt{\frac{2}{35}}$\\
		& $q=2$ & $c_{2,2,s}^{(+1)}$ & $\sqrt{\frac{2}{7}}$ & $-\sqrt{\frac{2}{21}}$ & $\sqrt{\frac{2}{105}}$\\[3ex]
		\multicolumn{6}{c}{M2}\\
		\ldelim\{{7.5}{3mm} & $q=-2$ & $c_{2,-2,s}^{(0)}$ & $\sqrt{\frac{1}{3}}$ & $-\sqrt{\frac{2}{3}}$ & $0$\\
		& $q=-1$ & $c_{2,-1,s}^{(0)}$ & $\sqrt{\frac{3}{6}}$ & $-\sqrt{\frac{1}{6}}$ & $-\sqrt{\frac{2}{6}}$\\
		$Y_{2}^{q-s}$\hspace*{5mm} & $q=0$ & $c_{2,0,s}^{(0)}$ & $\sqrt{\frac{1}{2}}$ & $0$ & $-\sqrt{\frac{1}{2}}$\\
		& $q=1$ & $c_{2,1,s}^{(0)}$ & $\sqrt{\frac{2}{6}}$ & $\sqrt{\frac{1}{6}}$ & $-\sqrt{\frac{3}{6}}$\\
		& $q=2$ & $c_{2,2,s}^{(-1)}$ & $0$ & $\sqrt{\frac{2}{3}}$ & $-\sqrt{\frac{1}{3}}$\\
		\hline \hline
	\end{tabularx}

	\label{tab:c_coeff}
\end{table}

We now discuss the transition amplitude's angular dependence described by the terms $\mathbf{Y}_{K,q}^{(\lambda)}(\hat{\mathbf{k}})\cdot\hat{\vec{\epsilon}}$.
Substituting the definitions of the vector spherical harmonics presented above into this expression, we obtain the expressions presented in Eqs.~\eqref{eq:electric_Y_decomp} and \eqref{eq:magnetic_Y_decomp} in the main text.

In Tab.~\ref{tab:c_coeff}, we present the explicit values of the decomposition coefficients $c_{K,q,s}^{(j)}$ for the first two orders corresponding to E1, M1, E2, and M2 transitions.

\begin{figure}
	\includegraphics{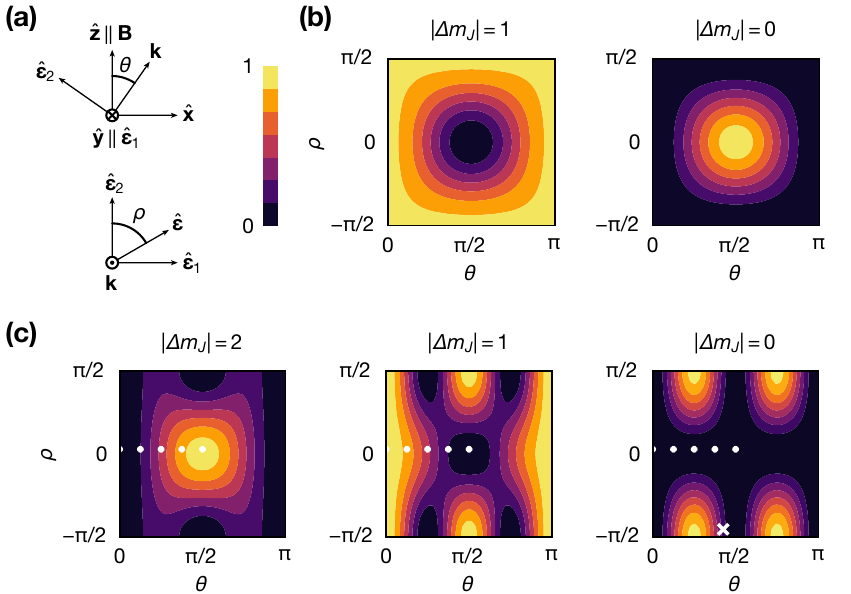}
	\caption{(a) The coordinate system defined in this section. $\mathbf{B}$ sets the $z$ axis, $\mathbf{k}$ and $\hat{\vec{\epsilon}}_2$ are in the $xz$ plane,  $\hat{\vec{\epsilon}}_1$ is along the $y$ axis. The polarization vector $\hat{\vec{\epsilon}}$ is found in the plane spanned by $\hat{\vec{\epsilon}}_{1}$ and $\hat{\vec{\epsilon}}_{2}$. (b) The total angular dependence of an E1 transition for linear polarization as a function of $\theta$ and $\rho$ as defined in panel (a), for the different possible $|\Delta m_J|$ values that this transition supports (since it is linear, it is independent of the sign of $\Delta m_J$). The dependence we observe is entirely due to the decomposition into the basis vectors. (c) As in panel (b), but for an M2 transition. Here, the dependence is more complicated and cannot be entirely attributed to the polarization decomposition. The dots mark the $(\theta,\rho)$ pairs corresponding to the measurements in Sec.~\ref{sec:bfield_angle}. The cross marks the values corresponding to the measurement in Sec.~\ref{sec:magnetic-field-insensitive}. The measurements in Sec.~\ref{sec:magnetic-field-sensitive} were performed for $\theta=0$.}
	\label{fig:si-angular-dependence}
\end{figure}

\paragraph*{Coordinate system. ---}
Here, we define the natural coordinate system used to calculate the transition amplitude's angular dependence.
This coordinate system is illustrated in Fig.~\ref{fig:si-angular-dependence}(a) and begins with
the $z$ axis that is already set by our choice of the quantization axis.
We find that choosing the direction of $x$ such that $\mathbf{k}$ is contained in the $xz$ plane is convenient, since then the spherical harmonics $Y_{K}^{q}(\theta,\phi=0)$ are real.
This choice results in the fact that explicitly complex terms only occur in the polarization decomposition terms $\hat{\mathbf{e}}_{s}\cdot\hat{\vec{\epsilon}}$.
Next, we decompose a linear polarization vector into the out-of-plane $\hat{\vec{\epsilon}}_1$ component (along $\hat{\mathbf{y}}$) and the in-plane component $\hat{\vec{\epsilon}}_2$ (in the $xz$ plane).
The orientation of $\mathbf{k}$ in the $xz$ plane is defined by $\theta$, and the orientation of $\hat{\vec{\epsilon}}$ in the plane spanned by $\hat{\vec{\epsilon}}_1$ and $\hat{\vec{\epsilon}}_2$ is defined by $\rho$.
Explicitly, we find
\begin{equation}
  \begin{aligned}
    \mathbf{B} & = B \hat{\mathbf{z}}, \\
    \mathbf{k} & = k (\cos{\theta}\hat{\mathbf{z}} + \sin{\theta}\hat{\mathbf{x}}), \\
    \hat{\vec{\epsilon}}_1 & = \hat{\mathbf{y}}, \\
    \hat{\vec{\epsilon}}_2 & =  \sin{\theta}\hat{\mathbf{z}} - \cos{\theta}\hat{\mathbf{x}}, \\
    \hat{\vec{\epsilon}} & = \sin{\rho}\hat{\vec{\epsilon}}_1 + \cos{\rho}\hat{\vec{\epsilon}}_2. \\
  \end{aligned}
\end{equation}

\paragraph*{Simplification of E1 transition amplitude's angular dependence. ---}
Considering the results in Tab.~\ref{tab:c_coeff}, it appears at first glance as if the behavior of the E1 transition is more complex than expected, with possible contributions from all polarization components to each transition.
However, we find that we can simplify the expressions to obtain the well-known picture presented in Fig.~\ref{fig:hamiltonian}(a).

Let us consider the scenario for a general polarization vector $\hat{\vec{\epsilon}}=\alpha\hat{\vec{\epsilon}}_1+\beta\hat{\vec{\epsilon}}_2$, such that $\alpha$ and $\beta$ are complex coefficients satisfying $|\alpha|^2+|\beta|^2=1$ (the Jones vector).
Then, the angular dependence is given by
\begin{align}
	|\mathbf{Y}_{1, 0} \cdot \hat{\vec{\epsilon}}|^{2} &=
	\frac{3}{8 \pi} \beta \beta^{*} \sin^{2}\theta, \\
	|\mathbf{Y}_{1, \pm1} \cdot \hat{\vec{\epsilon}} |^{2} &=
	\frac{3}{16 \pi} \left( \alpha \pm i \beta \cos \theta \right) \left( \alpha^{*} \mp i \beta^{*} \cos \theta \right).
\end{align}
However, we notice that
\begin{align}
	|\hat{\mathbf{e}}_{0}\cdot\hat{\vec{\epsilon}}|^{2} &=
	\beta \beta^{*} \sin^{2}\theta,
	\\
	|\hat{\mathbf{e}}_{\pm1}\cdot\hat{\vec{\epsilon}}|^{2} &=
	\frac{1}{2} \left( \alpha \pm i \beta \cos \theta \right) \left( \alpha^{*} \mp i \beta^{*} \cos \theta \right).
\end{align}
Hence, we can substitute the full expression for $\mathbf{Y}_{1, 0} \cdot \hat{\vec{\epsilon}}$ with $\sqrt{3/(8\pi)}\hat{\mathbf{e}}_{q}\cdot\hat{\vec{\epsilon}}$, and thus are justified to consider the E1 transition amplitude's angular dependence as presented in Fig.~\ref{fig:hamiltonian}(a).

\paragraph*{Complete transition amplitude's angular dependence. ---}

In Fig.~\ref{fig:si-angular-dependence}, we plot the complete transition amplitude's angular dependence, assuming a linear polarization, for E1 and M2 transitions in panels (b) and (c), respectively.
The complete angular dependence includes, on top of the explicit angular dependence discussed above, also the trivial dependence due to the decomposition into the polarization basis.
In this sense, the dependence is different from the information plotted in Fig.~\ref{fig:hamiltonian}, and similar to Fig.~\ref{fig:mstates}(b).

The coordinate system is defined such that $\rho=0$ means that the polarization is always in-plane relative to $\mathbf{B}$ and $\mathbf{k}$.
For this value of $\rho$ and for changing $\theta$ from 0 to $\pi/2$, the polarization starts from the sum of $\sigma^{\pm}$ with equal phases and ends with $\pi$.
Conversely, $\rho=\pi/2$ means that the polarization is fixed at a sum of $\sigma^{\pm}$ with opposite phases.

For an E1 transition, we once again can see the expected behavior as a function of the two angles.
For out-of-plane polarization, we drive only the $\Delta m_J=\pm1$ transition with a fixed amplitude.
For in-plane polarization, we change from driving the $\Delta m_J=\pm1$ transition to driving the $\Delta m_J=0$ transition, corresponding exactly to the change of polarization from the sum of $\sigma^{\pm}$ to $\pi$.
The sum of the $\sigma^{\pm}$ polarizations drives the $\Delta m_J=\pm1$ transitions with equal amplitude, regardless of the relative phase between them.

For the M2 transition, the pattern cannot be simplified by only considering the polarization projections.
We can once again identify the properties we understood from Fig.~\ref{fig:hamiltonian} in the main text.
Additionally, we observe a strong dependence on the relative phases between the contributions of the different polarization components.
For example, for $\Delta m_J=0$, we see that the in-plane polarization cannot drive the transition for any value of $\theta$.
We already knew that the $\pi$ polarization cannot drive this transition at all.
For the sum of $\sigma^{\pm}$, we attribute the vanishing transition amplitude to the destructive interference between the contributions of the individual $\sigma^{\pm}$ components.

We mark in Fig.~\ref{fig:si-angular-dependence}(c) the conditions of the measurements reported in Sections~\ref{sec:bfield_angle} and \ref{sec:magnetic-field-insensitive}.

Finally, we note that the complete angular dependence for electric and magnetic multipole transitions of the same order is identical, up to a shift of $\rho$ by $\pi/2$.
This can be intuitively understood by considering that the magnetic field orientation is simply perpendicular to the electric field within the polarization plane.

\section{Experimental setup}
\label{sec:si_setup}

\paragraph*{Optical transport. ---}
We transport the atoms from the magneto-optical trap (MOT) region into a second vacuum chamber.
For this purpose, we combine a traveling-wave optical lattice~\cite{schmid06} with a focus-tunable optical dipole trap~\cite{leonard14}, so that we move the lattice nodes and the focal position synchronously~\cite{bao22}.
The moving lattice allows for fast transport due to its deep longitudinal confinement, while the dipole trap supports the atoms against gravity.
With this setup, we can transport the atoms within \unit{600}{ms} over a distance of about \unit{55}{cm}.

\paragraph*{Vertical lattice. ---}
After the transport, we adiabatically load the atomic sample into the vertical 1D optical lattice.
We derive the vertical lattice beam from a high-power fiber amplifier and focus the beam to an estimated $1/e^{2}$ waist of \unit{140}{\mu m} at the position of the atoms.
With a typical power of \unit{5}{W}, the lattice has an estimated trap depth of \unit{33}{\mu K} and a measured axial trap frequency of \unit{72}{kHz} corresponding to a Lamb-Dicke parameter $\eta\simeq 0.21$~\cite{ido03}.
Motorized half-wave and quarter-wave plates allow us to dynamically adjust the lattice polarization, even during an experimental sequence.

Since the trap frequency is much larger than the $^{1}\mathrm{S}_0$-$^{3}\mathrm{P}_1$ linewidth, we can perform direct sideband cooling along the axial direction of the lattice.
The cooling beam is aligned collinearly with the lattice beam.
We apply a bias magnetic field of \unit{1}{G} pointing along the $-z$-axis to define the quantization axis.
To optimize the cooling efficiency, we set the lattice polarization to an ellipticity angle of \unit{0.17}{\pi}, making it magic for one $m_{J}$ state of the $^{1}\mathrm{S}_0$-$^{3}\mathrm{P}_1$ transition.
With this method, we typically reach a vertical temperature of $\sim$\unit{1.5}{\mu K}.

We typically load about $10^6$ $^{88}\mathrm{Sr}$ atoms in the lattice, corresponding to up to $40,000$ atoms per lattice layer.

\paragraph*{Spectroscopy laser. ---}
For the $^{1}\mathrm{S}_0$-$^{3}\mathrm{P}_2$ spectroscopy, we use a home-built diode laser operating at \unit{671}{nm}.
The laser is stabilized to a reference cavity with a finesse of $25,000$ using the Pound-Drever-Hall technique.
The cavity is kept under vacuum of \unit{10^{-8}}{mbar} and is temperature stabilized at the zero-crossing temperature of the ultra-low-expansion glass cavity spacer.
The polarization of the beam is fixed for all experiments reported here.
It is to a good approximation linear (ellipticity of $0.06\pi$), and oriented $5\degree$ with respect to the $x$ axis.

\paragraph*{Absolute frequency measurements. ---}
We use a few \unitonly{mW} of the spectroscopy laser power to beat it with a commercial frequency comb, which is referenced to a hydrogen maser.
The resulting beat frequency is measured with a counter.
By averaging about $1000$ counts, we obtain the beat frequency with an uncertainty of a few tens of \unitonly{Hz}, given by the standard error of the mean.

\paragraph*{Compression in light sheet. ---}
To improve the signal-to-noise ratio (SNR) in the local addressing measurement, we decrease the number of vertical lattice sites populated by atoms.
For this purpose, we spatially compress the atomic sample after the transport using a tightly focused optical dipole trap.
We adiabatically load the atoms from the transport lattice into the dipole trap propagating along the $y$ axis.
The dipole trap is formed by a \unit{1070}{nm} beam focused to an elliptical light sheet.
The light sheet has a $1/e^{2}$ waist of $\sim$\unit{15}{\mu m} ($\sim$\unit{300}{\mu m}) along the $z$-axis ($x$-axis) and a power of \unit{40}{W}, resulting in a vertical trap frequency of \unit{2.2}{kHz}.
We cool the atoms via Doppler cooling on the $^{1}\mathrm{S}_0$-$^{3}\mathrm{P}_1$ transition.
We reach near-magic condition for the cooling transition by setting the bias magnetic field of \unit{1}{G} along the $y$ axis and using circularly polarized light for the dipole trap.
After cooling, we typically obtain a vertical temperature of $\sim$\unit{0.6}{\mu K}.
Then, we adiabatically transfer the atoms from the sheet to the vertical lattice.
The compression stage results in about eight lattice layers being significantly populated.

\section{Spectroscopy}
\label{sec:spectroscopy}

\begin{figure}
	\includegraphics{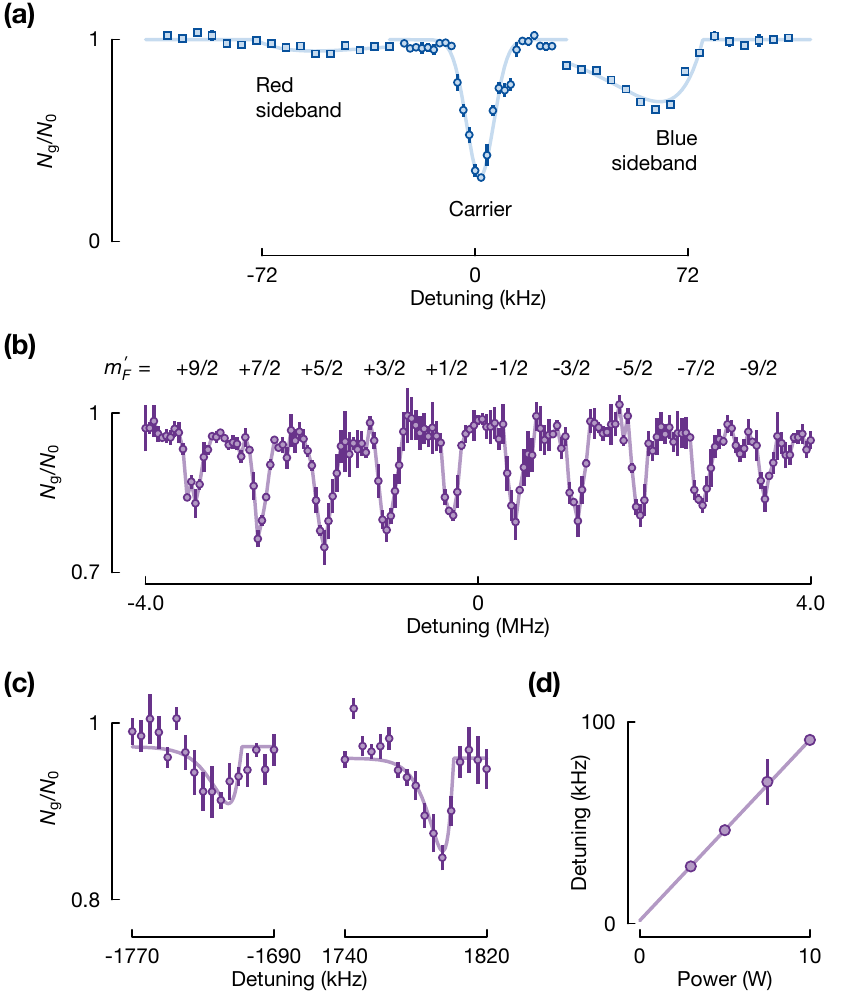}
	\caption{(a) Motional sideband spectrum of the $^{1}\mathrm{S}_0$-$^{3}\mathrm{P}_2$ $(m_{J}=1)$ transition in $^{88}\mathrm{Sr}$ in a magic 1D optical lattice. The sidebands (squares) are probed with a factor of ten longer interrogation time than the carrier (dots). From the relative amplitude of the blue sideband to the red sideband, we extract a ground state fraction of \unit{84}{\%}. Markers, error bars, and solid lines are as in Fig.~\ref{fig:spectroscopy}. (b) Spectrum of the $^{1}\mathrm{S}_0$-$^{3}\mathrm{P}_2$ $F=9/2$ transition in $^{87}\mathrm{Sr}$ in a bias magnetic field of \unit{3}{G} along the $-\it{z}$ axis. The markers represent an average of 3 measurements, and the error bars are the standard error of the mean. The solid lines serve as a guide to the eye. (c) Example of spectroscopic measurements for the determination of the $^{1}\mathrm{S}_0$-$^{3}\mathrm{P}_2$ transition frequency in $^{87}\mathrm{Sr}$. Here, we excite the atoms to the $m_{F}=\pm9/2$ states in an optical lattice with a power of \unit{5}{W}, and a bias magnetic field of \unit{1.5}{G}. Both magnetic field and the linear lattice polarization are along $(\hat{\mathbf{x}}+ \hat{\mathbf{y}})/\sqrt{2}$, creating close-to-magic conditions for these two transitions. The markers represent an average of 5 measurements, and the error bars are the standard error of the mean. The solid lines are fits to the expected ac-Stark-shift-broadened line shape. (d) Average detuning of the $^{1}\mathrm{S}_0$-$^{3}\mathrm{P}_2$, $F=9/2, m_F=\pm9/2$ transitions as a function of lattice power. The solid line is a linear fit, from which we extract the zero-crossing to obtain the absolute transition frequency.}
	\label{fig:si_spectrosocpy}
\end{figure}

\paragraph*{Spectral line shapes. ---}
At a finite atomic temperature, a differential light shift between two atomic states leads to an asymmetric broadening of the spectral line, as some atoms experience lower trapping intensity than others.
Similarly, motional sidebands are asymmetrically broadened toward the carrier -- even in a magic lattice -- due to some atoms experiencing lower trap frequency.

In both cases, at the limit of low temperatures the resulting line shape can be written as~\cite{blatt09,mcdonald15}
\begin{equation}
	y(\delta) = y_{0} - \alpha (\delta-\delta_{0}) e^{-\beta (\delta-\delta_{0})} \Theta(\delta - \delta_{0}),\label{eq:asymmetric_line}
\end{equation}
where $y_0$ is the baseline of the spectrum, $\delta$ is the detuning, $\delta_{0}$ is the resonance frequency without the effect of the light shift, and $\Theta$ is the Heaviside function.
The parameters $\alpha$ and $\beta$ determine the width and amplitude, according to
\begin{align}
	\delta_\mathrm{peak} & = 1/\beta + \delta_0, \\
	a & = \alpha\beta e^1 ,
\end{align}fitting
where $a$ is the line amplitude and $\delta_\mathrm{peak}$ is the detuning at the peak of the line.
This expression is valid for spectral lines broadened towards higher frequencies, that is, for the red sideband and for negative differential Stark shift.
By replacing $\delta-\delta_{0}\rightarrow\delta_{0}-\delta$ one can describe lines broadened to lower frequencies (blue sideband, positive differential Stark shift).

The fit parameters are $y_{0}$, $\alpha$, $\beta$, and $\delta_0$ (they can also be, equivalently, $y_{0}$, $a$, $\delta_0$, and $\delta_\mathrm{peak}$).
While in principle the width of the line is related to the axial and transverse temperatures of the atoms~\cite{blatt09,mcdonald15}, we do not use this information to constrain the fit.

This lineshape is used to describe both motional sidebands in Fig.~\ref{fig:spectroscopy} and Fig.~\ref{fig:si_spectrosocpy}(a), as well as the carrier in a strongly non-magic lattice in Fig.~\ref{fig:stark_shift}(b) and Fig.~\ref{fig:si_spectrosocpy}(c).

For lines where the dominant broadening mechanism is not the differential ac Stark shift -- that is, for carrier transitions close to magic conditions -- we instead use a symmetric function, taking the center as the resonance frequency.
For our experimental conditions, we find that these lines are best described by a Gaussian, as in the carrier in Fig.~\ref{fig:spectroscopy} and Fig.~\ref{fig:si_spectrosocpy}(a), as well as Fig.~\ref{fig:stark_shift}(c).

\paragraph*{Sideband spectra. ---}
We characterize the optical lattice by measuring the frequencies of the motional sidebands relative to the carrier  under magic conditions, as described in the main text.
By fitting the sidebands to Eq.~\eqref{eq:asymmetric_line}, we extract the sideband frequencies for various lattice powers.
Under the assumption of a full lattice contrast (that is, the full optical power contributes to the trap frequency), we can estimate the $1/e^2$ waist of the optical lattice to be \unit{{\sim}140}{\mu m}.
This waist is an upper bound since the lattice contrast is expected to be reduced by experimental imperfections, such as finite mirror reflectivities.
This value is in reasonable agreement with an estimate of the waist based on measuring the beam profile on a camera.

An example of a sideband spectrum is shown in Fig.~\ref{fig:spectroscopy} for the $^{1}\mathrm{S}_0$-$^{3}\mathrm{P}_2$ $m_{J} = 0$ transition.
In addition, we present in Fig.~\ref{fig:si_spectrosocpy}(a) a typical spectrum for the $^{1}\mathrm{S}_0$-$^{3}\mathrm{P}_2$ $m_{J} = 1$ transition in a magic \unit{5}{W} lattice.
Here, we achieve the magic condition as described in Sec.~\ref{sec:magnetic-field-sensitive}.
In this spectrum, the red sideband is strongly suppressed compared to the blue sideband, showing that the atomic sample is cooled close to the vibrational ground state of the optical lattice.
From the relative amplitudes of the two sidebands, we extract a ground state fraction of 84\%, consistent with a temperature of \unit{1.8}{\mu K}~\cite{blatt09,mcdonald15}.

\paragraph*{$^{88}{Sr}$ absolute frequency determination and error budget. ---}
We extract the absolute transition frequency in $^{88}\mathrm{Sr}$ from spectroscopy of the $m_J=0$ transition at the magic condition, as presented in Fig.~\ref{fig:spectroscopy}.
We beat the spectroscopy laser, set to the carrier resonance condition, with a frequency comb to acquire an absolute frequency measurement.
The statistical uncertainties from the estimation of the spectral line center and from the beat note with the comb are \unit{27}{Hz} and \unit{22}{Hz}, respectively, resulting in an overall statistical uncertainty of $\delta \nu_{\textrm{stat}} = \unit{35}{Hz}$.

We estimate the potential systematic shifts due to the atomic density with a set of additional measurements with the atom number reduced by an order of magnitude.
The systematic shift due to the quadratic Zeeman shift can be estimated using the known coefficient for the $^{3}\mathrm{P}_0$ state~\cite{takano17}.
Both density and quadratic Zeeman shifts are smaller than \unit{200}{Hz} and thus we neglect them and do not apply any additional correction to the absolute frequency measured by the frequency comb.

As the measurements were performed under our best estimate for the magic conditions, the expectation value for the light shift is zero.
  Nevertheless, we estimate the major source of systematic uncertainty to be due to the light shift, that is, from operating at imperfect magic conditions.
Based on typical day-to-day variations in the light shift that we have observed, we take an uncertainty of \unit{2}{kHz}~(1-$\sigma$) on the residual light shift as a conservative estimate.
We attribute this variation to the finite precision of the magic condition calibration, and to possible drifts in magnetic field and lattice light polarization over time.

In summary, we obtain an absolute transition frequency of $(446,647,242,704\pm 0.04_{\textrm{stat}} \pm 2_{\textrm{sys}})\,\textrm{kHz}$ for $^{88}\mathrm{Sr}$.

\paragraph*{$^{87}\mathrm{Sr}$ absolute frequency determination and error budget. ---}
Measuring the absolute frequency for $^{87}\mathrm{Sr}$ is more involved than for $^{88}\mathrm{Sr}$, due to the more complicated internal structure and the large number of transitions, as shown in the spectrum in Fig.~\ref{fig:si_spectrosocpy}(b).
To eliminate the influence of the magnetic field, we probe two $m_{F}$ states with opposite signs.
Finding magic conditions for two states simultaneously was beyond the scope of the present work, but we reached a condition close to magic for both states and performed a sequence of measurements with multiple lattice powers.
Taking the average of the two frequencies for each lattice power, and extrapolating it to zero lattice power allows us to extrapolate to zero light shift.
As for $^{88}\mathrm{Sr}$, we obtain the absolute transition frequency by beating the laser with a frequency comb.

Taking measurements of the $m_{F}=\pm9/2$ states in the $F=9/2$ manifold with the magnetic field parallel to the lattice polarization, shown in Fig~\ref{fig:si_spectrosocpy}(c), we observe the expected linear behavior as a function of the lattice power, as shown in Fig~\ref{fig:si_spectrosocpy}(d).
From the extrapolation to zero power, we estimate a statistical uncertainty of \unit{5}{kHz} on the intercept.
The uncertainty on the intercept already includes the fluctuations of Zeeman and light shifts.
The contribution from the beat note with the comb is once again <\unit{100}{Hz} and can be neglected.

We estimate the density correction by performing a set of measurements with a reduced number of atoms.
Unlike for $^{88}\mathrm{Sr}$, here we could observe effects on the scale of a few tens of \unitonly{kHz}.
The existence of density shifts in $^{87}\mathrm{Sr}$ and not $^{88}\mathrm{Sr}$ is consistent with the density shift arising from collisions between ground state atoms, as they are significantly stronger in $^{87}\mathrm{Sr}$ compared to $^{88}\mathrm{Sr}$~\cite{stellmer13}.
From these measurements, our estimate for the density shift is \unit{+20}{kHz} with a 1-$\sigma$ uncertainty of $\unit{40}{kHz}$.
The large uncertainty is the result of a significant scatter in the density correction measurements, and it is the dominant major systematic uncertainty.

After applying the density correction, we obtain for $^{87}\mathrm{Sr}$ a transition frequency to the $F=9/2$ state of $(446,647,798,443\pm 5_{\textrm{stat}} \pm 40_{\textrm{sys}})\,\textrm{kHz}$.
Our measurement is consistent with the value reported in Ref.~\cite{onishchenko19}, while improving the uncertainty by nearly three orders of magnitude.

The hyperfine structure of the $^{3}\mathrm{P}_2$ state in $^{87}\mathrm{Sr}$ was measured to better than \unit{10}{kHz} in Ref.~\cite{heider77}.
Using this data, we can estimate the transition frequencies to all other $F$ states, as well as the center of gravity of the lines.
We performed measurements of the transitions to the other hyperfine states with similar or larger uncertainties, and the frequencies agree with Ref.~\cite{heider77} within the error bars.

\paragraph*{Isotope shift. ---}
We can estimate the frequency difference between the transitions in the two isotopes simply by subtracting the frequency measurements performed with the frequency comb.
Additionally, we can directly compare the different frequency shifts we applied to the probe beam between subsequent measurements for the two isotopes.
Both estimates agree within a few \unitonly{kHz}.

The $F=9/2$ state is \unit{618.65}{MHz} blue detuned from the line's center of gravity~\cite{heider77}.
Therefore, the isotope shift is $\nu(^{88}\mathrm{Sr})-\nu(^{87}\mathrm{Sr}) = \unit{+62.91(4)}{MHz}$.

We can compare this value to isotope shifts measured for the other transitions from the ground state to the triplet manifold, that is, for $^{1}\mathrm{S}_0$-$^{3}\mathrm{P}_0$ and $^{1}\mathrm{S}_0$-$^{3}\mathrm{P}_1$.
They were measured to be \unit{+62.188(10)}{MHz}~\cite{takano17} and \unit{+62.187(12)}{MHz}~\cite{miyake19}, respectively.
Given that these two isotope shifts are almost identical, one naively might expect that the value for the $^{1}\mathrm{S}_0$-$^{3}\mathrm{P}_2$ transition should also be the same, rather than being about \unit{700}{kHz} larger, as we found.
However, the equality of those two specific isotope shifts appears to be a coincidence.
Generally, differences between isotope shifts on fine-structure states in the same manifold on the \unitonly{MHz} scale are typical, as seen in \emph{e.g.} Refs.~\cite{miyake19,stellmer14}.

Additionally, we note that the literature contains isotope shift measurements for the $^{3}\mathrm{P}_1$-$^{3}\mathrm{D}_2$ transition of \unit{+30.2(6)}{MHz}~\cite{iwata21}, and for the $^{3}\mathrm{P}_2$-$^{3}\mathrm{D}_2$ transition of \unit{+17(2)}{MHz}~\cite{stellmer14}.
Combining these values with the isotope shift for the $^{1}\mathrm{S}_0$-$^{3}\mathrm{P}_1$ transition, we expect an isotope shift for $^{3}\mathrm{P}_2$ of \unit{+75(2)}{MHz}.
Our measurement therefore uncovers a possible discrepancy in the literature data for strontium transition frequencies.

\section{Polarizability}
\label{sec:polarizability}

\begin{table}[h]
	\centering
	\caption{The required information for calculating the polarizability of $5s^{2}$ $^{1}\mathrm{S}_0$ state and $5s5p$ $^{3}\mathrm{P}_2$ state of $^{88}\mathrm{Sr}$. The energies and matrix elements of the different transitions are taken from Refs.~\cite{nistasd} and~\cite{safronova13,safronovapriv}, respectively. The contributions to scalar $\alpha_{i}^{\mathrm{s}}(\lambda_0)$, vector $\alpha_{i}^{\mathrm{v}}(\lambda_0)$ and tensor polarizability $\alpha_{i}^{\mathrm{t}}(\lambda_0)$ are calculated based on those values at $\lambda_0=$\unit{1064}{nm}. \textit{Other} refers to contributions to the polarizability from states which are not listed explicitly and \textit{Core} refers to the contribution to the polarizability from the closed electron shells of the atom. The units contain the electron charge $e$, the Bohr radius $a_0$, and the atomic unit of polarizability $1~\textrm{a.u.} = 4\pi \epsilon_0 a_0^3$, where $\epsilon_0$ is the vacuum permittivity.}
	\begin{tabularx}{\columnwidth}{%
      @{}
      >{\RaggedLeft}X
      @{}
      >{\setlength\hsize{0.165\columnwidth}\RaggedLeft}X
      @{}
      >{\setlength\hsize{0.165\columnwidth}\RaggedLeft}X
      @{}
      >{\setlength\hsize{0.165\columnwidth}\RaggedLeft}X
      @{}
      >{\setlength\hsize{0.165\columnwidth}\RaggedLeft}X
      @{}
      >{\setlength\hsize{0.165\columnwidth}\RaggedLeft}X
      @{}}
		\hline \hline
		\\
		State $k$ & $\Delta E_{ki} $ & $\Bra{k} D \Ket{i}$ & $\alpha_{i}^{\mathrm{s}}(\lambda_0)$ & $\alpha_{i}^{\mathrm{v}}(\lambda_0)$ & $\alpha_{i}^{\mathrm{t}}(\lambda_0)$\\
		& (\unitonly{cm^{-1}}) & ({$ea_0$}) & (\unitonly{a.u.}) & (\unitonly{a.u.}) & (\unitonly{a.u.}) \\
		\hline
		\\
		& \multicolumn{5}{c}{State $i = 5s^{2}$ $^{1}\mathrm{S}_{0}$} \\
		$5s5p$ $^{3}\mathrm{P}_{1}$ & 14504 &  0.158 & 0.40 & 0 & 0 \\
		$5s5p$ $^{1}\mathrm{P}_{1}$ & 21698 &  5.248 & 228.61 & 0 & 0\\
		$5s6p$ $^{3}\mathrm{P}_{1}$ & 33868 &  0.034 & 0.01 & 0 & 0\\
		$5s6p$ $^{1}\mathrm{P}_{1}$ & 34098 &  0.282 & 0.37 & 0 & 0\\
		Other & & & 5.8 & 0 & 0\\
      Core & & & 5.3 & 0 & 0\\
      \textbf{Total} & & & \textbf{240.49} & \textbf{0} & \textbf{0}\\
      & & & & & \\
		& \multicolumn{5}{c}{State $i = 5s5p$ $^{3}\mathrm{P}_{2}$} \\
		$5s4d$ $^{3}\mathrm{D}_{1}$ & 3260 &  0.6021 & $-$0.45 & 3.85 & 0.45\\
		$5s4d$ $^{3}\mathrm{D}_{2}$ & 3320 &  2.331 & $-$6.83 & 19.33 & $-$6.83\\
		$5s4d$ $^{3}\mathrm{D}_{3}$ & 3421 &  5.530 & $-$39.95 & $-$219.52 & 11.41\\
		$5s4d$ $^{1}\mathrm{D}_{2}$ & 5251 &  0.102 & $-$0.03 & 0.05 & $-$0.03\\
		$5s6s$ $^{3}\mathrm{S}_{1}$ & 14140 &  4.521 & 75.78 & $-$151.10 & $-$75.78\\
		$5s5d$ $^{1}\mathrm{D}_{2}$ & 19829 &  0.365 & 0.25 & $-$0.12 & 0.25\\
		$5s5d$ $^{3}\mathrm{D}_{1}$ & 20108 &  0.460 & 0.39 & $-$0.55 & $-$0.39\\
		$5s5d$ $^{3}\mathrm{D}_{2}$ & 20123 &  1.956 & 7.12 & $-$3.32 & 7.12\\
		$5s5d$ $^{3}\mathrm{D}_{3}$ & 20146 &  4.994 & 46.30 & 43.20 & $-$13.23\\
		$5p^{2}$ $^{3}\mathrm{P}_{1}$ & 20502 &  2.992 & 16.18 & $-$22.25 & $-$16.18\\
		$5p^{2}$ $^{3}\mathrm{P}_{2}$ & 20776 &  5.119 & 46.41 & $-$21.00 & 46.41\\
		$5p^{2}$ $^{1}\mathrm{D}_{2}$ & 22062 &  0.682 & 0.75 & $-$0.32 & 0.75\\
		$5s7s$ $^{3}\mathrm{S}_{1}$ & 22526 &  1.264 & 2.51 & $-$3.15 & $-$2.51\\
		Other & & & 43.1 & & 0.34\\
		Core & & & 5.6 &\\
        \textbf{Total} & & & \textbf{197.14} & \textbf{$-$354.90} & \textbf{$-$48.56} \\
		\hline \hline
	\end{tabularx}

	\label{tab:polarizability}
\end{table}

\paragraph*{General expression. ---}
Here, we theoretically investigate the polarizability of the $^{3}\mathrm{P}_2$ state.
The dynamic dipole polarizability $\alpha_{i}$ of an atomic state $\Ket{i}$ can be decomposed into a scalar polarizability $\alpha_{i}^{\mathrm{s}}$, a vector polarizability $\alpha_{i}^{\mathrm{v}}$, and a tensor polarizability $\alpha_{i}^{\mathrm{t}}$~\cite{lekien13, heinz20, cooper18}
\begin{align}
\begin{split}
\alpha_{i} = &\alpha_{i}^{\mathrm{s}}\\
&+ \alpha_{i}^{\mathrm{v}} \sin(2 \gamma) \frac{m_{i}}{2 J_{i}}\\
&+ \alpha_{i}^{\mathrm{t}} \frac{3 \cos^{2}(\beta) -1 }{2} \frac{3m_{J_{i}}^{2} - J_{i} (J_{i}+1)}{J_{i} (2J_{i} -1)},
\end{split}
\end{align}
where $\gamma$ is the ellipticity angle of the polarization~\cite{rosenbusch09, cooper18} and $\cos\beta$ is the projection of the polarization vector onto the quantization axis.
The ellipticity angle $\gamma$ is defined as ~\cite{rosenbusch09, cooper18}
\begin{equation}
\label{eq:polarization}
\hat{\vec{\epsilon}}_{\mathrm{l}} = \hat{\vec{\epsilon}}_{\mathrm{l,1}} \cos (\gamma) +i \hat{\vec{\epsilon}}_{\mathrm{l,2}} \sin (\gamma),
\end{equation}
where $\hat{\vec{\epsilon}}_{\mathrm{l,1}}$, $\hat{\vec{\epsilon}}_{\mathrm{l,2}}$ are orthogonal vectors spanning the plane perpendicular to lattice propagation direction $\mathbf{k}_{\mathrm{l}}$.
We note that rotating the polarization ellipse within this plane does not affect the ellipticity angle $\gamma$, and thus does not affect the polarizability.
Hence, we are not sensitive to the exact choice of $\hat{\vec{\epsilon}}_{\mathrm{l,1}}$, $\hat{\vec{\epsilon}}_{\mathrm{l,2}}$.
The quantization axis is assumed to be defined by a strong external magnetic field pointing along the $z$-axis.

The scalar polarizability of the state \Ket{i} with angular momentum $J_{i}$ at the light frequency $\omega$ can be calculated with ~\cite{lekien13, safronova15}
\begin{equation}
\alpha_{i}^{\mathrm{s}} = \frac{1}{3 (2J_{i} +1)} \sum_{k} \frac{2}{\hbar}\frac{| \Bra{k} D \Ket{i} |^2 \omega_{ki}}{\omega_{ki}^{2} - \omega^{2}} + \alpha_{i}^{\mathrm{c}}.
\end{equation}
We sum over the dipole-allowed transitions to states \Ket{k} with the corresponding dipole matrix element $\Bra{k} D \Ket{i},$ and the transition frequency $\omega_{ki} \equiv (E_k - E_i)/\hbar = \Delta E_{ki}/\hbar$.
Here, $\alpha_{i}^{\mathrm{c}}$ is the contribution of the core electrons to the scalar polarizability, where Ref.~\cite{safronovapriv} provided us with a calculated value listed in Tab.~\ref{tab:polarizability}.
The vector and the tensor polarizabilities are given by
\begin{align}
\begin{split}
\alpha_{i}^{\mathrm{v}} = - \sqrt{\frac{6J_{i}}{(J_{i}+1)(2J_{i}+1)}} \sum_k (-1)^{J_{i} + J_{k}} \\
\times
\begin{Bmatrix}
1 & 1 & 1\\
J_{i} & J_{k} & J_{i} \\
\end{Bmatrix}
\frac{| \Bra{k} D \Ket{i} | ^2}{\hbar} \Big( \frac{1}{\omega_{ki} - \omega} - \frac{1}{\omega_{ki} + \omega} \Big),
\end{split} \\
\begin{split}
\alpha_{i}^{\mathrm{t}} = - \sqrt{\frac{10J_{i} (2J_{i} -1)}{3(J_{i}+1)(2J_{i}+1)(2J_{i} +3)}}  \\
\times \sum_k (-1)^{J_{i} + J_{k} + 1}
\begin{Bmatrix}
1 & 2 & 1\\
J_{i} & J_{k} & J_{i} \\
\end{Bmatrix}
\frac{2}{\hbar}\frac{| \Bra{k} D \Ket{i} | ^2 \omega_{ki}}{\omega_{ki}^2 - \omega^2}.
\end{split}
\end{align}

\begin{figure}
	\centering
	\includegraphics{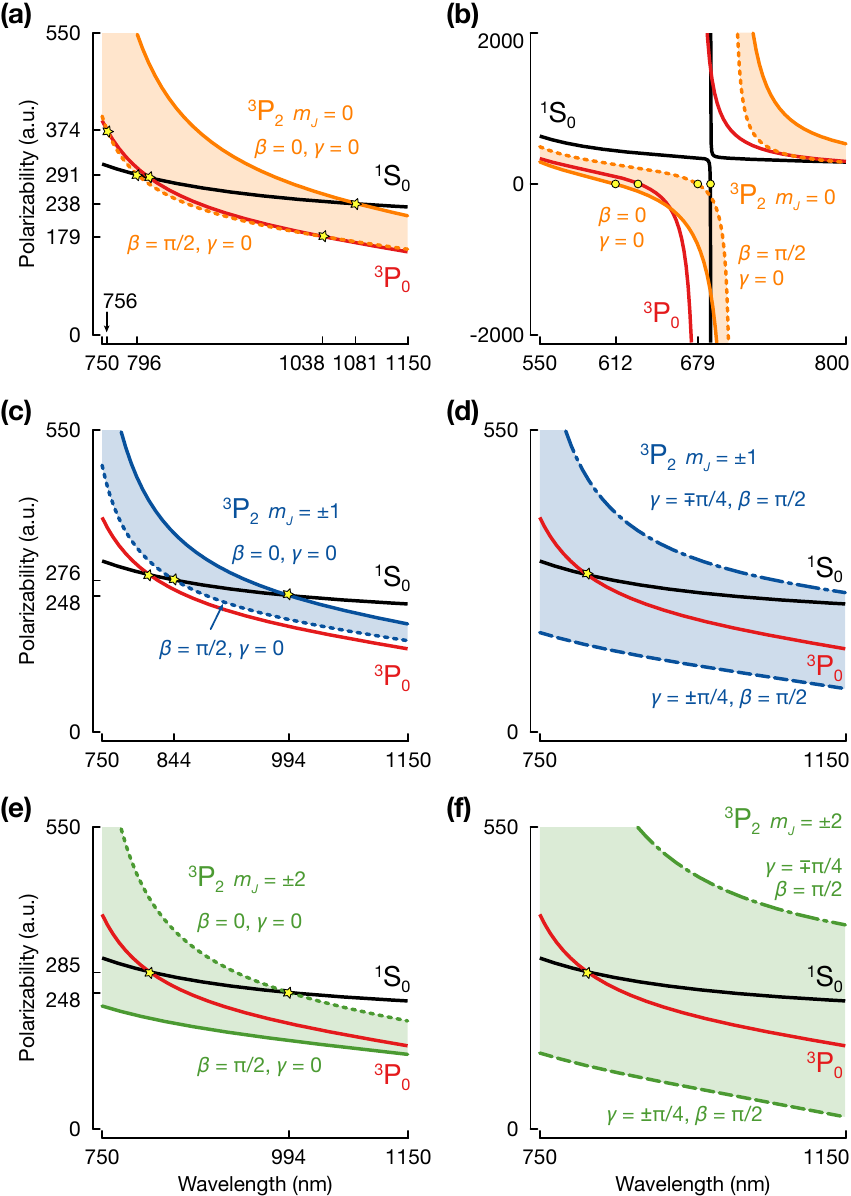}
	\caption{
        Calculated polarizability of the ground state $^{1}\mathrm{S}_0$, excited clock state $^{3}\mathrm{P}_0$, and the $^{3}\mathrm{P}_2$ Zeeman sublevels for different trapping wavelengths and polarizations. The polarizability of the ground and excited clock states is independent of the orientation of the magnetic field with respect to the quantization axis and the polarization ellipticity, while the polarizability of the different Zeeman sublevels of $^{3}\mathrm{P}_2$ depends strongly on these conditions. The shaded areas indicate the dynamic range of polarizability tuning for the Zeeman sublevel under consideration.
        (a) Tensor polarizability tuning of the $^{3}\mathrm{P}_2$ $m_{J}=0$ state, assuming linear trap polarization ($\gamma=0$). The curves are the limiting case of $\pi$ ($\beta=0$) and horizontal ($\beta=\pi/2$) polarizations, while the shaded area is the range accessible by tuning the angle between polarization and magnetic field. Specific magic wavelengths are indicated by stars. At the magic wavelength for the clock states, the $m_J=0$ state can be made to be magic as well.
        (b) The same conditions as in panel (a), but plotted in the wavelength range that covers multiple tune-out conditions (indicated by circles). Around \unit{633}{nm}, the polarizability of both $^{3}\mathrm{P}_0$ and $^{3}\mathrm{P}_2$ $m_J=0$ can be tuned out simultaneously.
        (c) Tensor polarizability tuning for the $^{3}\mathrm{P}_2$ $m_{J}=\pm1$ states.
        (d) Vector polarizability tuning for the $^{3}\mathrm{P}_2$ $m_{J}=\pm1$ states. The curves are the limiting case of $\sigma^\pm$ ($\gamma=\pm\pi/4$) and $\sigma^\mp$ ($\gamma=\mp\pi/4$) polarizations. The shaded area is the range accessible by tuning the ellipticity angle of the polarization while maintaining propagation along the magnetic field ($\beta=\pi/2$). In contrast to panel (c), we can find a magic wavelength for all three states.
        (e) Tensor polarizability tuning for the $^{3}\mathrm{P}_2$ $m_{J}=\pm2$ states.
        (f) Vector polarizability tuning for the $^{3}\mathrm{P}_2$ $m_{J}=\pm2$ states.}
	\label{fig:si-polarizability}
\end{figure}

\paragraph*{Calculated magic and tune-out wavelengths. ---}
In Fig.~\ref{fig:si-polarizability}, we plot the polarizabilities of the Zeeman sublevels of the $^{3}\mathrm{P}_2$ state under different polarization conditions as a function of the trap wavelength.
We also plot the polarizability of the ground $^{1}\mathrm{S}_0$ and excited clock $^{3}\mathrm{P}_0$ states for reference.
For the $^{1}\mathrm{S}_0$ and $^{3}\mathrm{P}_2$ states, we use the matrix elements listed in Tab.~\ref{tab:polarizability}.
For the $^{3}\mathrm{P}_0$ state, we use the matrix elements reported in Ref.~\cite{park22}.

We look for magic conditions between the $^{3}\mathrm{P}_2$ states and the ground or excited clock state in the near-infrared, which is usually the wavelength range in which deep, far-detuned traps can be realized.

In Fig.~\ref{fig:si-polarizability}(a), we show that we can achieve magic conditions between the $m_J = 0$ state and the ground state for wavelengths between \unit{796}{nm} (horizontal polarization, \emph{i.e.} linear polarization perpendicular to the quantization axis) and \unit{1081}{nm} ($\pi$ polarization).
We see that for horizontal polarization, the differential polarizability between $m_J= 0$ and the excited clock state is very small, reaching up to about \unit{12}{a.u.} at most, over the entire examined wavelength range.
Based on those calculations, magic conditions for the clock and the $^{3}\mathrm{P}_2$ $m_J= 0$ states can be achieved between \unit{756}{nm} and \unit{1038}{nm} (both limiting cases with horizontal polarization).

Similarly, we see in Fig.~\ref{fig:si-polarizability}(c) that, using tensor polarizability tuning, we can achieve magic conditions between the ground and the $m_J= \pm1$ states between \unit{844}{nm} (horizontal polarization) and \unit{994}{nm} ($\pi$ polarization).
For these states, it is impossible to reach magic condition with the clock state using only tensor polarizability tunings.
Conversely, as seen in Fig.~\ref{fig:si-polarizability}(e), magic conditions between the ground and the $m_J= \pm2$ states in the near-infrared can be found for wavelengths of up to \unit{994}{nm} (horizontal polarization), and between the $m_J= \pm2$ states and the clock state thoughout the entire examined wavelength range.
In both cases, the vector polarizability is large enough to enable reaching magic conditions with either the ground and clock states across the examined wavelength range, as seen in Fig.~\ref{fig:si-polarizability}(d) and~(f).

Finally, in Fig.~\ref{fig:si-polarizability}(b), we examine the polarizability of the $m_J= 0$ state in a wavelength range in the visible, where resonant transitions can be found.
These resonances lead to the existence of tune-out wavelengths~\cite{heinz20}, where the state's polarizability vanishes.
The tune-out condition can be used to generate highly state-dependent optical traps with applications in quantum computing and quantum simulation~\cite{daley08,daley11}.
We find that tune-out conditions can be found for wavelengths between \unit{612}{nm} ($\pi$ polarization) and \unit{679}{nm} (horizontal polarization).

The uncertainty estimates on the calculated polarizability values are estimated to be about 1\%.
Due to the shallow crossings in the near infrared wavelength range, this corresponds to uncertainties of up to \unit{10}{nm} on the calculated magic wavelengths between the ground states and the $^{3}\mathrm{P}_2$ states.
For the magic conditions between the clock and the $^{3}\mathrm{P}_2$ $m_{J}=0$ states, the uncertainty is even larger, due to the two polarizability curves nearly overlapping.
Additional experimental information, such as the magic conditions at \unit{1064}{nm} that we found in this work, can help to constrain these values better.

\paragraph*{Calculated ``triple-magic'' conditions for $^{1}\mathrm{S}_0$, $^{3}\mathrm{P}_0$, $^{3}\mathrm{P}_2$. ---}
To take full advantage of both ultranarrow transitions available between the ground state and the triplet manifold, it is useful to find a condition where all three relevant states have the same polarizability -- a ``triple-magic'' condition.
Such a condition requires operating at the $^{1}\mathrm{S}_0$-$^{3}\mathrm{P}_0$ magic wavelength, because the polarizabilities of these states can only be tuned via the trap wavelength.
We will consider the most common and useful magic wavelength of this transition, at \unit{813}{nm}.
The question of existence of the triple-magic condition is then equivalent to the question of whether, for the desired Zeeman sublevel of $^{3}\mathrm{P}_2$ and the polarizability tuning mechanism under consideration, \unit{813}{nm} is within the range of possible magic wavelengths with the ground state.

For the magnetically-insensitive $^{3}\mathrm{P}_2$ $m_J=0$ state, \unit{813}{nm} is indeed found within this range.
Given the uncertainty estimates of the magic wavelengths discussed above, the probability of a sufficiently large error such that a triple-magic condition cannot be met is very small.
Quantitatively, we estimate that a triple-magic condition can be achieved for $\beta_{\mathrm{triple}}^{m_J=0} = 0.42 \, \pi$.

For the magnetically sensitive states, $^{3}\mathrm{P}_2$ $m_J=\pm1$ and $^{3}\mathrm{P}_2$ $m_J=\pm2$, it is possible to achieve a triple-magic condition by tuning the vector polarizability.
For the $^{3}\mathrm{P}_2$ $m_J=\pm2$ states, it is also possible to achieve a triple-magic condition by tuning the tensor polarizability, at $\beta_{\mathrm{triple}}^{m_J=\pm2} = 0.22 \, \pi$.

\paragraph*{Calculated magic conditions at \unit{1064}{nm}. ---}
In this work, we did not vary the trap wavelength, but rather modified the atomic-frame trap polarization.
This method allows finding magic conditions while choosing a trapping wavelength based on other technical considerations.
In our case, we use \unit{1064}{nm}, a wavelength in which lasers with high optical power and low intensity noise are readily available.

For this reason, we are interested in the differential polarizability between the $^{1}\mathrm{S}_0$ state ($g$) and a certain $^{3}\mathrm{P}_2$ $m_J$ state ($e$), given by
\begin{align}
	\begin{split}
		\Delta \alpha = \alpha_{g}^{\mathrm{s}} - & \Bigl(\alpha_{e}^{\mathrm{s}} + \frac{m_J}{2J}\sin(2 \gamma_{0})\alpha_{e}^{\mathrm{v}} \\
		& +\frac{3 \cos^{2}(\beta) -1 }{2} \frac{3m_{J}^{2} - J (J+1)}{J (2J -1)} \alpha_{e}^{\mathrm{t}}\Bigr),
	\end{split}
	\label{eq:differential_polarizability}
\end{align}
where $J=2$ and $m_{J}$ refer to the angular momentum of $e$.
Using the values from Tab.~\ref{tab:polarizability}, we can thus extract the magic conditions for the scenarios of interest.

First, we consider the $^{3}\mathrm{P}_2$ $m_J=0$ state.
For this state, the vector polarizability vanishes.
We can therefore use a linear polarization, and tune its orientation relative to quantization axis ($\beta$) to modify the tensor polarizability.
We find that the polarizability vanishes for the magic angle value of $\beta_{0} = 0.089 \, \pi$.

Second, we consider the $^{3}\mathrm{P}_2$ $m_J=1$ state, and assume that the trapping light propagates parallel to the quantization axis.
As described in Sec.~\ref{sec:magnetic-field-sensitive}, these are the conditions required for local addressing.
Here, $\beta = \pi/2$, and the contribution of the tensor polarizability is constant.
Hence, we tune the ellipticity angle $\gamma$ and thereby the vector polarizability to find the magic condition.
The polarizability vanishes for the magic ellipticity value of $\gamma_{0} = 0.108 \, \pi$.

\paragraph*{Experimental determination of the magic ellipticity. ---}
Here we describe the determination of the magic ellipticity shown in Fig.~\ref{fig:stark_shift}.
In our experimental characterization of the vector ac Stark shift, we can only measure the differential ac Stark shift between the $^{1}\mathrm{S}_0$ state and the $^{3}\mathrm{P}_2$ state.
The differential ac Stark shift $\Delta \nu$ and the differential polarizability $\Delta \alpha$ are connected by
\begin{equation}
\Delta \nu_\mathrm{ac}  = \frac{1}{2 \epsilon_{0} c h} \Delta \alpha\, I \propto \Delta\alpha\,P,
\end{equation}
where $c$ is the speed of light, and $I$ is the trapping light intensity, which in turn depends linearly on the beam power $P$.

We measure spectra for each value of the ellipticity angle at lattice powers of \unit{2.5}{W}, \unit{5}{W}, \unit{7.5}{W}, and \unit{10}{W}, with some examples shown in Figs.~\ref{fig:stark_shift}(b) and (c).
From each spectrum, we extract the transition frequency, and fit all the measured points to a function with independent slopes for each ellipticity value, but a common intercept, with part of the data shown in the inset of Fig.~\ref{fig:stark_shift}(d).

We then fit the slopes $d\Delta\nu_\mathrm{ac}/dP$ as a function of $\gamma$ with
\begin{equation}
\frac{d\Delta\nu_\mathrm{ac}}{dP} = a_{0} \left[\sin(2 \gamma) - \sin(2 \gamma_{0})\right],
\end{equation}
where $a_{0}$ is the amplitude of the curve and $\gamma_0$ is the magic ellipticity, where the differential Stark shift changes sign, as shown in Fig.~\ref{fig:stark_shift}(d).
In this parametrization, all the experimental uncertainties related to the intensity calibration are contained in $a_0$, and do not influence $\gamma_0$.
For this reason, we rely on the magic condition $\gamma_0$ for comparison
with atomic structure calculations, as is the standard practice in the
field~\cite{safronova13}.

The largest uncertainty in the estimation of $\gamma_0$ arises from the uncertainty in the measured ellipticity values, determined with a commercial polarimeter.
Based on the day-to-day variations we observed between measurements under ostensibly identical conditions, we conservatively estimate the uncertainty in the ellipticity as $\Delta \gamma = \pm 0.01 \pi$.
We attribute this uncertainty to a combination of polarimeter precision and repeatability of the  motorized waveplate rotation mounts.
The fit uncertainty leading to $\gamma_0$ is thus mostly determined by the error on the $x$-axis and not by errors on the $y$-axis.


\begin{thebibliography}{94}%
\makeatletter
\providecommand \@ifxundefined [1]{%
 \@ifx{#1\undefined}
}%
\providecommand \@ifnum [1]{%
 \ifnum #1\expandafter \@firstoftwo
 \else \expandafter \@secondoftwo
 \fi
}%
\providecommand \@ifx [1]{%
 \ifx #1\expandafter \@firstoftwo
 \else \expandafter \@secondoftwo
 \fi
}%
\providecommand \natexlab [1]{#1}%
\providecommand \enquote  [1]{``#1''}%
\providecommand \bibnamefont  [1]{#1}%
\providecommand \bibfnamefont [1]{#1}%
\providecommand \citenamefont [1]{#1}%
\providecommand \href@noop [0]{\@secondoftwo}%
\providecommand \href [0]{\begingroup \@sanitize@url \@href}%
\providecommand \@href[1]{\@@startlink{#1}\@@href}%
\providecommand \@@href[1]{\endgroup#1\@@endlink}%
\providecommand \@sanitize@url [0]{\catcode `\\12\catcode `\$12\catcode
  `\&12\catcode `\#12\catcode `\^12\catcode `\_12\catcode `\%12\relax}%
\providecommand \@@startlink[1]{}%
\providecommand \@@endlink[0]{}%
\providecommand \url  [0]{\begingroup\@sanitize@url \@url }%
\providecommand \@url [1]{\endgroup\@href {#1}{\urlprefix }}%
\providecommand \urlprefix  [0]{URL }%
\providecommand \Eprint [0]{\href }%
\providecommand \doibase [0]{https://doi.org/}%
\providecommand \selectlanguage [0]{\@gobble}%
\providecommand \bibinfo  [0]{\@secondoftwo}%
\providecommand \bibfield  [0]{\@secondoftwo}%
\providecommand \translation [1]{[#1]}%
\providecommand \BibitemOpen [0]{}%
\providecommand \bibitemStop [0]{}%
\providecommand \bibitemNoStop [0]{.\EOS\space}%
\providecommand \EOS [0]{\spacefactor3000\relax}%
\providecommand \BibitemShut  [1]{\csname bibitem#1\endcsname}%
\let\auto@bib@innerbib\@empty
\bibitem [{\citenamefont {Takamoto}\ \emph {et~al.}(2005)\citenamefont
  {Takamoto}, \citenamefont {Hong}, \citenamefont {Higashi},\ and\
  \citenamefont {Katori}}]{takamoto05}%
  \BibitemOpen
  \bibfield  {author} {\bibinfo {author} {\bibfnamefont {M.}~\bibnamefont
  {Takamoto}}, \bibinfo {author} {\bibfnamefont {F.-L.}\ \bibnamefont {Hong}},
  \bibinfo {author} {\bibfnamefont {R.}~\bibnamefont {Higashi}},\ and\ \bibinfo
  {author} {\bibfnamefont {H.}~\bibnamefont {Katori}},\ }\bibfield  {title}
  {\bibinfo {title} {An optical lattice clock},\ }\href
  {https://doi.org/10.1038/nature03541} {\bibfield  {journal} {\bibinfo
  {journal} {Nature}\ }\textbf {\bibinfo {volume} {435}},\ \bibinfo {pages}
  {321} (\bibinfo {year} {2005})}\BibitemShut {NoStop}%
\bibitem [{\citenamefont {Ludlow}\ \emph {et~al.}(2006)\citenamefont {Ludlow},
  \citenamefont {Boyd}, \citenamefont {Zelevinsky}, \citenamefont {Foreman},
  \citenamefont {Blatt}, \citenamefont {Notcutt}, \citenamefont {Ido},\ and\
  \citenamefont {Ye}}]{ludlow06}%
  \BibitemOpen
  \bibfield  {author} {\bibinfo {author} {\bibfnamefont {A.~D.}\ \bibnamefont
  {Ludlow}}, \bibinfo {author} {\bibfnamefont {M.~M.}\ \bibnamefont {Boyd}},
  \bibinfo {author} {\bibfnamefont {T.}~\bibnamefont {Zelevinsky}}, \bibinfo
  {author} {\bibfnamefont {S.~M.}\ \bibnamefont {Foreman}}, \bibinfo {author}
  {\bibfnamefont {S.}~\bibnamefont {Blatt}}, \bibinfo {author} {\bibfnamefont
  {M.}~\bibnamefont {Notcutt}}, \bibinfo {author} {\bibfnamefont
  {T.}~\bibnamefont {Ido}},\ and\ \bibinfo {author} {\bibfnamefont
  {J.}~\bibnamefont {Ye}},\ }\bibfield  {title} {\bibinfo {title} {Systematic
  {S}tudy of the $^{87}\mathrm{Sr}$ {C}lock {T}ransition in an {O}ptical
  {L}attice},\ }\href {https://doi.org/10.1103/PhysRevLett.96.033003}
  {\bibfield  {journal} {\bibinfo  {journal} {Physical Review Letters}\
  }\textbf {\bibinfo {volume} {96}},\ \bibinfo {pages} {033003} (\bibinfo
  {year} {2006})}\BibitemShut {NoStop}%
\bibitem [{\citenamefont {Ludlow}\ \emph {et~al.}(2015)\citenamefont {Ludlow},
  \citenamefont {Boyd}, \citenamefont {Ye}, \citenamefont {Peik},\ and\
  \citenamefont {Schmidt}}]{ludlow15}%
  \BibitemOpen
  \bibfield  {author} {\bibinfo {author} {\bibfnamefont {A.~D.}\ \bibnamefont
  {Ludlow}}, \bibinfo {author} {\bibfnamefont {M.~M.}\ \bibnamefont {Boyd}},
  \bibinfo {author} {\bibfnamefont {J.}~\bibnamefont {Ye}}, \bibinfo {author}
  {\bibfnamefont {E.}~\bibnamefont {Peik}},\ and\ \bibinfo {author}
  {\bibfnamefont {P.~O.}\ \bibnamefont {Schmidt}},\ }\bibfield  {title}
  {\bibinfo {title} {Optical atomic clocks},\ }\href
  {https://doi.org/10.1103/RevModPhys.87.637} {\bibfield  {journal} {\bibinfo
  {journal} {Rev. Mod. Phys.}\ }\textbf {\bibinfo {volume} {87}},\ \bibinfo
  {pages} {637} (\bibinfo {year} {2015})}\BibitemShut {NoStop}%
\bibitem [{\citenamefont {Bothwell}\ \emph {et~al.}(2022)\citenamefont
  {Bothwell}, \citenamefont {Kennedy}, \citenamefont {Aeppli}, \citenamefont
  {Kedar}, \citenamefont {Robinson}, \citenamefont {Oelker}, \citenamefont
  {Staron},\ and\ \citenamefont {Ye}}]{bothwell22}%
  \BibitemOpen
  \bibfield  {author} {\bibinfo {author} {\bibfnamefont {T.}~\bibnamefont
  {Bothwell}}, \bibinfo {author} {\bibfnamefont {C.~J.}\ \bibnamefont
  {Kennedy}}, \bibinfo {author} {\bibfnamefont {A.}~\bibnamefont {Aeppli}},
  \bibinfo {author} {\bibfnamefont {D.}~\bibnamefont {Kedar}}, \bibinfo
  {author} {\bibfnamefont {J.~M.}\ \bibnamefont {Robinson}}, \bibinfo {author}
  {\bibfnamefont {E.}~\bibnamefont {Oelker}}, \bibinfo {author} {\bibfnamefont
  {A.}~\bibnamefont {Staron}},\ and\ \bibinfo {author} {\bibfnamefont
  {J.}~\bibnamefont {Ye}},\ }\bibfield  {title} {\bibinfo {title} {Resolving
  the gravitational redshift across a millimetre-scale atomic sample},\ }\href
  {https://doi.org/10.1038/s41586-021-04349-7} {\bibfield  {journal} {\bibinfo
  {journal} {Nature}\ }\textbf {\bibinfo {volume} {602}},\ \bibinfo {pages}
  {420} (\bibinfo {year} {2022})}\BibitemShut {NoStop}%
\bibitem [{\citenamefont {Zheng}\ \emph {et~al.}(2022)\citenamefont {Zheng},
  \citenamefont {Dolde}, \citenamefont {Lochab}, \citenamefont {Merriman},
  \citenamefont {Li},\ and\ \citenamefont {Kolkowitz}}]{zheng22}%
  \BibitemOpen
  \bibfield  {author} {\bibinfo {author} {\bibfnamefont {X.}~\bibnamefont
  {Zheng}}, \bibinfo {author} {\bibfnamefont {J.}~\bibnamefont {Dolde}},
  \bibinfo {author} {\bibfnamefont {V.}~\bibnamefont {Lochab}}, \bibinfo
  {author} {\bibfnamefont {B.~N.}\ \bibnamefont {Merriman}}, \bibinfo {author}
  {\bibfnamefont {H.}~\bibnamefont {Li}},\ and\ \bibinfo {author}
  {\bibfnamefont {S.}~\bibnamefont {Kolkowitz}},\ }\bibfield  {title} {\bibinfo
  {title} {Differential clock comparisons with a multiplexed optical lattice
  clock},\ }\href {https://doi.org/10.1038/s41586-021-04344-y} {\bibfield
  {journal} {\bibinfo  {journal} {Nature}\ }\textbf {\bibinfo {volume} {602}},\
  \bibinfo {pages} {425} (\bibinfo {year} {2022})}\BibitemShut {NoStop}%
\bibitem [{\citenamefont {Sonderhouse}\ \emph {et~al.}(2020)\citenamefont
  {Sonderhouse}, \citenamefont {Sanner}, \citenamefont {Hutson}, \citenamefont
  {Goban}, \citenamefont {Bilitewski}, \citenamefont {Yan}, \citenamefont
  {Milner}, \citenamefont {Rey},\ and\ \citenamefont {Ye}}]{sonderhouse20}%
  \BibitemOpen
  \bibfield  {author} {\bibinfo {author} {\bibfnamefont {L.}~\bibnamefont
  {Sonderhouse}}, \bibinfo {author} {\bibfnamefont {C.}~\bibnamefont {Sanner}},
  \bibinfo {author} {\bibfnamefont {R.~B.}\ \bibnamefont {Hutson}}, \bibinfo
  {author} {\bibfnamefont {A.}~\bibnamefont {Goban}}, \bibinfo {author}
  {\bibfnamefont {T.}~\bibnamefont {Bilitewski}}, \bibinfo {author}
  {\bibfnamefont {L.}~\bibnamefont {Yan}}, \bibinfo {author} {\bibfnamefont
  {W.~R.}\ \bibnamefont {Milner}}, \bibinfo {author} {\bibfnamefont {A.~M.}\
  \bibnamefont {Rey}},\ and\ \bibinfo {author} {\bibfnamefont {J.}~\bibnamefont
  {Ye}},\ }\bibfield  {title} {\bibinfo {title} {Thermodynamics of a deeply
  degenerate {SU} ({N})-symmetric {F}ermi gas},\ }\href
  {https://doi.org/10.1038/s41567-020-0986-6} {\bibfield  {journal} {\bibinfo
  {journal} {Nature Physics}\ }\textbf {\bibinfo {volume} {16}},\ \bibinfo
  {pages} {1216} (\bibinfo {year} {2020})}\BibitemShut {NoStop}%
\bibitem [{\citenamefont {Ozawa}\ \emph {et~al.}(2018)\citenamefont {Ozawa},
  \citenamefont {Taie}, \citenamefont {Takasu},\ and\ \citenamefont
  {Takahashi}}]{ozawa18}%
  \BibitemOpen
  \bibfield  {author} {\bibinfo {author} {\bibfnamefont {H.}~\bibnamefont
  {Ozawa}}, \bibinfo {author} {\bibfnamefont {S.}~\bibnamefont {Taie}},
  \bibinfo {author} {\bibfnamefont {Y.}~\bibnamefont {Takasu}},\ and\ \bibinfo
  {author} {\bibfnamefont {Y.}~\bibnamefont {Takahashi}},\ }\bibfield  {title}
  {\bibinfo {title} {Antiferromagnetic {S}pin {C}orrelation of
  $\mathrm{SU}(\mathcal{N})$ {F}ermi {G}as in an {O}ptical {S}uperlattice},\
  }\href {https://doi.org/10.1103/PhysRevLett.121.225303} {\bibfield  {journal}
  {\bibinfo  {journal} {Physical Review Letters}\ }\textbf {\bibinfo {volume}
  {121}},\ \bibinfo {pages} {225303} (\bibinfo {year} {2018})}\BibitemShut
  {NoStop}%
\bibitem [{\citenamefont {Sch{\"a}fer}\ \emph {et~al.}(2020)\citenamefont
  {Sch{\"a}fer}, \citenamefont {Fukuhara}, \citenamefont {Sugawa},
  \citenamefont {Takasu},\ and\ \citenamefont {Takahashi}}]{schafer20}%
  \BibitemOpen
  \bibfield  {author} {\bibinfo {author} {\bibfnamefont {F.}~\bibnamefont
  {Sch{\"a}fer}}, \bibinfo {author} {\bibfnamefont {T.}~\bibnamefont
  {Fukuhara}}, \bibinfo {author} {\bibfnamefont {S.}~\bibnamefont {Sugawa}},
  \bibinfo {author} {\bibfnamefont {Y.}~\bibnamefont {Takasu}},\ and\ \bibinfo
  {author} {\bibfnamefont {Y.}~\bibnamefont {Takahashi}},\ }\bibfield  {title}
  {\bibinfo {title} {Tools for quantum simulation with ultracold atoms in
  optical lattices},\ }\href {https://doi.org/10.1038/s42254-020-0195-3}
  {\bibfield  {journal} {\bibinfo  {journal} {Nature Reviews Physics}\ }\textbf
  {\bibinfo {volume} {2}},\ \bibinfo {pages} {411} (\bibinfo {year}
  {2020})}\BibitemShut {NoStop}%
\bibitem [{\citenamefont {Young}\ \emph {et~al.}(2022)\citenamefont {Young},
  \citenamefont {Eckner}, \citenamefont {Schine}, \citenamefont {Childs},\ and\
  \citenamefont {Kaufman}}]{young22}%
  \BibitemOpen
  \bibfield  {author} {\bibinfo {author} {\bibfnamefont {A.~W.}\ \bibnamefont
  {Young}}, \bibinfo {author} {\bibfnamefont {W.~J.}\ \bibnamefont {Eckner}},
  \bibinfo {author} {\bibfnamefont {N.}~\bibnamefont {Schine}}, \bibinfo
  {author} {\bibfnamefont {A.~M.}\ \bibnamefont {Childs}},\ and\ \bibinfo
  {author} {\bibfnamefont {A.~M.}\ \bibnamefont {Kaufman}},\ }\bibfield
  {title} {\bibinfo {title} {Tweezer-programmable 2d quantum walks in a
  hubbard-regime lattice},\ }\href {https://doi.org/10.1126/science.abo0608}
  {\bibfield  {journal} {\bibinfo  {journal} {Science}\ }\textbf {\bibinfo
  {volume} {377}},\ \bibinfo {pages} {885} (\bibinfo {year}
  {2022})}\BibitemShut {NoStop}%
\bibitem [{\citenamefont {Schine}\ \emph {et~al.}(2022)\citenamefont {Schine},
  \citenamefont {Young}, \citenamefont {Eckner}, \citenamefont {Martin},\ and\
  \citenamefont {Kaufman}}]{schine22}%
  \BibitemOpen
  \bibfield  {author} {\bibinfo {author} {\bibfnamefont {N.}~\bibnamefont
  {Schine}}, \bibinfo {author} {\bibfnamefont {A.~W.}\ \bibnamefont {Young}},
  \bibinfo {author} {\bibfnamefont {W.~J.}\ \bibnamefont {Eckner}}, \bibinfo
  {author} {\bibfnamefont {M.~J.}\ \bibnamefont {Martin}},\ and\ \bibinfo
  {author} {\bibfnamefont {A.~M.}\ \bibnamefont {Kaufman}},\ }\bibfield
  {title} {\bibinfo {title} {Long-lived {B}ell states in an array of optical
  clock qubits},\ }\href {https://doi.org/10.1038/s41567-022-01678-w}
  {\bibfield  {journal} {\bibinfo  {journal} {Nature Physics}\ }\textbf
  {\bibinfo {volume} {18}},\ \bibinfo {pages} {1067} (\bibinfo {year}
  {2022})}\BibitemShut {NoStop}%
\bibitem [{\citenamefont {Madjarov}\ \emph {et~al.}(2020)\citenamefont
  {Madjarov}, \citenamefont {Covey}, \citenamefont {Shaw}, \citenamefont
  {Choi}, \citenamefont {Kale}, \citenamefont {Cooper}, \citenamefont
  {Pichler}, \citenamefont {Schkolnik}, \citenamefont {Williams},\ and\
  \citenamefont {Endres}}]{madjarov20}%
  \BibitemOpen
  \bibfield  {author} {\bibinfo {author} {\bibfnamefont {I.~S.}\ \bibnamefont
  {Madjarov}}, \bibinfo {author} {\bibfnamefont {J.~P.}\ \bibnamefont {Covey}},
  \bibinfo {author} {\bibfnamefont {A.~L.}\ \bibnamefont {Shaw}}, \bibinfo
  {author} {\bibfnamefont {J.}~\bibnamefont {Choi}}, \bibinfo {author}
  {\bibfnamefont {A.}~\bibnamefont {Kale}}, \bibinfo {author} {\bibfnamefont
  {A.}~\bibnamefont {Cooper}}, \bibinfo {author} {\bibfnamefont
  {H.}~\bibnamefont {Pichler}}, \bibinfo {author} {\bibfnamefont
  {V.}~\bibnamefont {Schkolnik}}, \bibinfo {author} {\bibfnamefont {J.~R.}\
  \bibnamefont {Williams}},\ and\ \bibinfo {author} {\bibfnamefont
  {M.}~\bibnamefont {Endres}},\ }\bibfield  {title} {\bibinfo {title}
  {High-fidelity entanglement and detection of alkaline-earth {R}ydberg
  atoms},\ }\href {https://doi.org/10.1038/s41567-020-0903-z} {\bibfield
  {journal} {\bibinfo  {journal} {Nature Physics}\ }\textbf {\bibinfo {volume}
  {16}},\ \bibinfo {pages} {857} (\bibinfo {year} {2020})}\BibitemShut
  {NoStop}%
\bibitem [{\citenamefont {Ma}\ \emph {et~al.}(2022)\citenamefont {Ma},
  \citenamefont {Burgers}, \citenamefont {Liu}, \citenamefont {Wilson},
  \citenamefont {Zhang},\ and\ \citenamefont {Thompson}}]{ma22}%
  \BibitemOpen
  \bibfield  {author} {\bibinfo {author} {\bibfnamefont {S.}~\bibnamefont
  {Ma}}, \bibinfo {author} {\bibfnamefont {A.~P.}\ \bibnamefont {Burgers}},
  \bibinfo {author} {\bibfnamefont {G.}~\bibnamefont {Liu}}, \bibinfo {author}
  {\bibfnamefont {J.}~\bibnamefont {Wilson}}, \bibinfo {author} {\bibfnamefont
  {B.}~\bibnamefont {Zhang}},\ and\ \bibinfo {author} {\bibfnamefont {J.~D.}\
  \bibnamefont {Thompson}},\ }\bibfield  {title} {\bibinfo {title} {Universal
  gate operations on nuclear spin qubits in an optical tweezer array of
  $^{171}\mathrm{Yb}$ atoms},\ }\href
  {https://doi.org/10.1103/PhysRevX.12.021028} {\bibfield  {journal} {\bibinfo
  {journal} {Phys. Rev. X}\ }\textbf {\bibinfo {volume} {12}},\ \bibinfo
  {pages} {021028} (\bibinfo {year} {2022})}\BibitemShut {NoStop}%
\bibitem [{\citenamefont {Jenkins}\ \emph {et~al.}(2022)\citenamefont
  {Jenkins}, \citenamefont {Lis}, \citenamefont {Senoo}, \citenamefont
  {McGrew},\ and\ \citenamefont {Kaufman}}]{jenkins22}%
  \BibitemOpen
  \bibfield  {author} {\bibinfo {author} {\bibfnamefont {A.}~\bibnamefont
  {Jenkins}}, \bibinfo {author} {\bibfnamefont {J.~W.}\ \bibnamefont {Lis}},
  \bibinfo {author} {\bibfnamefont {A.}~\bibnamefont {Senoo}}, \bibinfo
  {author} {\bibfnamefont {W.~F.}\ \bibnamefont {McGrew}},\ and\ \bibinfo
  {author} {\bibfnamefont {A.~M.}\ \bibnamefont {Kaufman}},\ }\bibfield
  {title} {\bibinfo {title} {Ytterbium nuclear-spin qubits in an optical
  tweezer array},\ }\href {https://doi.org/10.1103/PhysRevX.12.021027}
  {\bibfield  {journal} {\bibinfo  {journal} {Phys. Rev. X}\ }\textbf {\bibinfo
  {volume} {12}},\ \bibinfo {pages} {021027} (\bibinfo {year}
  {2022})}\BibitemShut {NoStop}%
\bibitem [{\citenamefont {Sobelman}(2012)}]{sobelman12}%
  \BibitemOpen
  \bibfield  {author} {\bibinfo {author} {\bibfnamefont {I.~I.}\ \bibnamefont
  {Sobelman}},\ }\href@noop {} {\emph {\bibinfo {title} {Atomic spectra and
  radiative transitions}}}\ (\bibinfo  {publisher} {Springer Science \&
  Business Media},\ \bibinfo {year} {2012})\BibitemShut {NoStop}%
\bibitem [{\citenamefont {Boyd}\ \emph {et~al.}(2007)\citenamefont {Boyd},
  \citenamefont {Zelevinsky}, \citenamefont {Ludlow}, \citenamefont {Blatt},
  \citenamefont {Zanon-Willette}, \citenamefont {Foreman},\ and\ \citenamefont
  {Ye}}]{boyd07}%
  \BibitemOpen
  \bibfield  {author} {\bibinfo {author} {\bibfnamefont {M.~M.}\ \bibnamefont
  {Boyd}}, \bibinfo {author} {\bibfnamefont {T.}~\bibnamefont {Zelevinsky}},
  \bibinfo {author} {\bibfnamefont {A.~D.}\ \bibnamefont {Ludlow}}, \bibinfo
  {author} {\bibfnamefont {S.}~\bibnamefont {Blatt}}, \bibinfo {author}
  {\bibfnamefont {T.}~\bibnamefont {Zanon-Willette}}, \bibinfo {author}
  {\bibfnamefont {S.~M.}\ \bibnamefont {Foreman}},\ and\ \bibinfo {author}
  {\bibfnamefont {J.}~\bibnamefont {Ye}},\ }\bibfield  {title} {\bibinfo
  {title} {Nuclear spin effects in optical lattice clocks},\ }\href
  {https://doi.org/10.1103/PhysRevA.76.022510} {\bibfield  {journal} {\bibinfo
  {journal} {Physical Review A}\ }\textbf {\bibinfo {volume} {76}},\ \bibinfo
  {pages} {022510} (\bibinfo {year} {2007})}\BibitemShut {NoStop}%
\bibitem [{\citenamefont {Taichenachev}\ \emph {et~al.}(2006)\citenamefont
  {Taichenachev}, \citenamefont {Yudin}, \citenamefont {Oates}, \citenamefont
  {Hoyt}, \citenamefont {Barber},\ and\ \citenamefont
  {Hollberg}}]{taichenachev06}%
  \BibitemOpen
  \bibfield  {author} {\bibinfo {author} {\bibfnamefont {A.~V.}\ \bibnamefont
  {Taichenachev}}, \bibinfo {author} {\bibfnamefont {V.~I.}\ \bibnamefont
  {Yudin}}, \bibinfo {author} {\bibfnamefont {C.~W.}\ \bibnamefont {Oates}},
  \bibinfo {author} {\bibfnamefont {C.~W.}\ \bibnamefont {Hoyt}}, \bibinfo
  {author} {\bibfnamefont {Z.~W.}\ \bibnamefont {Barber}},\ and\ \bibinfo
  {author} {\bibfnamefont {L.}~\bibnamefont {Hollberg}},\ }\bibfield  {title}
  {\bibinfo {title} {Magnetic {F}ield-{I}nduced {S}pectroscopy of {F}orbidden
  {O}ptical {T}ransitions with {A}pplication to {L}attice-{B}ased {O}ptical
  {A}tomic {C}locks},\ }\href {https://doi.org/10.1103/PhysRevLett.96.083001}
  {\bibfield  {journal} {\bibinfo  {journal} {Physical Review Letters}\
  }\textbf {\bibinfo {volume} {96}},\ \bibinfo {pages} {083001} (\bibinfo
  {year} {2006})}\BibitemShut {NoStop}%
\bibitem [{\citenamefont {Garstang}(1967)}]{garstang67}%
  \BibitemOpen
  \bibfield  {author} {\bibinfo {author} {\bibfnamefont {R.~H.}\ \bibnamefont
  {Garstang}},\ }\bibfield  {title} {\bibinfo {title} {Magnetic {Q}uadrupole
  {L}ine {I}ntensities},\ }\href {https://doi.org/10.1086/149179} {\bibfield
  {journal} {\bibinfo  {journal} {Astrophysical Journal}\ }\textbf {\bibinfo
  {volume} {148}},\ \bibinfo {pages} {579} (\bibinfo {year}
  {1967})}\BibitemShut {NoStop}%
\bibitem [{\citenamefont {Heinz}\ \emph {et~al.}(2020)\citenamefont {Heinz},
  \citenamefont {Park}, \citenamefont {\v{S}anti{\'c}}, \citenamefont
  {Trautmann}, \citenamefont {Porsev}, \citenamefont {Safronova}, \citenamefont
  {Bloch},\ and\ \citenamefont {Blatt}}]{heinz20}%
  \BibitemOpen
  \bibfield  {author} {\bibinfo {author} {\bibfnamefont {A.}~\bibnamefont
  {Heinz}}, \bibinfo {author} {\bibfnamefont {A.~J.}\ \bibnamefont {Park}},
  \bibinfo {author} {\bibfnamefont {N.}~\bibnamefont {\v{S}anti{\'c}}},
  \bibinfo {author} {\bibfnamefont {J.}~\bibnamefont {Trautmann}}, \bibinfo
  {author} {\bibfnamefont {S.~G.}\ \bibnamefont {Porsev}}, \bibinfo {author}
  {\bibfnamefont {M.~S.}\ \bibnamefont {Safronova}}, \bibinfo {author}
  {\bibfnamefont {I.}~\bibnamefont {Bloch}},\ and\ \bibinfo {author}
  {\bibfnamefont {S.}~\bibnamefont {Blatt}},\ }\bibfield  {title} {\bibinfo
  {title} {State-dependent optical lattices for the strontium optical qubit},\
  }\href@noop {} {\bibfield  {journal} {\bibinfo  {journal} {Physical Review
  Letters}\ }\textbf {\bibinfo {volume} {124}},\ \bibinfo {pages} {203201}
  (\bibinfo {year} {2020})}\BibitemShut {NoStop}%
\bibitem [{\citenamefont {Takano}\ \emph {et~al.}(2017)\citenamefont {Takano},
  \citenamefont {Mizushima},\ and\ \citenamefont {Katori}}]{takano17}%
  \BibitemOpen
  \bibfield  {author} {\bibinfo {author} {\bibfnamefont {T.}~\bibnamefont
  {Takano}}, \bibinfo {author} {\bibfnamefont {R.}~\bibnamefont {Mizushima}},\
  and\ \bibinfo {author} {\bibfnamefont {H.}~\bibnamefont {Katori}},\
  }\bibfield  {title} {\bibinfo {title} {Precise determination of the isotope
  shift of 88sr-87sr optical lattice clock by sharing perturbations},\ }\href
  {https://doi.org/10.7567/apex.10.072801} {\bibfield  {journal} {\bibinfo
  {journal} {Applied Physics Express}\ }\textbf {\bibinfo {volume} {10}},\
  \bibinfo {pages} {072801} (\bibinfo {year} {2017})}\BibitemShut {NoStop}%
\bibitem [{\citenamefont {Riehle}\ \emph {et~al.}(2018)\citenamefont {Riehle},
  \citenamefont {Gill}, \citenamefont {Arias},\ and\ \citenamefont
  {Robertsson}}]{riehle18}%
  \BibitemOpen
  \bibfield  {author} {\bibinfo {author} {\bibfnamefont {F.}~\bibnamefont
  {Riehle}}, \bibinfo {author} {\bibfnamefont {P.}~\bibnamefont {Gill}},
  \bibinfo {author} {\bibfnamefont {F.}~\bibnamefont {Arias}},\ and\ \bibinfo
  {author} {\bibfnamefont {L.}~\bibnamefont {Robertsson}},\ }\bibfield  {title}
  {\bibinfo {title} {The {CIPM} list of recommended frequency standard values:
  guidelines and procedures},\ }\href
  {https://doi.org/10.1088/1681-7575/aaa302} {\bibfield  {journal} {\bibinfo
  {journal} {Metrologia}\ }\textbf {\bibinfo {volume} {55}},\ \bibinfo {pages}
  {188} (\bibinfo {year} {2018})}\BibitemShut {NoStop}%
\bibitem [{\citenamefont {Ferrari}\ \emph {et~al.}(2003)\citenamefont
  {Ferrari}, \citenamefont {Cancio}, \citenamefont {Drullinger}, \citenamefont
  {Giusfredi}, \citenamefont {Poli}, \citenamefont {Prevedelli}, \citenamefont
  {Toninelli},\ and\ \citenamefont {Tino}}]{ferrari03}%
  \BibitemOpen
  \bibfield  {author} {\bibinfo {author} {\bibfnamefont {G.}~\bibnamefont
  {Ferrari}}, \bibinfo {author} {\bibfnamefont {P.}~\bibnamefont {Cancio}},
  \bibinfo {author} {\bibfnamefont {R.}~\bibnamefont {Drullinger}}, \bibinfo
  {author} {\bibfnamefont {G.}~\bibnamefont {Giusfredi}}, \bibinfo {author}
  {\bibfnamefont {N.}~\bibnamefont {Poli}}, \bibinfo {author} {\bibfnamefont
  {M.}~\bibnamefont {Prevedelli}}, \bibinfo {author} {\bibfnamefont
  {C.}~\bibnamefont {Toninelli}},\ and\ \bibinfo {author} {\bibfnamefont
  {G.~M.}\ \bibnamefont {Tino}},\ }\bibfield  {title} {\bibinfo {title}
  {Precision frequency measurement of visible intercombination lines of
  strontium},\ }\href {https://doi.org/10.1103/PhysRevLett.91.243002}
  {\bibfield  {journal} {\bibinfo  {journal} {Physical Review Letters}\
  }\textbf {\bibinfo {volume} {91}},\ \bibinfo {pages} {243002} (\bibinfo
  {year} {2003})}\BibitemShut {NoStop}%
\bibitem [{\citenamefont {Nicholson}\ \emph {et~al.}(2015)\citenamefont
  {Nicholson}, \citenamefont {Campbell}, \citenamefont {Hutson}, \citenamefont
  {Marti}, \citenamefont {Bloom}, \citenamefont {McNally}, \citenamefont
  {Zhang}, \citenamefont {Barrett}, \citenamefont {Safronova}, \citenamefont
  {Strouse}, \citenamefont {Tew},\ and\ \citenamefont {Ye}}]{nicholson15}%
  \BibitemOpen
  \bibfield  {author} {\bibinfo {author} {\bibfnamefont {T.}~\bibnamefont
  {Nicholson}}, \bibinfo {author} {\bibfnamefont {S.}~\bibnamefont {Campbell}},
  \bibinfo {author} {\bibfnamefont {R.}~\bibnamefont {Hutson}}, \bibinfo
  {author} {\bibfnamefont {G.}~\bibnamefont {Marti}}, \bibinfo {author}
  {\bibfnamefont {B.}~\bibnamefont {Bloom}}, \bibinfo {author} {\bibfnamefont
  {R.}~\bibnamefont {McNally}}, \bibinfo {author} {\bibfnamefont
  {W.}~\bibnamefont {Zhang}}, \bibinfo {author} {\bibfnamefont
  {M.}~\bibnamefont {Barrett}}, \bibinfo {author} {\bibfnamefont
  {M.}~\bibnamefont {Safronova}}, \bibinfo {author} {\bibfnamefont
  {G.}~\bibnamefont {Strouse}}, \bibinfo {author} {\bibfnamefont
  {W.}~\bibnamefont {Tew}},\ and\ \bibinfo {author} {\bibfnamefont
  {J.}~\bibnamefont {Ye}},\ }\bibfield  {title} {\bibinfo {title} {Systematic
  evaluation of an atomic clock at {$2\times 10^{-18}$} total uncertainty},\
  }\href {https://doi.org/10.1038/ncomms7896} {\bibfield  {journal} {\bibinfo
  {journal} {Nat. Commun.}\ }\textbf {\bibinfo {volume} {6}},\ \bibinfo {pages}
  {6896} (\bibinfo {year} {2015})}\BibitemShut {NoStop}%
\bibitem [{\citenamefont {Campbell}\ \emph {et~al.}(2017)\citenamefont
  {Campbell}, \citenamefont {Hutson}, \citenamefont {Marti}, \citenamefont
  {Goban}, \citenamefont {Darkwah~Oppong}, \citenamefont {McNally},
  \citenamefont {Sonderhouse}, \citenamefont {Robinson}, \citenamefont {Zhang},
  \citenamefont {Bloom} \emph {et~al.}}]{campbell17}%
  \BibitemOpen
  \bibfield  {author} {\bibinfo {author} {\bibfnamefont {S.~L.}\ \bibnamefont
  {Campbell}}, \bibinfo {author} {\bibfnamefont {R.}~\bibnamefont {Hutson}},
  \bibinfo {author} {\bibfnamefont {G.}~\bibnamefont {Marti}}, \bibinfo
  {author} {\bibfnamefont {A.}~\bibnamefont {Goban}}, \bibinfo {author}
  {\bibfnamefont {N.}~\bibnamefont {Darkwah~Oppong}}, \bibinfo {author}
  {\bibfnamefont {R.}~\bibnamefont {McNally}}, \bibinfo {author} {\bibfnamefont
  {L.}~\bibnamefont {Sonderhouse}}, \bibinfo {author} {\bibfnamefont
  {J.}~\bibnamefont {Robinson}}, \bibinfo {author} {\bibfnamefont
  {W.}~\bibnamefont {Zhang}}, \bibinfo {author} {\bibfnamefont
  {B.}~\bibnamefont {Bloom}}, \emph {et~al.},\ }\bibfield  {title} {\bibinfo
  {title} {A fermi-degenerate three-dimensional optical lattice clock},\ }\href
  {https://doi.org/10.1126/science.aam5538} {\bibfield  {journal} {\bibinfo
  {journal} {Science}\ }\textbf {\bibinfo {volume} {358}},\ \bibinfo {pages}
  {90} (\bibinfo {year} {2017})}\BibitemShut {NoStop}%
\bibitem [{\citenamefont {Young}\ \emph {et~al.}(2020)\citenamefont {Young},
  \citenamefont {Eckner}, \citenamefont {Milner}, \citenamefont {Kedar},
  \citenamefont {Norcia}, \citenamefont {Oelker}, \citenamefont {Schine},
  \citenamefont {Ye},\ and\ \citenamefont {Kaufman}}]{young20}%
  \BibitemOpen
  \bibfield  {author} {\bibinfo {author} {\bibfnamefont {A.~W.}\ \bibnamefont
  {Young}}, \bibinfo {author} {\bibfnamefont {W.~J.}\ \bibnamefont {Eckner}},
  \bibinfo {author} {\bibfnamefont {W.~R.}\ \bibnamefont {Milner}}, \bibinfo
  {author} {\bibfnamefont {D.}~\bibnamefont {Kedar}}, \bibinfo {author}
  {\bibfnamefont {M.~A.}\ \bibnamefont {Norcia}}, \bibinfo {author}
  {\bibfnamefont {E.}~\bibnamefont {Oelker}}, \bibinfo {author} {\bibfnamefont
  {N.}~\bibnamefont {Schine}}, \bibinfo {author} {\bibfnamefont
  {J.}~\bibnamefont {Ye}},\ and\ \bibinfo {author} {\bibfnamefont {A.~M.}\
  \bibnamefont {Kaufman}},\ }\bibfield  {title} {\bibinfo {title}
  {Half-minute-scale atomic coherence and high relative stability in a tweezer
  clock},\ }\href {https://doi.org/10.1038/s41586-020-3009-y} {\bibfield
  {journal} {\bibinfo  {journal} {Nature}\ }\textbf {\bibinfo {volume} {588}},\
  \bibinfo {pages} {408} (\bibinfo {year} {2020})}\BibitemShut {NoStop}%
\bibitem [{\citenamefont {Brusch}\ \emph {et~al.}(2006)\citenamefont {Brusch},
  \citenamefont {Le~Targat}, \citenamefont {Baillard}, \citenamefont
  {Fouch\'e},\ and\ \citenamefont {Lemonde}}]{brusch06}%
  \BibitemOpen
  \bibfield  {author} {\bibinfo {author} {\bibfnamefont {A.}~\bibnamefont
  {Brusch}}, \bibinfo {author} {\bibfnamefont {R.}~\bibnamefont {Le~Targat}},
  \bibinfo {author} {\bibfnamefont {X.}~\bibnamefont {Baillard}}, \bibinfo
  {author} {\bibfnamefont {M.}~\bibnamefont {Fouch\'e}},\ and\ \bibinfo
  {author} {\bibfnamefont {P.}~\bibnamefont {Lemonde}},\ }\bibfield  {title}
  {\bibinfo {title} {Hyperpolarizability {E}ffects in a {S}r {O}ptical
  {L}attice {C}lock},\ }\href {https://doi.org/10.1103/PhysRevLett.96.103003}
  {\bibfield  {journal} {\bibinfo  {journal} {Physical Review Letters}\
  }\textbf {\bibinfo {volume} {96}},\ \bibinfo {pages} {103003} (\bibinfo
  {year} {2006})}\BibitemShut {NoStop}%
\bibitem [{\citenamefont {Bishof}\ \emph {et~al.}(2011)\citenamefont {Bishof},
  \citenamefont {Martin}, \citenamefont {Swallows}, \citenamefont {Benko},
  \citenamefont {Lin}, \citenamefont {Qu\'em\'ener}, \citenamefont {Rey},\ and\
  \citenamefont {Ye}}]{bishof11}%
  \BibitemOpen
  \bibfield  {author} {\bibinfo {author} {\bibfnamefont {M.}~\bibnamefont
  {Bishof}}, \bibinfo {author} {\bibfnamefont {M.~J.}\ \bibnamefont {Martin}},
  \bibinfo {author} {\bibfnamefont {M.~D.}\ \bibnamefont {Swallows}}, \bibinfo
  {author} {\bibfnamefont {C.}~\bibnamefont {Benko}}, \bibinfo {author}
  {\bibfnamefont {Y.}~\bibnamefont {Lin}}, \bibinfo {author} {\bibfnamefont
  {G.}~\bibnamefont {Qu\'em\'ener}}, \bibinfo {author} {\bibfnamefont {A.~M.}\
  \bibnamefont {Rey}},\ and\ \bibinfo {author} {\bibfnamefont {J.}~\bibnamefont
  {Ye}},\ }\bibfield  {title} {\bibinfo {title} {Inelastic collisions and
  density-dependent excitation suppression in a ${}^{87}${S}r optical lattice
  clock},\ }\href {https://doi.org/10.1103/PhysRevA.84.052716} {\bibfield
  {journal} {\bibinfo  {journal} {Physical Review A}\ }\textbf {\bibinfo
  {volume} {84}},\ \bibinfo {pages} {052716} (\bibinfo {year}
  {2011})}\BibitemShut {NoStop}%
\bibitem [{\citenamefont {Muniz}\ \emph {et~al.}(2021)\citenamefont {Muniz},
  \citenamefont {Young}, \citenamefont {Cline},\ and\ \citenamefont
  {Thompson}}]{muniz21}%
  \BibitemOpen
  \bibfield  {author} {\bibinfo {author} {\bibfnamefont {J.~A.}\ \bibnamefont
  {Muniz}}, \bibinfo {author} {\bibfnamefont {D.~J.}\ \bibnamefont {Young}},
  \bibinfo {author} {\bibfnamefont {J.~R.~K.}\ \bibnamefont {Cline}},\ and\
  \bibinfo {author} {\bibfnamefont {J.~K.}\ \bibnamefont {Thompson}},\
  }\bibfield  {title} {\bibinfo {title} {Cavity-{QED} measurements of the
  $^{87}\mathrm{Sr}$ millihertz optical clock transition and determination of
  its natural linewidth},\ }\href
  {https://doi.org/10.1103/PhysRevResearch.3.023152} {\bibfield  {journal}
  {\bibinfo  {journal} {Physical Review Research}\ }\textbf {\bibinfo {volume}
  {3}},\ \bibinfo {pages} {023152} (\bibinfo {year} {2021})}\BibitemShut
  {NoStop}%
\bibitem [{\citenamefont {McGrew}\ \emph {et~al.}(2018)\citenamefont {McGrew},
  \citenamefont {Zhang}, \citenamefont {Fasano}, \citenamefont {Sch{\"a}ffer},
  \citenamefont {Beloy}, \citenamefont {Nicolodi}, \citenamefont {Brown},
  \citenamefont {Hinkley}, \citenamefont {Milani}, \citenamefont {Schioppo},
  \citenamefont {Yoon},\ and\ \citenamefont {Ludlow}}]{mcgrew18}%
  \BibitemOpen
  \bibfield  {author} {\bibinfo {author} {\bibfnamefont {W.~F.}\ \bibnamefont
  {McGrew}}, \bibinfo {author} {\bibfnamefont {X.}~\bibnamefont {Zhang}},
  \bibinfo {author} {\bibfnamefont {R.~J.}\ \bibnamefont {Fasano}}, \bibinfo
  {author} {\bibfnamefont {S.~A.}\ \bibnamefont {Sch{\"a}ffer}}, \bibinfo
  {author} {\bibfnamefont {K.}~\bibnamefont {Beloy}}, \bibinfo {author}
  {\bibfnamefont {D.}~\bibnamefont {Nicolodi}}, \bibinfo {author}
  {\bibfnamefont {R.~C.}\ \bibnamefont {Brown}}, \bibinfo {author}
  {\bibfnamefont {N.}~\bibnamefont {Hinkley}}, \bibinfo {author} {\bibfnamefont
  {G.}~\bibnamefont {Milani}}, \bibinfo {author} {\bibfnamefont
  {M.}~\bibnamefont {Schioppo}}, \bibinfo {author} {\bibfnamefont {T.~H.}\
  \bibnamefont {Yoon}},\ and\ \bibinfo {author} {\bibfnamefont {A.~D.}\
  \bibnamefont {Ludlow}},\ }\bibfield  {title} {\bibinfo {title} {Atomic clock
  performance enabling geodesy below the centimetre level},\ }\href
  {https://doi.org/10.1038/s41586-018-0738-2} {\bibfield  {journal} {\bibinfo
  {journal} {Nature}\ }\textbf {\bibinfo {volume} {564}},\ \bibinfo {pages}
  {87} (\bibinfo {year} {2018})}\BibitemShut {NoStop}%
\bibitem [{\citenamefont {Pizzocaro}\ \emph {et~al.}(2020)\citenamefont
  {Pizzocaro}, \citenamefont {Bregolin}, \citenamefont {Barbieri},
  \citenamefont {Rauf}, \citenamefont {Levi},\ and\ \citenamefont
  {Calonico}}]{pizzocaro20}%
  \BibitemOpen
  \bibfield  {author} {\bibinfo {author} {\bibfnamefont {M.}~\bibnamefont
  {Pizzocaro}}, \bibinfo {author} {\bibfnamefont {F.}~\bibnamefont {Bregolin}},
  \bibinfo {author} {\bibfnamefont {P.}~\bibnamefont {Barbieri}}, \bibinfo
  {author} {\bibfnamefont {B.}~\bibnamefont {Rauf}}, \bibinfo {author}
  {\bibfnamefont {F.}~\bibnamefont {Levi}},\ and\ \bibinfo {author}
  {\bibfnamefont {D.}~\bibnamefont {Calonico}},\ }\bibfield  {title} {\bibinfo
  {title} {Absolute frequency measurement of the
  ${}^{1}${S}${}_{0}$-${}^{3}${P}${}_{0}$ transition of $^{171}\mathrm{Yb}$
  with a link to international atomic time},\ }\href
  {https://doi.org/10.1088/1681-7575/ab50e8} {\bibfield  {journal} {\bibinfo
  {journal} {Metrologia}\ }\textbf {\bibinfo {volume} {57}},\ \bibinfo {pages}
  {035007} (\bibinfo {year} {2020})}\BibitemShut {NoStop}%
\bibitem [{\citenamefont {Kobayashi}\ \emph {et~al.}(2022)\citenamefont
  {Kobayashi}, \citenamefont {Takamizawa}, \citenamefont {Akamatsu},
  \citenamefont {Kawasaki}, \citenamefont {Nishiyama}, \citenamefont {Hosaka},
  \citenamefont {Hisai}, \citenamefont {Wada}, \citenamefont {Inaba},
  \citenamefont {Tanabe},\ and\ \citenamefont {Yasuda}}]{kobayashi22}%
  \BibitemOpen
  \bibfield  {author} {\bibinfo {author} {\bibfnamefont {T.}~\bibnamefont
  {Kobayashi}}, \bibinfo {author} {\bibfnamefont {A.}~\bibnamefont
  {Takamizawa}}, \bibinfo {author} {\bibfnamefont {D.}~\bibnamefont
  {Akamatsu}}, \bibinfo {author} {\bibfnamefont {A.}~\bibnamefont {Kawasaki}},
  \bibinfo {author} {\bibfnamefont {A.}~\bibnamefont {Nishiyama}}, \bibinfo
  {author} {\bibfnamefont {K.}~\bibnamefont {Hosaka}}, \bibinfo {author}
  {\bibfnamefont {Y.}~\bibnamefont {Hisai}}, \bibinfo {author} {\bibfnamefont
  {M.}~\bibnamefont {Wada}}, \bibinfo {author} {\bibfnamefont {H.}~\bibnamefont
  {Inaba}}, \bibinfo {author} {\bibfnamefont {T.}~\bibnamefont {Tanabe}},\ and\
  \bibinfo {author} {\bibfnamefont {M.}~\bibnamefont {Yasuda}},\ }\bibfield
  {title} {\bibinfo {title} {Search for ultralight dark matter from long-term
  frequency comparisons of optical and microwave atomic clocks},\ }\href
  {https://doi.org/10.1103/PhysRevLett.129.241301} {\bibfield  {journal}
  {\bibinfo  {journal} {Phys. Rev. Lett.}\ }\textbf {\bibinfo {volume} {129}},\
  \bibinfo {pages} {241301} (\bibinfo {year} {2022})}\BibitemShut {NoStop}%
\bibitem [{\citenamefont {Tyumenev}\ \emph {et~al.}(2016)\citenamefont
  {Tyumenev}, \citenamefont {Favier}, \citenamefont {Bilicki}, \citenamefont
  {Bookjans}, \citenamefont {{Le Targat}}, \citenamefont {Lodewyck},
  \citenamefont {Nicolodi}, \citenamefont {{Le Coq}}, \citenamefont {Abgrall},
  \citenamefont {Guéna}, \citenamefont {{De Sarlo}},\ and\ \citenamefont
  {Bize}}]{tyumenev16}%
  \BibitemOpen
  \bibfield  {author} {\bibinfo {author} {\bibfnamefont {R.}~\bibnamefont
  {Tyumenev}}, \bibinfo {author} {\bibfnamefont {M.}~\bibnamefont {Favier}},
  \bibinfo {author} {\bibfnamefont {S.}~\bibnamefont {Bilicki}}, \bibinfo
  {author} {\bibfnamefont {E.}~\bibnamefont {Bookjans}}, \bibinfo {author}
  {\bibfnamefont {R.}~\bibnamefont {{Le Targat}}}, \bibinfo {author}
  {\bibfnamefont {J.}~\bibnamefont {Lodewyck}}, \bibinfo {author}
  {\bibfnamefont {D.}~\bibnamefont {Nicolodi}}, \bibinfo {author}
  {\bibfnamefont {Y.}~\bibnamefont {{Le Coq}}}, \bibinfo {author}
  {\bibfnamefont {M.}~\bibnamefont {Abgrall}}, \bibinfo {author} {\bibfnamefont
  {J.}~\bibnamefont {Guéna}}, \bibinfo {author} {\bibfnamefont
  {L.}~\bibnamefont {{De Sarlo}}},\ and\ \bibinfo {author} {\bibfnamefont
  {S.}~\bibnamefont {Bize}},\ }\bibfield  {title} {\bibinfo {title} {Comparing
  a mercury optical lattice clock with microwave and optical frequency
  standards},\ }\href {https://doi.org/10.1088/1367-2630/18/11/113002}
  {\bibfield  {journal} {\bibinfo  {journal} {New Journal of Physics}\ }\textbf
  {\bibinfo {volume} {18}},\ \bibinfo {pages} {113002} (\bibinfo {year}
  {2016})}\BibitemShut {NoStop}%
\bibitem [{\citenamefont {Ohmae}\ \emph {et~al.}(2020)\citenamefont {Ohmae},
  \citenamefont {Bregolin}, \citenamefont {Nemitz},\ and\ \citenamefont
  {Katori}}]{ohmae20}%
  \BibitemOpen
  \bibfield  {author} {\bibinfo {author} {\bibfnamefont {N.}~\bibnamefont
  {Ohmae}}, \bibinfo {author} {\bibfnamefont {F.}~\bibnamefont {Bregolin}},
  \bibinfo {author} {\bibfnamefont {N.}~\bibnamefont {Nemitz}},\ and\ \bibinfo
  {author} {\bibfnamefont {H.}~\bibnamefont {Katori}},\ }\bibfield  {title}
  {\bibinfo {title} {Direct measurement of the frequency ratio for {H}g and
  {Y}b optical lattice clocks and closure of the {H}g/{Y}b/{S}r loop},\ }\href
  {https://doi.org/10.1364/OE.391602} {\bibfield  {journal} {\bibinfo
  {journal} {Opt. Express}\ }\textbf {\bibinfo {volume} {28}},\ \bibinfo
  {pages} {15112} (\bibinfo {year} {2020})}\BibitemShut {NoStop}%
\bibitem [{\citenamefont {Mizushima}(1964)}]{mizushima64}%
  \BibitemOpen
  \bibfield  {author} {\bibinfo {author} {\bibfnamefont {M.}~\bibnamefont
  {Mizushima}},\ }\bibfield  {title} {\bibinfo {title} {$\ensuremath{\Delta
  {s}=\pm1}$ {M}agnetic {M}ultipole {R}adiative {T}ransitions},\ }\href
  {https://doi.org/10.1103/PhysRev.134.A883} {\bibfield  {journal} {\bibinfo
  {journal} {Physical Review}\ }\textbf {\bibinfo {volume} {134}},\ \bibinfo
  {pages} {A883} (\bibinfo {year} {1964})}\BibitemShut {NoStop}%
\bibitem [{\citenamefont {Mizushima}(1966)}]{mizushima66}%
  \BibitemOpen
  \bibfield  {author} {\bibinfo {author} {\bibfnamefont {M.}~\bibnamefont
  {Mizushima}},\ }\bibfield  {title} {\bibinfo {title} {$\delta$ s=$\pm$1
  {M}agnetic {Q}uadrupole {R}adiative {T}ransitions in {A}toms and
  {M}olecules},\ }\href {https://doi.org/10.1143/JPSJ.21.2335} {\bibfield
  {journal} {\bibinfo  {journal} {Journal of the Physical Society of Japan}\
  }\textbf {\bibinfo {volume} {21}},\ \bibinfo {pages} {2335} (\bibinfo {year}
  {1966})}\BibitemShut {NoStop}%
\bibitem [{\citenamefont {Bromley}\ \emph {et~al.}(2016)\citenamefont
  {Bromley}, \citenamefont {Zhu}, \citenamefont {Bishof}, \citenamefont
  {Zhang}, \citenamefont {Bothwell}, \citenamefont {Schachenmayer},
  \citenamefont {Nicholson}, \citenamefont {Kaiser}, \citenamefont {Yelin},
  \citenamefont {Lukin}, \citenamefont {Rey},\ and\ \citenamefont
  {Ye}}]{bromley16}%
  \BibitemOpen
  \bibfield  {author} {\bibinfo {author} {\bibfnamefont {S.~L.}\ \bibnamefont
  {Bromley}}, \bibinfo {author} {\bibfnamefont {B.}~\bibnamefont {Zhu}},
  \bibinfo {author} {\bibfnamefont {M.}~\bibnamefont {Bishof}}, \bibinfo
  {author} {\bibfnamefont {X.}~\bibnamefont {Zhang}}, \bibinfo {author}
  {\bibfnamefont {T.}~\bibnamefont {Bothwell}}, \bibinfo {author}
  {\bibfnamefont {J.}~\bibnamefont {Schachenmayer}}, \bibinfo {author}
  {\bibfnamefont {T.~L.}\ \bibnamefont {Nicholson}}, \bibinfo {author}
  {\bibfnamefont {R.}~\bibnamefont {Kaiser}}, \bibinfo {author} {\bibfnamefont
  {S.~F.}\ \bibnamefont {Yelin}}, \bibinfo {author} {\bibfnamefont {M.~D.}\
  \bibnamefont {Lukin}}, \bibinfo {author} {\bibfnamefont {A.~M.}\ \bibnamefont
  {Rey}},\ and\ \bibinfo {author} {\bibfnamefont {J.}~\bibnamefont {Ye}},\
  }\bibfield  {title} {\bibinfo {title} {Collective atomic scattering and
  motional effects in a dense coherent medium},\ }\bibfield  {journal}
  {\bibinfo  {journal} {Nature Communications}\ }\textbf {\bibinfo {volume}
  {7}},\ \href {https://doi.org/10.1038/ncomms11039} {10.1038/ncomms11039}
  (\bibinfo {year} {2016})\BibitemShut {NoStop}%
\bibitem [{\citenamefont {Ido}\ and\ \citenamefont {Katori}(2003)}]{ido03}%
  \BibitemOpen
  \bibfield  {author} {\bibinfo {author} {\bibfnamefont {T.}~\bibnamefont
  {Ido}}\ and\ \bibinfo {author} {\bibfnamefont {H.}~\bibnamefont {Katori}},\
  }\bibfield  {title} {\bibinfo {title} {Recoil-{F}ree {S}pectroscopy of
  {N}eutral {S}r {A}toms in the {L}amb-{D}icke {R}egime},\ }\href
  {https://doi.org/10.1103/PhysRevLett.91.053001} {\bibfield  {journal}
  {\bibinfo  {journal} {Physical Review Letters}\ }\textbf {\bibinfo {volume}
  {91}},\ \bibinfo {pages} {053001} (\bibinfo {year} {2003})}\BibitemShut
  {NoStop}%
\bibitem [{\citenamefont {Cooper}\ \emph {et~al.}(2018)\citenamefont {Cooper},
  \citenamefont {Covey}, \citenamefont {Madjarov}, \citenamefont {Porsev},
  \citenamefont {Safronova},\ and\ \citenamefont {Endres}}]{cooper18}%
  \BibitemOpen
  \bibfield  {author} {\bibinfo {author} {\bibfnamefont {A.}~\bibnamefont
  {Cooper}}, \bibinfo {author} {\bibfnamefont {J.~P.}\ \bibnamefont {Covey}},
  \bibinfo {author} {\bibfnamefont {I.~S.}\ \bibnamefont {Madjarov}}, \bibinfo
  {author} {\bibfnamefont {S.~G.}\ \bibnamefont {Porsev}}, \bibinfo {author}
  {\bibfnamefont {M.~S.}\ \bibnamefont {Safronova}},\ and\ \bibinfo {author}
  {\bibfnamefont {M.}~\bibnamefont {Endres}},\ }\bibfield  {title} {\bibinfo
  {title} {Alkaline-earth atoms in optical tweezers},\ }\href
  {https://doi.org/10.1103/PhysRevX.8.041055} {\bibfield  {journal} {\bibinfo
  {journal} {Physical Review X}\ }\textbf {\bibinfo {volume} {8}},\ \bibinfo
  {pages} {041055} (\bibinfo {year} {2018})}\BibitemShut {NoStop}%
\bibitem [{\citenamefont {Norcia}\ \emph {et~al.}(2018)\citenamefont {Norcia},
  \citenamefont {Young},\ and\ \citenamefont {Kaufman}}]{norcia18}%
  \BibitemOpen
  \bibfield  {author} {\bibinfo {author} {\bibfnamefont {M.~A.}\ \bibnamefont
  {Norcia}}, \bibinfo {author} {\bibfnamefont {A.~W.}\ \bibnamefont {Young}},\
  and\ \bibinfo {author} {\bibfnamefont {A.~M.}\ \bibnamefont {Kaufman}},\
  }\bibfield  {title} {\bibinfo {title} {Microscopic {C}ontrol and {D}etection
  of {U}ltracold {S}trontium in {O}ptical-{T}weezer {A}rrays},\ }\href
  {https://doi.org/10.1103/PhysRevX.8.041054} {\bibfield  {journal} {\bibinfo
  {journal} {Physical Review X}\ }\textbf {\bibinfo {volume} {8}},\ \bibinfo
  {pages} {041054} (\bibinfo {year} {2018})}\BibitemShut {NoStop}%
\bibitem [{\citenamefont {Yasuda}\ and\ \citenamefont
  {Katori}(2004)}]{yasuda04}%
  \BibitemOpen
  \bibfield  {author} {\bibinfo {author} {\bibfnamefont {M.}~\bibnamefont
  {Yasuda}}\ and\ \bibinfo {author} {\bibfnamefont {H.}~\bibnamefont
  {Katori}},\ }\bibfield  {title} {\bibinfo {title} {Lifetime measurement of
  the $^3\mathrm{P}_2$ metastable state of strontium atoms},\ }\href
  {https://doi.org/10.1103/PhysRevLett.92.153004} {\bibfield  {journal}
  {\bibinfo  {journal} {Physical Review Letters}\ }\textbf {\bibinfo {volume}
  {92}},\ \bibinfo {pages} {153004} (\bibinfo {year} {2004})}\BibitemShut
  {NoStop}%
\bibitem [{\citenamefont {Daley}\ \emph {et~al.}(2008)\citenamefont {Daley},
  \citenamefont {Boyd}, \citenamefont {Ye},\ and\ \citenamefont
  {Zoller}}]{daley08}%
  \BibitemOpen
  \bibfield  {author} {\bibinfo {author} {\bibfnamefont {A.~J.}\ \bibnamefont
  {Daley}}, \bibinfo {author} {\bibfnamefont {M.~M.}\ \bibnamefont {Boyd}},
  \bibinfo {author} {\bibfnamefont {J.}~\bibnamefont {Ye}},\ and\ \bibinfo
  {author} {\bibfnamefont {P.}~\bibnamefont {Zoller}},\ }\bibfield  {title}
  {\bibinfo {title} {Quantum computing with alkaline-earth-metal atoms},\
  }\href {https://doi.org/10.1103/PhysRevLett.101.170504} {\bibfield  {journal}
  {\bibinfo  {journal} {Physical Review Letters}\ }\textbf {\bibinfo {volume}
  {101}},\ \bibinfo {pages} {170504} (\bibinfo {year} {2008})}\BibitemShut
  {NoStop}%
\bibitem [{\citenamefont {Shibata}\ \emph {et~al.}(2009)\citenamefont
  {Shibata}, \citenamefont {Kato}, \citenamefont {Yamaguchi}, \citenamefont
  {Uetake},\ and\ \citenamefont {Takahashi}}]{shibata09}%
  \BibitemOpen
  \bibfield  {author} {\bibinfo {author} {\bibfnamefont {K.}~\bibnamefont
  {Shibata}}, \bibinfo {author} {\bibfnamefont {S.}~\bibnamefont {Kato}},
  \bibinfo {author} {\bibfnamefont {A.}~\bibnamefont {Yamaguchi}}, \bibinfo
  {author} {\bibfnamefont {S.}~\bibnamefont {Uetake}},\ and\ \bibinfo {author}
  {\bibfnamefont {Y.}~\bibnamefont {Takahashi}},\ }\bibfield  {title} {\bibinfo
  {title} {A scalable quantum computer with ultranarrow optical transition of
  ultracold neutral atoms in an optical lattice},\ }\href
  {https://doi.org/10.1007/s00340-009-3696-4} {\bibfield  {journal} {\bibinfo
  {journal} {Applied Physics B}\ }\textbf {\bibinfo {volume} {97}},\ \bibinfo
  {pages} {753} (\bibinfo {year} {2009})}\BibitemShut {NoStop}%
\bibitem [{\citenamefont {Kato}\ \emph {et~al.}(2012)\citenamefont {Kato},
  \citenamefont {Shibata}, \citenamefont {Yamamoto}, \citenamefont
  {Yoshikawa},\ and\ \citenamefont {Takahashi}}]{kato12}%
  \BibitemOpen
  \bibfield  {author} {\bibinfo {author} {\bibfnamefont {S.}~\bibnamefont
  {Kato}}, \bibinfo {author} {\bibfnamefont {K.}~\bibnamefont {Shibata}},
  \bibinfo {author} {\bibfnamefont {R.}~\bibnamefont {Yamamoto}}, \bibinfo
  {author} {\bibfnamefont {Y.}~\bibnamefont {Yoshikawa}},\ and\ \bibinfo
  {author} {\bibfnamefont {Y.}~\bibnamefont {Takahashi}},\ }\bibfield  {title}
  {\bibinfo {title} {Optical magnetic resonance imaging with an ultra-narrow
  optical transition},\ }\href {https://doi.org/10.1007/s00340-012-4893-0}
  {\bibfield  {journal} {\bibinfo  {journal} {Applied Physics B}\ }\textbf
  {\bibinfo {volume} {108}},\ \bibinfo {pages} {31} (\bibinfo {year}
  {2012})}\BibitemShut {NoStop}%
\bibitem [{\citenamefont {Shibata}\ \emph {et~al.}(2014)\citenamefont
  {Shibata}, \citenamefont {Yamamoto}, \citenamefont {Seki},\ and\
  \citenamefont {Takahashi}}]{shibata14}%
  \BibitemOpen
  \bibfield  {author} {\bibinfo {author} {\bibfnamefont {K.}~\bibnamefont
  {Shibata}}, \bibinfo {author} {\bibfnamefont {R.}~\bibnamefont {Yamamoto}},
  \bibinfo {author} {\bibfnamefont {Y.}~\bibnamefont {Seki}},\ and\ \bibinfo
  {author} {\bibfnamefont {Y.}~\bibnamefont {Takahashi}},\ }\bibfield  {title}
  {\bibinfo {title} {Optical spectral imaging of a single layer of a quantum
  gas with an ultranarrow optical transition},\ }\href
  {https://doi.org/10.1103/PhysRevA.89.031601} {\bibfield  {journal} {\bibinfo
  {journal} {Physical Review A}\ }\textbf {\bibinfo {volume} {89}},\ \bibinfo
  {pages} {031601} (\bibinfo {year} {2014})}\BibitemShut {NoStop}%
\bibitem [{\citenamefont {Yamamoto}\ \emph {et~al.}(2016)\citenamefont
  {Yamamoto}, \citenamefont {Kobayashi}, \citenamefont {Kuno}, \citenamefont
  {Kato},\ and\ \citenamefont {Takahashi}}]{yamamoto16}%
  \BibitemOpen
  \bibfield  {author} {\bibinfo {author} {\bibfnamefont {R.}~\bibnamefont
  {Yamamoto}}, \bibinfo {author} {\bibfnamefont {J.}~\bibnamefont {Kobayashi}},
  \bibinfo {author} {\bibfnamefont {T.}~\bibnamefont {Kuno}}, \bibinfo {author}
  {\bibfnamefont {K.}~\bibnamefont {Kato}},\ and\ \bibinfo {author}
  {\bibfnamefont {Y.}~\bibnamefont {Takahashi}},\ }\bibfield  {title} {\bibinfo
  {title} {An ytterbium quantum gas microscope with narrow-line laser
  cooling},\ }\href {https://doi.org/10.1088/1367-2630/18/2/023016} {\bibfield
  {journal} {\bibinfo  {journal} {New Journal of Physics}\ }\textbf {\bibinfo
  {volume} {18}},\ \bibinfo {pages} {023016} (\bibinfo {year}
  {2016})}\BibitemShut {NoStop}%
\bibitem [{\citenamefont {Bakr}\ \emph {et~al.}(2009)\citenamefont {Bakr},
  \citenamefont {Gillen}, \citenamefont {Peng}, \citenamefont {F{\"o}lling},\
  and\ \citenamefont {Greiner}}]{bakr09}%
  \BibitemOpen
  \bibfield  {author} {\bibinfo {author} {\bibfnamefont {W.~S.}\ \bibnamefont
  {Bakr}}, \bibinfo {author} {\bibfnamefont {J.~I.}\ \bibnamefont {Gillen}},
  \bibinfo {author} {\bibfnamefont {A.}~\bibnamefont {Peng}}, \bibinfo {author}
  {\bibfnamefont {S.}~\bibnamefont {F{\"o}lling}},\ and\ \bibinfo {author}
  {\bibfnamefont {M.}~\bibnamefont {Greiner}},\ }\bibfield  {title} {\bibinfo
  {title} {A quantum gas microscope for detecting single atoms in a
  {H}ubbard-regime optical lattice},\ }\href
  {https://doi.org/10.1038/nature08482} {\bibfield  {journal} {\bibinfo
  {journal} {Nature}\ }\textbf {\bibinfo {volume} {462}},\ \bibinfo {pages}
  {74} (\bibinfo {year} {2009})}\BibitemShut {NoStop}%
\bibitem [{\citenamefont {Sherson}\ \emph {et~al.}(2010)\citenamefont
  {Sherson}, \citenamefont {Weitenberg}, \citenamefont {Endres}, \citenamefont
  {Cheneau}, \citenamefont {Bloch},\ and\ \citenamefont {Kuhr}}]{sherson10}%
  \BibitemOpen
  \bibfield  {author} {\bibinfo {author} {\bibfnamefont {J.~F.}\ \bibnamefont
  {Sherson}}, \bibinfo {author} {\bibfnamefont {C.}~\bibnamefont {Weitenberg}},
  \bibinfo {author} {\bibfnamefont {M.}~\bibnamefont {Endres}}, \bibinfo
  {author} {\bibfnamefont {M.}~\bibnamefont {Cheneau}}, \bibinfo {author}
  {\bibfnamefont {I.}~\bibnamefont {Bloch}},\ and\ \bibinfo {author}
  {\bibfnamefont {S.}~\bibnamefont {Kuhr}},\ }\bibfield  {title} {\bibinfo
  {title} {Single-atom-resolved fluorescence imaging of an atomic {M}ott
  insulator},\ }\href {https://doi.org/10.1038/nature09378} {\bibfield
  {journal} {\bibinfo  {journal} {Nature}\ }\textbf {\bibinfo {volume} {467}},\
  \bibinfo {pages} {68} (\bibinfo {year} {2010})}\BibitemShut {NoStop}%
\bibitem [{\citenamefont {Kato}\ \emph {et~al.}(2013)\citenamefont {Kato},
  \citenamefont {Sugawa}, \citenamefont {Shibata}, \citenamefont {Yamamoto},\
  and\ \citenamefont {Takahashi}}]{kato13}%
  \BibitemOpen
  \bibfield  {author} {\bibinfo {author} {\bibfnamefont {S.}~\bibnamefont
  {Kato}}, \bibinfo {author} {\bibfnamefont {S.}~\bibnamefont {Sugawa}},
  \bibinfo {author} {\bibfnamefont {K.}~\bibnamefont {Shibata}}, \bibinfo
  {author} {\bibfnamefont {R.}~\bibnamefont {Yamamoto}},\ and\ \bibinfo
  {author} {\bibfnamefont {Y.}~\bibnamefont {Takahashi}},\ }\bibfield  {title}
  {\bibinfo {title} {Control of resonant interaction between electronic ground
  and excited states},\ }\href {https://doi.org/10.1103/PhysRevLett.110.173201}
  {\bibfield  {journal} {\bibinfo  {journal} {Physical Review Letters}\
  }\textbf {\bibinfo {volume} {110}},\ \bibinfo {pages} {173201} (\bibinfo
  {year} {2013})}\BibitemShut {NoStop}%
\bibitem [{\citenamefont {Yamaguchi}\ \emph {et~al.}(2008)\citenamefont
  {Yamaguchi}, \citenamefont {Uetake}, \citenamefont {Hashimoto}, \citenamefont
  {Doyle},\ and\ \citenamefont {Takahashi}}]{yamaguchi08}%
  \BibitemOpen
  \bibfield  {author} {\bibinfo {author} {\bibfnamefont {A.}~\bibnamefont
  {Yamaguchi}}, \bibinfo {author} {\bibfnamefont {S.}~\bibnamefont {Uetake}},
  \bibinfo {author} {\bibfnamefont {D.}~\bibnamefont {Hashimoto}}, \bibinfo
  {author} {\bibfnamefont {J.~M.}\ \bibnamefont {Doyle}},\ and\ \bibinfo
  {author} {\bibfnamefont {Y.}~\bibnamefont {Takahashi}},\ }\bibfield  {title}
  {\bibinfo {title} {Inelastic {C}ollisions in {O}ptically {T}rapped
  {U}ltracold {M}etastable {Y}tterbium},\ }\href
  {https://doi.org/10.1103/PhysRevLett.101.233002} {\bibfield  {journal}
  {\bibinfo  {journal} {Physical Review Letters}\ }\textbf {\bibinfo {volume}
  {101}},\ \bibinfo {pages} {233002} (\bibinfo {year} {2008})}\BibitemShut
  {NoStop}%
\bibitem [{\citenamefont {Okuno}\ \emph {et~al.}(2022)\citenamefont {Okuno},
  \citenamefont {Nakamura}, \citenamefont {Kusano}, \citenamefont {Takasu},
  \citenamefont {Takei}, \citenamefont {Konishi},\ and\ \citenamefont
  {Takahashi}}]{okuno22}%
  \BibitemOpen
  \bibfield  {author} {\bibinfo {author} {\bibfnamefont {D.}~\bibnamefont
  {Okuno}}, \bibinfo {author} {\bibfnamefont {Y.}~\bibnamefont {Nakamura}},
  \bibinfo {author} {\bibfnamefont {T.}~\bibnamefont {Kusano}}, \bibinfo
  {author} {\bibfnamefont {Y.}~\bibnamefont {Takasu}}, \bibinfo {author}
  {\bibfnamefont {N.}~\bibnamefont {Takei}}, \bibinfo {author} {\bibfnamefont
  {H.}~\bibnamefont {Konishi}},\ and\ \bibinfo {author} {\bibfnamefont
  {Y.}~\bibnamefont {Takahashi}},\ }\bibfield  {title} {\bibinfo {title}
  {High-resolution spectroscopy and single-photon rydberg excitation of
  reconfigurable ytterbium atom tweezer arrays utilizing a metastable state},\
  }\href {https://doi.org/10.7566/JPSJ.91.084301} {\bibfield  {journal}
  {\bibinfo  {journal} {Journal of the Physical Society of Japan}\ }\textbf
  {\bibinfo {volume} {91}},\ \bibinfo {pages} {084301} (\bibinfo {year}
  {2022})}\BibitemShut {NoStop}%
\bibitem [{\citenamefont {Onishchenko}\ \emph {et~al.}(2019)\citenamefont
  {Onishchenko}, \citenamefont {Pyatchenkov}, \citenamefont {Urech},
  \citenamefont {Chen}, \citenamefont {Bennetts}, \citenamefont {Siviloglou},\
  and\ \citenamefont {Schreck}}]{onishchenko19}%
  \BibitemOpen
  \bibfield  {author} {\bibinfo {author} {\bibfnamefont {O.}~\bibnamefont
  {Onishchenko}}, \bibinfo {author} {\bibfnamefont {S.}~\bibnamefont
  {Pyatchenkov}}, \bibinfo {author} {\bibfnamefont {A.}~\bibnamefont {Urech}},
  \bibinfo {author} {\bibfnamefont {C.~C.}\ \bibnamefont {Chen}}, \bibinfo
  {author} {\bibfnamefont {S.}~\bibnamefont {Bennetts}}, \bibinfo {author}
  {\bibfnamefont {G.~A.}\ \bibnamefont {Siviloglou}},\ and\ \bibinfo {author}
  {\bibfnamefont {F.}~\bibnamefont {Schreck}},\ }\bibfield  {title} {\bibinfo
  {title} {Frequency of the ultranarrow {$^{1}\mathrm{S}_{0}$ -
  $^{3}\mathrm{P}_{2}$} transition in {$^{87}$Sr}},\ }\href
  {https://doi.org/10.1103/PhysRevA.99.052503} {\bibfield  {journal} {\bibinfo
  {journal} {Physical Review A}\ }\textbf {\bibinfo {volume} {99}},\ \bibinfo
  {pages} {052503} (\bibinfo {year} {2019})}\BibitemShut {NoStop}%
\bibitem [{\citenamefont {Hertel}\ and\ \citenamefont
  {Schulz}(2014)}]{hertel14}%
  \BibitemOpen
  \bibfield  {author} {\bibinfo {author} {\bibfnamefont {I.~V.}\ \bibnamefont
  {Hertel}}\ and\ \bibinfo {author} {\bibfnamefont {C.-P.}\ \bibnamefont
  {Schulz}},\ }\href@noop {} {\emph {\bibinfo {title} {Atoms, molecules and
  optical physics}}}\ (\bibinfo  {publisher} {Springer},\ \bibinfo {year}
  {2014})\BibitemShut {NoStop}%
\bibitem [{\citenamefont {Johnson}(2007)}]{johnson07}%
  \BibitemOpen
  \bibfield  {author} {\bibinfo {author} {\bibfnamefont {W.~R.}\ \bibnamefont
  {Johnson}},\ }\href@noop {} {\emph {\bibinfo {title} {Atomic structure
  theory}}}\ (\bibinfo  {publisher} {Springer},\ \bibinfo {year}
  {2007})\BibitemShut {NoStop}%
\bibitem [{\citenamefont {Raab}(1975)}]{raab75}%
  \BibitemOpen
  \bibfield  {author} {\bibinfo {author} {\bibfnamefont {R.~E.}\ \bibnamefont
  {Raab}},\ }\bibfield  {title} {\bibinfo {title} {Magnetic multipole
  moments},\ }\href {https://doi.org/10.1080/00268977500101151} {\bibfield
  {journal} {\bibinfo  {journal} {Molecular Physics}\ }\textbf {\bibinfo
  {volume} {29}},\ \bibinfo {pages} {1323} (\bibinfo {year}
  {1975})}\BibitemShut {NoStop}%
\bibitem [{\citenamefont {Lamb}\ \emph {et~al.}(1987)\citenamefont {Lamb},
  \citenamefont {Schlicher},\ and\ \citenamefont {Scully}}]{lamb87}%
  \BibitemOpen
  \bibfield  {author} {\bibinfo {author} {\bibfnamefont {W.~E.}\ \bibnamefont
  {Lamb}}, \bibinfo {author} {\bibfnamefont {R.~R.}\ \bibnamefont
  {Schlicher}},\ and\ \bibinfo {author} {\bibfnamefont {M.~O.}\ \bibnamefont
  {Scully}},\ }\bibfield  {title} {\bibinfo {title} {Matter-field interaction
  in atomic physics and quantum optics},\ }\href
  {https://doi.org/10.1103/PhysRevA.36.2763} {\bibfield  {journal} {\bibinfo
  {journal} {Physical Review A}\ }\textbf {\bibinfo {volume} {36}},\ \bibinfo
  {pages} {2763} (\bibinfo {year} {1987})}\BibitemShut {NoStop}%
\bibitem [{\citenamefont {Edmonds}(1960)}]{edmonds60}%
  \BibitemOpen
  \bibfield  {author} {\bibinfo {author} {\bibfnamefont {A.~R.}\ \bibnamefont
  {Edmonds}},\ }\href@noop {} {\emph {\bibinfo {title} {Angular Momentum in
  Quantum Mechanics}}},\ \bibinfo {edition} {2nd}\ ed.\ (\bibinfo  {publisher}
  {Princeton University Press},\ \bibinfo {year} {1960})\BibitemShut {NoStop}%
\bibitem [{\citenamefont {Varshalovich}\ \emph {et~al.}(1988)\citenamefont
  {Varshalovich}, \citenamefont {Moskalev},\ and\ \citenamefont
  {Khersonskii}}]{varshalovich88}%
  \BibitemOpen
  \bibfield  {author} {\bibinfo {author} {\bibfnamefont {D.~A.}\ \bibnamefont
  {Varshalovich}}, \bibinfo {author} {\bibfnamefont {A.~N.}\ \bibnamefont
  {Moskalev}},\ and\ \bibinfo {author} {\bibfnamefont {V.~K.}\ \bibnamefont
  {Khersonskii}},\ }\href@noop {} {\emph {\bibinfo {title} {Quantum Theory of
  Angular Momentum}}}\ (\bibinfo  {publisher} {World Scientific, Singapore},\
  \bibinfo {year} {1988})\BibitemShut {NoStop}%
\bibitem [{\citenamefont {Auzinsh}\ \emph {et~al.}(2010)\citenamefont
  {Auzinsh}, \citenamefont {Budker},\ and\ \citenamefont
  {Rochester}}]{auzinsh10}%
  \BibitemOpen
  \bibfield  {author} {\bibinfo {author} {\bibfnamefont {M.}~\bibnamefont
  {Auzinsh}}, \bibinfo {author} {\bibfnamefont {D.}~\bibnamefont {Budker}},\
  and\ \bibinfo {author} {\bibfnamefont {S.}~\bibnamefont {Rochester}},\
  }\href@noop {} {\emph {\bibinfo {title} {Optically polarized atoms:
  understanding light-atom interactions}}}\ (\bibinfo  {publisher} {Oxford
  University Press},\ \bibinfo {year} {2010})\BibitemShut {NoStop}%
\bibitem [{\citenamefont {Lodahl}\ \emph {et~al.}(2017)\citenamefont {Lodahl},
  \citenamefont {Mahmoodian}, \citenamefont {Stobbe}, \citenamefont
  {Rauschenbeutel}, \citenamefont {Schneeweiss}, \citenamefont {Volz},
  \citenamefont {Pichler},\ and\ \citenamefont {Zoller}}]{lodahl17}%
  \BibitemOpen
  \bibfield  {author} {\bibinfo {author} {\bibfnamefont {P.}~\bibnamefont
  {Lodahl}}, \bibinfo {author} {\bibfnamefont {S.}~\bibnamefont {Mahmoodian}},
  \bibinfo {author} {\bibfnamefont {S.}~\bibnamefont {Stobbe}}, \bibinfo
  {author} {\bibfnamefont {A.}~\bibnamefont {Rauschenbeutel}}, \bibinfo
  {author} {\bibfnamefont {P.}~\bibnamefont {Schneeweiss}}, \bibinfo {author}
  {\bibfnamefont {J.}~\bibnamefont {Volz}}, \bibinfo {author} {\bibfnamefont
  {H.}~\bibnamefont {Pichler}},\ and\ \bibinfo {author} {\bibfnamefont
  {P.}~\bibnamefont {Zoller}},\ }\bibfield  {title} {\bibinfo {title} {Chiral
  quantum optics},\ }\href {https://doi.org/10.1038/nature21037} {\bibfield
  {journal} {\bibinfo  {journal} {Nature}\ }\textbf {\bibinfo {volume} {541}},\
  \bibinfo {pages} {473} (\bibinfo {year} {2017})}\BibitemShut {NoStop}%
\bibitem [{\citenamefont {Schulz}\ \emph {et~al.}(2020)\citenamefont {Schulz},
  \citenamefont {Peshkov}, \citenamefont {M\"uller}, \citenamefont {Lange},
  \citenamefont {Huntemann}, \citenamefont {Tamm}, \citenamefont {Peik},\ and\
  \citenamefont {Surzhykov}}]{schulz20}%
  \BibitemOpen
  \bibfield  {author} {\bibinfo {author} {\bibfnamefont {S.~A.-L.}\
  \bibnamefont {Schulz}}, \bibinfo {author} {\bibfnamefont {A.~A.}\
  \bibnamefont {Peshkov}}, \bibinfo {author} {\bibfnamefont {R.~A.}\
  \bibnamefont {M\"uller}}, \bibinfo {author} {\bibfnamefont {R.}~\bibnamefont
  {Lange}}, \bibinfo {author} {\bibfnamefont {N.}~\bibnamefont {Huntemann}},
  \bibinfo {author} {\bibfnamefont {C.}~\bibnamefont {Tamm}}, \bibinfo {author}
  {\bibfnamefont {E.}~\bibnamefont {Peik}},\ and\ \bibinfo {author}
  {\bibfnamefont {A.}~\bibnamefont {Surzhykov}},\ }\bibfield  {title} {\bibinfo
  {title} {Generalized excitation of atomic multipole transitions by twisted
  light modes},\ }\href {https://doi.org/10.1103/PhysRevA.102.012812}
  {\bibfield  {journal} {\bibinfo  {journal} {Physical Review A}\ }\textbf
  {\bibinfo {volume} {102}},\ \bibinfo {pages} {012812} (\bibinfo {year}
  {2020})}\BibitemShut {NoStop}%
\bibitem [{\citenamefont {Schmiegelow}\ and\ \citenamefont
  {Schmidt-Kaler}(2012)}]{schmiegelow12}%
  \BibitemOpen
  \bibfield  {author} {\bibinfo {author} {\bibfnamefont {C.~T.}\ \bibnamefont
  {Schmiegelow}}\ and\ \bibinfo {author} {\bibfnamefont {F.}~\bibnamefont
  {Schmidt-Kaler}},\ }\bibfield  {title} {\bibinfo {title} {Light with orbital
  angular momentum interacting with trapped ions},\ }\href
  {https://doi.org/10.1140/epjd/e2012-20730-4} {\bibfield  {journal} {\bibinfo
  {journal} {The European Physical Journal D}\ }\textbf {\bibinfo {volume}
  {66}},\ \bibinfo {pages} {1} (\bibinfo {year} {2012})}\BibitemShut {NoStop}%
\bibitem [{\citenamefont {Schmiegelow}\ \emph {et~al.}(2016)\citenamefont
  {Schmiegelow}, \citenamefont {Schulz}, \citenamefont {Kaufmann},
  \citenamefont {Ruster}, \citenamefont {Poschinger},\ and\ \citenamefont
  {Schmidt-Kaler}}]{schmiegelow16}%
  \BibitemOpen
  \bibfield  {author} {\bibinfo {author} {\bibfnamefont {C.~T.}\ \bibnamefont
  {Schmiegelow}}, \bibinfo {author} {\bibfnamefont {J.}~\bibnamefont {Schulz}},
  \bibinfo {author} {\bibfnamefont {H.}~\bibnamefont {Kaufmann}}, \bibinfo
  {author} {\bibfnamefont {T.}~\bibnamefont {Ruster}}, \bibinfo {author}
  {\bibfnamefont {U.~G.}\ \bibnamefont {Poschinger}},\ and\ \bibinfo {author}
  {\bibfnamefont {F.}~\bibnamefont {Schmidt-Kaler}},\ }\bibfield  {title}
  {\bibinfo {title} {Transfer of optical orbital angular momentum to a bound
  electron},\ }\href {https://doi.org/10.1038/ncomms12998} {\bibfield
  {journal} {\bibinfo  {journal} {Nature communications}\ }\textbf {\bibinfo
  {volume} {7}},\ \bibinfo {pages} {1} (\bibinfo {year} {2016})}\BibitemShut
  {NoStop}%
\bibitem [{\citenamefont {Snigirev}\ \emph {et~al.}(2019)\citenamefont
  {Snigirev}, \citenamefont {Park}, \citenamefont {Heinz}, \citenamefont
  {Bloch},\ and\ \citenamefont {Blatt}}]{snigirev19}%
  \BibitemOpen
  \bibfield  {author} {\bibinfo {author} {\bibfnamefont {S.}~\bibnamefont
  {Snigirev}}, \bibinfo {author} {\bibfnamefont {A.~J.}\ \bibnamefont {Park}},
  \bibinfo {author} {\bibfnamefont {A.}~\bibnamefont {Heinz}}, \bibinfo
  {author} {\bibfnamefont {I.}~\bibnamefont {Bloch}},\ and\ \bibinfo {author}
  {\bibfnamefont {S.}~\bibnamefont {Blatt}},\ }\bibfield  {title} {\bibinfo
  {title} {Fast and dense magneto-optical traps for strontium},\ }\href
  {https://doi.org/10.1103/PhysRevA.99.063421} {\bibfield  {journal} {\bibinfo
  {journal} {Physical Review A}\ }\textbf {\bibinfo {volume} {99}},\ \bibinfo
  {pages} {063421} (\bibinfo {year} {2019})}\BibitemShut {NoStop}%
\bibitem [{\citenamefont {Park}\ \emph {et~al.}(2022)\citenamefont {Park},
  \citenamefont {Trautmann.}, \citenamefont {{\v{S}anti\'{c}}}, \citenamefont
  {Kl\"{u}sener}, \citenamefont {Heinz}, \citenamefont {Bloch},\ and\
  \citenamefont {Blatt}}]{park22}%
  \BibitemOpen
  \bibfield  {author} {\bibinfo {author} {\bibfnamefont {A.~J.}\ \bibnamefont
  {Park}}, \bibinfo {author} {\bibfnamefont {J.}~\bibnamefont {Trautmann.}},
  \bibinfo {author} {\bibfnamefont {N.}~\bibnamefont {{\v{S}anti\'{c}}}},
  \bibinfo {author} {\bibfnamefont {V.}~\bibnamefont {Kl\"{u}sener}}, \bibinfo
  {author} {\bibfnamefont {A.}~\bibnamefont {Heinz}}, \bibinfo {author}
  {\bibfnamefont {I.}~\bibnamefont {Bloch}},\ and\ \bibinfo {author}
  {\bibfnamefont {S.}~\bibnamefont {Blatt}},\ }\bibfield  {title} {\bibinfo
  {title} {Cavity-enhanced optical lattices for scaling neutral atom quantum
  technologies to higher qubit numbers},\ }\href
  {https://doi.org/10.1103/PRXQuantum.3.030314} {\bibfield  {journal} {\bibinfo
   {journal} {PRX Quantum}\ }\textbf {\bibinfo {volume} {3}},\ \bibinfo {pages}
  {030314} (\bibinfo {year} {2022})}\BibitemShut {NoStop}%
\bibitem [{\citenamefont {Schmid}\ \emph {et~al.}(2006)\citenamefont {Schmid},
  \citenamefont {Thalhammer}, \citenamefont {Winkler}, \citenamefont {Lang},\
  and\ \citenamefont {Denschlag}}]{schmid06}%
  \BibitemOpen
  \bibfield  {author} {\bibinfo {author} {\bibfnamefont {S.}~\bibnamefont
  {Schmid}}, \bibinfo {author} {\bibfnamefont {G.}~\bibnamefont {Thalhammer}},
  \bibinfo {author} {\bibfnamefont {K.}~\bibnamefont {Winkler}}, \bibinfo
  {author} {\bibfnamefont {F.}~\bibnamefont {Lang}},\ and\ \bibinfo {author}
  {\bibfnamefont {J.~H.}\ \bibnamefont {Denschlag}},\ }\bibfield  {title}
  {\bibinfo {title} {Long distance transport of ultracold atoms using a 1{D}
  optical lattice},\ }\href {https://doi.org/10.1088/1367-2630/8/8/159}
  {\bibfield  {journal} {\bibinfo  {journal} {New Journal of Physics}\ }\textbf
  {\bibinfo {volume} {8}},\ \bibinfo {pages} {159} (\bibinfo {year}
  {2006})}\BibitemShut {NoStop}%
\bibitem [{\citenamefont {Blatt}\ \emph {et~al.}(2009)\citenamefont {Blatt},
  \citenamefont {Thomsen}, \citenamefont {Campbell}, \citenamefont {Ludlow},
  \citenamefont {Swallows}, \citenamefont {Martin}, \citenamefont {Boyd},\ and\
  \citenamefont {Ye}}]{blatt09}%
  \BibitemOpen
  \bibfield  {author} {\bibinfo {author} {\bibfnamefont {S.}~\bibnamefont
  {Blatt}}, \bibinfo {author} {\bibfnamefont {J.~W.}\ \bibnamefont {Thomsen}},
  \bibinfo {author} {\bibfnamefont {G.~K.}\ \bibnamefont {Campbell}}, \bibinfo
  {author} {\bibfnamefont {A.~D.}\ \bibnamefont {Ludlow}}, \bibinfo {author}
  {\bibfnamefont {M.~D.}\ \bibnamefont {Swallows}}, \bibinfo {author}
  {\bibfnamefont {M.~J.}\ \bibnamefont {Martin}}, \bibinfo {author}
  {\bibfnamefont {M.~M.}\ \bibnamefont {Boyd}},\ and\ \bibinfo {author}
  {\bibfnamefont {J.}~\bibnamefont {Ye}},\ }\bibfield  {title} {\bibinfo
  {title} {Rabi spectroscopy and excitation inhomogeneity in a one-dimensional
  optical lattice clock},\ }\href {https://doi.org/10.1103/PhysRevA.80.052703}
  {\bibfield  {journal} {\bibinfo  {journal} {Physical Review A}\ }\textbf
  {\bibinfo {volume} {80}},\ \bibinfo {pages} {052703} (\bibinfo {year}
  {2009})}\BibitemShut {NoStop}%
\bibitem [{\citenamefont {Traverso}\ \emph {et~al.}(2009)\citenamefont
  {Traverso}, \citenamefont {Chakraborty}, \citenamefont {Martinez~de Escobar},
  \citenamefont {Mickelson}, \citenamefont {Nagel}, \citenamefont {Yan},\ and\
  \citenamefont {Killian}}]{traverso09}%
  \BibitemOpen
  \bibfield  {author} {\bibinfo {author} {\bibfnamefont {A.}~\bibnamefont
  {Traverso}}, \bibinfo {author} {\bibfnamefont {R.}~\bibnamefont
  {Chakraborty}}, \bibinfo {author} {\bibfnamefont {Y.~N.}\ \bibnamefont
  {Martinez~de Escobar}}, \bibinfo {author} {\bibfnamefont {P.~G.}\
  \bibnamefont {Mickelson}}, \bibinfo {author} {\bibfnamefont {S.~B.}\
  \bibnamefont {Nagel}}, \bibinfo {author} {\bibfnamefont {M.}~\bibnamefont
  {Yan}},\ and\ \bibinfo {author} {\bibfnamefont {T.~C.}\ \bibnamefont
  {Killian}},\ }\bibfield  {title} {\bibinfo {title} {Inelastic and elastic
  collision rates for triplet states of ultracold strontium},\ }\href
  {https://doi.org/10.1103/PhysRevA.79.060702} {\bibfield  {journal} {\bibinfo
  {journal} {Physical Review A}\ }\textbf {\bibinfo {volume} {79}},\ \bibinfo
  {pages} {060702} (\bibinfo {year} {2009})}\BibitemShut {NoStop}%
\bibitem [{\citenamefont {Lisdat}\ \emph {et~al.}(2009)\citenamefont {Lisdat},
  \citenamefont {Winfred}, \citenamefont {Middelmann}, \citenamefont {Riehle},\
  and\ \citenamefont {Sterr}}]{lisdat09}%
  \BibitemOpen
  \bibfield  {author} {\bibinfo {author} {\bibfnamefont {C.}~\bibnamefont
  {Lisdat}}, \bibinfo {author} {\bibfnamefont {J.~S. R.~V.}\ \bibnamefont
  {Winfred}}, \bibinfo {author} {\bibfnamefont {T.}~\bibnamefont {Middelmann}},
  \bibinfo {author} {\bibfnamefont {F.}~\bibnamefont {Riehle}},\ and\ \bibinfo
  {author} {\bibfnamefont {U.}~\bibnamefont {Sterr}},\ }\bibfield  {title}
  {\bibinfo {title} {Collisional {L}osses, {D}ecoherence, and {F}requency
  {S}hifts in {O}ptical {L}attice {C}locks with {B}osons},\ }\href
  {https://doi.org/10.1103/PhysRevLett.103.090801} {\bibfield  {journal}
  {\bibinfo  {journal} {Physical Review Letters}\ }\textbf {\bibinfo {volume}
  {103}},\ \bibinfo {pages} {090801} (\bibinfo {year} {2009})}\BibitemShut
  {NoStop}%
\bibitem [{\citenamefont {Safronova}\ \emph {et~al.}(2013)\citenamefont
  {Safronova}, \citenamefont {Porsev}, \citenamefont {Safronova}, \citenamefont
  {Kozlov},\ and\ \citenamefont {Clark}}]{safronova13}%
  \BibitemOpen
  \bibfield  {author} {\bibinfo {author} {\bibfnamefont {M.~S.}\ \bibnamefont
  {Safronova}}, \bibinfo {author} {\bibfnamefont {S.~G.}\ \bibnamefont
  {Porsev}}, \bibinfo {author} {\bibfnamefont {U.~I.}\ \bibnamefont
  {Safronova}}, \bibinfo {author} {\bibfnamefont {M.~G.}\ \bibnamefont
  {Kozlov}},\ and\ \bibinfo {author} {\bibfnamefont {C.~W.}\ \bibnamefont
  {Clark}},\ }\bibfield  {title} {\bibinfo {title} {Blackbody-radiation shift
  in the {Sr} optical atomic clock},\ }\href
  {https://doi.org/10.1103/PhysRevA.87.012509} {\bibfield  {journal} {\bibinfo
  {journal} {Physical Review A}\ }\textbf {\bibinfo {volume} {87}},\ \bibinfo
  {pages} {012509} (\bibinfo {year} {2013})}\BibitemShut {NoStop}%
\bibitem [{\citenamefont {Safronova}\ \emph {et~al.}(2015)\citenamefont
  {Safronova}, \citenamefont {Zuhrianda}, \citenamefont {Safronova},\ and\
  \citenamefont {Clark}}]{safronova15}%
  \BibitemOpen
  \bibfield  {author} {\bibinfo {author} {\bibfnamefont {M.~S.}\ \bibnamefont
  {Safronova}}, \bibinfo {author} {\bibfnamefont {Z.}~\bibnamefont
  {Zuhrianda}}, \bibinfo {author} {\bibfnamefont {U.~I.}\ \bibnamefont
  {Safronova}},\ and\ \bibinfo {author} {\bibfnamefont {C.~W.}\ \bibnamefont
  {Clark}},\ }\bibfield  {title} {\bibinfo {title} {Extracting transition rates
  from zero-polarizability spectroscopy},\ }\href
  {https://doi.org/10.1103/PhysRevA.92.040501} {\bibfield  {journal} {\bibinfo
  {journal} {Physical Review A}\ }\textbf {\bibinfo {volume} {92}},\ \bibinfo
  {pages} {040501} (\bibinfo {year} {2015})}\BibitemShut {NoStop}%
\bibitem [{\citenamefont {Kien}\ \emph {et~al.}(2013)\citenamefont {Kien},
  \citenamefont {Schneeweiss},\ and\ \citenamefont
  {Rauschenbeutel}}]{lekien13}%
  \BibitemOpen
  \bibfield  {author} {\bibinfo {author} {\bibfnamefont {F.~L.}\ \bibnamefont
  {Kien}}, \bibinfo {author} {\bibfnamefont {P.}~\bibnamefont {Schneeweiss}},\
  and\ \bibinfo {author} {\bibfnamefont {A.}~\bibnamefont {Rauschenbeutel}},\
  }\bibfield  {title} {\bibinfo {title} {Dynamical polarizability of atoms in
  arbitrary light fields: general theory and application to cesium},\ }\href
  {https://doi.org/10.1140/epjd/e2013-30729-x} {\bibfield  {journal} {\bibinfo
  {journal} {The European Physical Journal D}\ }\textbf {\bibinfo {volume}
  {67}},\ \bibinfo {pages} {1} (\bibinfo {year} {2013})}\BibitemShut {NoStop}%
\bibitem [{\citenamefont {Rosenbusch}\ \emph {et~al.}(2009)\citenamefont
  {Rosenbusch}, \citenamefont {Ghezali}, \citenamefont {Dzuba}, \citenamefont
  {Flambaum}, \citenamefont {Beloy},\ and\ \citenamefont
  {Derevianko}}]{rosenbusch09}%
  \BibitemOpen
  \bibfield  {author} {\bibinfo {author} {\bibfnamefont {P.}~\bibnamefont
  {Rosenbusch}}, \bibinfo {author} {\bibfnamefont {S.}~\bibnamefont {Ghezali}},
  \bibinfo {author} {\bibfnamefont {V.~A.}\ \bibnamefont {Dzuba}}, \bibinfo
  {author} {\bibfnamefont {V.~V.}\ \bibnamefont {Flambaum}}, \bibinfo {author}
  {\bibfnamefont {K.}~\bibnamefont {Beloy}},\ and\ \bibinfo {author}
  {\bibfnamefont {A.}~\bibnamefont {Derevianko}},\ }\bibfield  {title}
  {\bibinfo {title} {ac {S}tark shift of the {C}s microwave atomic clock
  transitions},\ }\href {https://doi.org/10.1103/PhysRevA.79.013404} {\bibfield
   {journal} {\bibinfo  {journal} {Physical Review A}\ }\textbf {\bibinfo
  {volume} {79}},\ \bibinfo {pages} {013404} (\bibinfo {year}
  {2009})}\BibitemShut {NoStop}%
\bibitem [{\citenamefont {Leibfried}\ \emph {et~al.}(2003)\citenamefont
  {Leibfried}, \citenamefont {Blatt}, \citenamefont {Monroe},\ and\
  \citenamefont {Wineland}}]{leibfried03}%
  \BibitemOpen
  \bibfield  {author} {\bibinfo {author} {\bibfnamefont {D.}~\bibnamefont
  {Leibfried}}, \bibinfo {author} {\bibfnamefont {R.}~\bibnamefont {Blatt}},
  \bibinfo {author} {\bibfnamefont {C.}~\bibnamefont {Monroe}},\ and\ \bibinfo
  {author} {\bibfnamefont {D.}~\bibnamefont {Wineland}},\ }\bibfield  {title}
  {\bibinfo {title} {Quantum dynamics of single trapped ions},\ }\href
  {https://doi.org/10.1103/RevModPhys.75.281} {\bibfield  {journal} {\bibinfo
  {journal} {Rev. Mod. Phys.}\ }\textbf {\bibinfo {volume} {75}},\ \bibinfo
  {pages} {281} (\bibinfo {year} {2003})}\BibitemShut {NoStop}%
\bibitem [{\citenamefont {Heider}\ and\ \citenamefont
  {Brink}(1977)}]{heider77}%
  \BibitemOpen
  \bibfield  {author} {\bibinfo {author} {\bibfnamefont {S.~M.}\ \bibnamefont
  {Heider}}\ and\ \bibinfo {author} {\bibfnamefont {G.~O.}\ \bibnamefont
  {Brink}},\ }\bibfield  {title} {\bibinfo {title} {Hyperfine structure of
  $^{87}\mathrm{Sr}$ in the $^{3}\mathrm{{P}}_{2}$ metastable state},\ }\href
  {https://doi.org/10.1103/PhysRevA.16.1371} {\bibfield  {journal} {\bibinfo
  {journal} {Physical Review A}\ }\textbf {\bibinfo {volume} {16}},\ \bibinfo
  {pages} {1371} (\bibinfo {year} {1977})}\BibitemShut {NoStop}%
\bibitem [{\citenamefont {Weitenberg}\ \emph {et~al.}(2011)\citenamefont
  {Weitenberg}, \citenamefont {Endres}, \citenamefont {Sherson}, \citenamefont
  {Cheneau}, \citenamefont {Schau{\ss}}, \citenamefont {Fukuhara},
  \citenamefont {Bloch},\ and\ \citenamefont {Kuhr}}]{weitenberg11}%
  \BibitemOpen
  \bibfield  {author} {\bibinfo {author} {\bibfnamefont {C.}~\bibnamefont
  {Weitenberg}}, \bibinfo {author} {\bibfnamefont {M.}~\bibnamefont {Endres}},
  \bibinfo {author} {\bibfnamefont {J.~F.}\ \bibnamefont {Sherson}}, \bibinfo
  {author} {\bibfnamefont {M.}~\bibnamefont {Cheneau}}, \bibinfo {author}
  {\bibfnamefont {P.}~\bibnamefont {Schau{\ss}}}, \bibinfo {author}
  {\bibfnamefont {T.}~\bibnamefont {Fukuhara}}, \bibinfo {author}
  {\bibfnamefont {I.}~\bibnamefont {Bloch}},\ and\ \bibinfo {author}
  {\bibfnamefont {S.}~\bibnamefont {Kuhr}},\ }\bibfield  {title} {\bibinfo
  {title} {Single-spin addressing in an atomic {M}ott insulator},\ }\href
  {https://doi.org/10.1038/nature09827} {\bibfield  {journal} {\bibinfo
  {journal} {Nature}\ }\textbf {\bibinfo {volume} {471}},\ \bibinfo {pages}
  {319} (\bibinfo {year} {2011})}\BibitemShut {NoStop}%
\bibitem [{\citenamefont {Daley}(2011)}]{daley11}%
  \BibitemOpen
  \bibfield  {author} {\bibinfo {author} {\bibfnamefont {A.~J.}\ \bibnamefont
  {Daley}},\ }\bibfield  {title} {\bibinfo {title} {Quantum computing and
  quantum simulation with group-{II} atoms},\ }\href
  {https://doi.org/10.1007/s11128-011-0293-3} {\bibfield  {journal} {\bibinfo
  {journal} {Quantum Information Processing}\ }\textbf {\bibinfo {volume}
  {10}},\ \bibinfo {pages} {865} (\bibinfo {year} {2011})}\BibitemShut
  {NoStop}%
\bibitem [{\citenamefont {Gross}\ and\ \citenamefont {Bakr}(2021)}]{gross21}%
  \BibitemOpen
  \bibfield  {author} {\bibinfo {author} {\bibfnamefont {C.}~\bibnamefont
  {Gross}}\ and\ \bibinfo {author} {\bibfnamefont {W.~S.}\ \bibnamefont
  {Bakr}},\ }\bibfield  {title} {\bibinfo {title} {Quantum gas microscopy for
  single atom and spin detection},\ }\href
  {https://doi.org/10.1038/s41567-021-01370-5} {\bibfield  {journal} {\bibinfo
  {journal} {Nature Physics}\ }\textbf {\bibinfo {volume} {17}},\ \bibinfo
  {pages} {1316} (\bibinfo {year} {2021})}\BibitemShut {NoStop}%
\bibitem [{\citenamefont {Born}\ and\ \citenamefont {Wolf}(2013)}]{born13}%
  \BibitemOpen
  \bibfield  {author} {\bibinfo {author} {\bibfnamefont {M.}~\bibnamefont
  {Born}}\ and\ \bibinfo {author} {\bibfnamefont {E.}~\bibnamefont {Wolf}},\
  }\href@noop {} {\emph {\bibinfo {title} {Principles of {O}ptics:
  {E}lectromagnetic {T}heory of {P}ropagation, {I}nterference and {D}iffraction
  of {L}ight}}}\ (\bibinfo  {publisher} {Elsevier},\ \bibinfo {year}
  {2013})\BibitemShut {NoStop}%
\bibitem [{\citenamefont {Sansonetti}\ and\ \citenamefont
  {Nave}(2010)}]{sansonetti10}%
  \BibitemOpen
  \bibfield  {author} {\bibinfo {author} {\bibfnamefont {J.}~\bibnamefont
  {Sansonetti}}\ and\ \bibinfo {author} {\bibfnamefont {G.}~\bibnamefont
  {Nave}},\ }\bibfield  {title} {\bibinfo {title} {Wavelengths, transition
  probabilities, and energy levels for the spectrum of neutral strontium ({S}r
  {I})},\ }\href {https://doi.org/10.1063/1.3449176} {\bibfield  {journal}
  {\bibinfo  {journal} {Journal of Physical and Chemical Reference Data}\
  }\textbf {\bibinfo {volume} {39}},\ \bibinfo {pages} {033103} (\bibinfo
  {year} {2010})}\BibitemShut {NoStop}%
\bibitem [{\citenamefont {Haller}\ \emph {et~al.}(2015)\citenamefont {Haller},
  \citenamefont {Hudson}, \citenamefont {Kelly}, \citenamefont {Cotta},
  \citenamefont {Peaudecerf}, \citenamefont {Bruce},\ and\ \citenamefont
  {Kuhr}}]{haller15}%
  \BibitemOpen
  \bibfield  {author} {\bibinfo {author} {\bibfnamefont {E.}~\bibnamefont
  {Haller}}, \bibinfo {author} {\bibfnamefont {J.}~\bibnamefont {Hudson}},
  \bibinfo {author} {\bibfnamefont {A.}~\bibnamefont {Kelly}}, \bibinfo
  {author} {\bibfnamefont {D.~A.}\ \bibnamefont {Cotta}}, \bibinfo {author}
  {\bibfnamefont {B.}~\bibnamefont {Peaudecerf}}, \bibinfo {author}
  {\bibfnamefont {G.~D.}\ \bibnamefont {Bruce}},\ and\ \bibinfo {author}
  {\bibfnamefont {S.}~\bibnamefont {Kuhr}},\ }\bibfield  {title} {\bibinfo
  {title} {Single-atom imaging of fermions in a quantum-gas microscope},\
  }\href {https://doi.org/10.1038/nphys3403} {\bibfield  {journal} {\bibinfo
  {journal} {Nature Physics}\ }\textbf {\bibinfo {volume} {11}},\ \bibinfo
  {pages} {738} (\bibinfo {year} {2015})}\BibitemShut {NoStop}%
\bibitem [{\citenamefont {Pagano}\ \emph {et~al.}(2022)\citenamefont {Pagano},
  \citenamefont {Weber}, \citenamefont {Jaschke}, \citenamefont {Pfau},
  \citenamefont {Meinert}, \citenamefont {Montangero},\ and\ \citenamefont
  {B\"uchler}}]{pagano22}%
  \BibitemOpen
  \bibfield  {author} {\bibinfo {author} {\bibfnamefont {A.}~\bibnamefont
  {Pagano}}, \bibinfo {author} {\bibfnamefont {S.}~\bibnamefont {Weber}},
  \bibinfo {author} {\bibfnamefont {D.}~\bibnamefont {Jaschke}}, \bibinfo
  {author} {\bibfnamefont {T.}~\bibnamefont {Pfau}}, \bibinfo {author}
  {\bibfnamefont {F.}~\bibnamefont {Meinert}}, \bibinfo {author} {\bibfnamefont
  {S.}~\bibnamefont {Montangero}},\ and\ \bibinfo {author} {\bibfnamefont
  {H.~P.}\ \bibnamefont {B\"uchler}},\ }\bibfield  {title} {\bibinfo {title}
  {Error budgeting for a controlled-phase gate with strontium-88 {R}ydberg
  atoms},\ }\href {https://doi.org/10.1103/PhysRevResearch.4.033019} {\bibfield
   {journal} {\bibinfo  {journal} {Physical Review Research}\ }\textbf
  {\bibinfo {volume} {4}},\ \bibinfo {pages} {033019} (\bibinfo {year}
  {2022})}\BibitemShut {NoStop}%
\bibitem [{\citenamefont {Cohen-Tannoudji}\ \emph {et~al.}(2004)\citenamefont
  {Cohen-Tannoudji}, \citenamefont {Dupont-Roc},\ and\ \citenamefont
  {Grynberg}}]{cohen-tannoudji04}%
  \BibitemOpen
  \bibfield  {author} {\bibinfo {author} {\bibfnamefont {C.}~\bibnamefont
  {Cohen-Tannoudji}}, \bibinfo {author} {\bibfnamefont {J.}~\bibnamefont
  {Dupont-Roc}},\ and\ \bibinfo {author} {\bibfnamefont {G.}~\bibnamefont
  {Grynberg}},\ }\href@noop {} {\emph {\bibinfo {title} {Photons \& Atoms --
  Introduction to Quantum Electrodynamics}}}\ (\bibinfo  {publisher} {Wiley},\
  \bibinfo {year} {2004})\BibitemShut {NoStop}%
\bibitem [{\citenamefont {Loudon}(2000)}]{loudon00}%
  \BibitemOpen
  \bibfield  {author} {\bibinfo {author} {\bibfnamefont {R.}~\bibnamefont
  {Loudon}},\ }\href@noop {} {\emph {\bibinfo {title} {The Quantum Theory of
  Light}}},\ \bibinfo {edition} {3rd}\ ed.\ (\bibinfo  {publisher} {Oxford
  Science Publications},\ \bibinfo {year} {2000})\BibitemShut {NoStop}%
\bibitem [{\citenamefont {Bransden}\ and\ \citenamefont
  {Joachain}(2003)}]{bransden03}%
  \BibitemOpen
  \bibfield  {author} {\bibinfo {author} {\bibfnamefont {B.~H.}\ \bibnamefont
  {Bransden}}\ and\ \bibinfo {author} {\bibfnamefont {C.~J.}\ \bibnamefont
  {Joachain}},\ }\href@noop {} {\emph {\bibinfo {title} {Physics of Atoms and
  Molecules}}},\ \bibinfo {edition} {2nd}\ ed.\ (\bibinfo  {publisher}
  {Prentice-Hall},\ \bibinfo {year} {2003})\BibitemShut {NoStop}%
\bibitem [{\citenamefont {Jackson}(1999)}]{jackson99}%
  \BibitemOpen
  \bibfield  {author} {\bibinfo {author} {\bibfnamefont {J.~D.}\ \bibnamefont
  {Jackson}},\ }\href@noop {} {\emph {\bibinfo {title} {Classical
  electrodynamics}}},\ \bibinfo {edition} {3rd}\ ed.\ (\bibinfo  {publisher}
  {Wiley},\ \bibinfo {year} {1999})\BibitemShut {NoStop}%
\bibitem [{\citenamefont {Blatt}\ and\ \citenamefont
  {Weisskopf}(1979)}]{blatt79}%
  \BibitemOpen
  \bibfield  {author} {\bibinfo {author} {\bibfnamefont {J.~M.}\ \bibnamefont
  {Blatt}}\ and\ \bibinfo {author} {\bibfnamefont {V.~F.}\ \bibnamefont
  {Weisskopf}},\ }\href@noop {} {\emph {\bibinfo {title} {Theoretical nuclear
  physics}}}\ (\bibinfo  {publisher} {Springer},\ \bibinfo {year}
  {1979})\BibitemShut {NoStop}%
\bibitem [{\citenamefont {L{\'e}onard}\ \emph {et~al.}(2014)\citenamefont
  {L{\'e}onard}, \citenamefont {Lee}, \citenamefont {Morales}, \citenamefont
  {Karg}, \citenamefont {Esslinger},\ and\ \citenamefont {Donner}}]{leonard14}%
  \BibitemOpen
  \bibfield  {author} {\bibinfo {author} {\bibfnamefont {J.}~\bibnamefont
  {L{\'e}onard}}, \bibinfo {author} {\bibfnamefont {M.}~\bibnamefont {Lee}},
  \bibinfo {author} {\bibfnamefont {A.}~\bibnamefont {Morales}}, \bibinfo
  {author} {\bibfnamefont {T.~M.}\ \bibnamefont {Karg}}, \bibinfo {author}
  {\bibfnamefont {T.}~\bibnamefont {Esslinger}},\ and\ \bibinfo {author}
  {\bibfnamefont {T.}~\bibnamefont {Donner}},\ }\bibfield  {title} {\bibinfo
  {title} {Optical transport and manipulation of an ultracold atomic cloud
  using focus-tunable lenses},\ }\href
  {https://doi.org/10.1088/1367-2630/16/9/093028} {\bibfield  {journal}
  {\bibinfo  {journal} {New Journal of Physics}\ }\textbf {\bibinfo {volume}
  {16}},\ \bibinfo {pages} {093028} (\bibinfo {year} {2014})}\BibitemShut
  {NoStop}%
\bibitem [{\citenamefont {Bao}\ \emph {et~al.}(2022)\citenamefont {Bao},
  \citenamefont {Yu}, \citenamefont {Anderegg}, \citenamefont {Burchesky},
  \citenamefont {Gonzalez-Acevedo}, \citenamefont {Chae}, \citenamefont
  {Ketterle}, \citenamefont {Ni},\ and\ \citenamefont {Doyle}}]{bao22}%
  \BibitemOpen
  \bibfield  {author} {\bibinfo {author} {\bibfnamefont {Y.}~\bibnamefont
  {Bao}}, \bibinfo {author} {\bibfnamefont {S.~S.}\ \bibnamefont {Yu}},
  \bibinfo {author} {\bibfnamefont {L.}~\bibnamefont {Anderegg}}, \bibinfo
  {author} {\bibfnamefont {S.}~\bibnamefont {Burchesky}}, \bibinfo {author}
  {\bibfnamefont {D.}~\bibnamefont {Gonzalez-Acevedo}}, \bibinfo {author}
  {\bibfnamefont {E.}~\bibnamefont {Chae}}, \bibinfo {author} {\bibfnamefont
  {W.}~\bibnamefont {Ketterle}}, \bibinfo {author} {\bibfnamefont {K.-K.}\
  \bibnamefont {Ni}},\ and\ \bibinfo {author} {\bibfnamefont {J.~M.}\
  \bibnamefont {Doyle}},\ }\href {https://doi.org/10.1088/1367-2630/ac900f}
  {\bibinfo {title} {Fast optical transport of ultracold molecules over long
  distances}} (\bibinfo {year} {2022})\BibitemShut {NoStop}%
\bibitem [{\citenamefont {McDonald}\ \emph {et~al.}(2015)\citenamefont
  {McDonald}, \citenamefont {McGuyer}, \citenamefont {Iwata},\ and\
  \citenamefont {Zelevinsky}}]{mcdonald15}%
  \BibitemOpen
  \bibfield  {author} {\bibinfo {author} {\bibfnamefont {M.}~\bibnamefont
  {McDonald}}, \bibinfo {author} {\bibfnamefont {B.~H.}\ \bibnamefont
  {McGuyer}}, \bibinfo {author} {\bibfnamefont {G.~Z.}\ \bibnamefont {Iwata}},\
  and\ \bibinfo {author} {\bibfnamefont {T.}~\bibnamefont {Zelevinsky}},\
  }\bibfield  {title} {\bibinfo {title} {Thermometry via {L}ight {S}hifts in
  {O}ptical {L}attices},\ }\href
  {https://doi.org/10.1103/PhysRevLett.114.023001} {\bibfield  {journal}
  {\bibinfo  {journal} {Physical Review Letters}\ }\textbf {\bibinfo {volume}
  {114}},\ \bibinfo {pages} {023001} (\bibinfo {year} {2015})}\BibitemShut
  {NoStop}%
\bibitem [{\citenamefont {Stellmer}\ \emph {et~al.}(2013)\citenamefont
  {Stellmer}, \citenamefont {Grimm},\ and\ \citenamefont
  {Schreck}}]{stellmer13}%
  \BibitemOpen
  \bibfield  {author} {\bibinfo {author} {\bibfnamefont {S.}~\bibnamefont
  {Stellmer}}, \bibinfo {author} {\bibfnamefont {R.}~\bibnamefont {Grimm}},\
  and\ \bibinfo {author} {\bibfnamefont {F.}~\bibnamefont {Schreck}},\
  }\bibfield  {title} {\bibinfo {title} {Production of quantum-degenerate
  strontium gases},\ }\href {https://doi.org/10.1103/PhysRevA.87.013611}
  {\bibfield  {journal} {\bibinfo  {journal} {Phys. Rev. A}\ }\textbf {\bibinfo
  {volume} {87}},\ \bibinfo {pages} {013611} (\bibinfo {year}
  {2013})}\BibitemShut {NoStop}%
\bibitem [{\citenamefont {Miyake}\ \emph {et~al.}(2019)\citenamefont {Miyake},
  \citenamefont {Pisenti}, \citenamefont {Elgee}, \citenamefont {Sitaram},\
  and\ \citenamefont {Campbell}}]{miyake19}%
  \BibitemOpen
  \bibfield  {author} {\bibinfo {author} {\bibfnamefont {H.}~\bibnamefont
  {Miyake}}, \bibinfo {author} {\bibfnamefont {N.~C.}\ \bibnamefont {Pisenti}},
  \bibinfo {author} {\bibfnamefont {P.~K.}\ \bibnamefont {Elgee}}, \bibinfo
  {author} {\bibfnamefont {A.}~\bibnamefont {Sitaram}},\ and\ \bibinfo {author}
  {\bibfnamefont {G.~K.}\ \bibnamefont {Campbell}},\ }\bibfield  {title}
  {\bibinfo {title} {Isotope-shift spectroscopy of the
  {$^{1}S_{0}\ensuremath{\rightarrow}^{3}P_{1}$ and
  $^{1}S_{0}\ensuremath{\rightarrow}^{3}P_{0}$} transitions in strontium},\
  }\href {https://doi.org/10.1103/PhysRevResearch.1.033113} {\bibfield
  {journal} {\bibinfo  {journal} {Physical Review Research}\ }\textbf {\bibinfo
  {volume} {1}},\ \bibinfo {pages} {033113} (\bibinfo {year}
  {2019})}\BibitemShut {NoStop}%
\bibitem [{\citenamefont {Stellmer}\ and\ \citenamefont
  {Schreck}(2014)}]{stellmer14}%
  \BibitemOpen
  \bibfield  {author} {\bibinfo {author} {\bibfnamefont {S.}~\bibnamefont
  {Stellmer}}\ and\ \bibinfo {author} {\bibfnamefont {F.}~\bibnamefont
  {Schreck}},\ }\bibfield  {title} {\bibinfo {title} {Reservoir spectroscopy of
  $5s5p$ ${}^{3}p{}_{2}$--$5snd$ ${}^{3}{D}_{1,2,3}$ transitions in
  strontium},\ }\href {https://doi.org/10.1103/physreva.90.022512} {\bibfield
  {journal} {\bibinfo  {journal} {Physical Review A}\ }\textbf {\bibinfo
  {volume} {90}},\ \bibinfo {pages} {022512} (\bibinfo {year}
  {2014})}\BibitemShut {NoStop}%
\bibitem [{\citenamefont {Iwata}\ \emph {et~al.}(2021)\citenamefont {Iwata},
  \citenamefont {Miyabe}, \citenamefont {Akaoka},\ and\ \citenamefont
  {Wakaida}}]{iwata21}%
  \BibitemOpen
  \bibfield  {author} {\bibinfo {author} {\bibfnamefont {Y.}~\bibnamefont
  {Iwata}}, \bibinfo {author} {\bibfnamefont {M.}~\bibnamefont {Miyabe}},
  \bibinfo {author} {\bibfnamefont {K.}~\bibnamefont {Akaoka}},\ and\ \bibinfo
  {author} {\bibfnamefont {I.}~\bibnamefont {Wakaida}},\ }\bibfield  {title}
  {\bibinfo {title} {Isotope shift and hyperfine structure measurements on
  triple resonance excitation to the autoionizing rydberg state of atomic
  strontium},\ }\href {https://doi.org/10.1016/j.jqsrt.2021.107882} {\bibfield
  {journal} {\bibinfo  {journal} {Journal of Quantitative Spectroscopy and
  Radiative Transfer}\ }\textbf {\bibinfo {volume} {275}},\ \bibinfo {pages}
  {107882} (\bibinfo {year} {2021})}\BibitemShut {NoStop}%
\bibitem [{\citenamefont {Kramida}\ \emph {et~al.}(2020)\citenamefont
  {Kramida}, \citenamefont {{Yu.~Ralchenko}}, \citenamefont {Reader},\ and\
  \citenamefont {{and NIST ASD Team}}}]{nistasd}%
  \BibitemOpen
  \bibfield  {author} {\bibinfo {author} {\bibfnamefont {A.}~\bibnamefont
  {Kramida}}, \bibinfo {author} {\bibnamefont {{Yu.~Ralchenko}}}, \bibinfo
  {author} {\bibfnamefont {J.}~\bibnamefont {Reader}},\ and\ \bibinfo {author}
  {\bibnamefont {{and NIST ASD Team}}},\ }\href
  {https://doi.org/10.18434/T4W30F} {}\bibinfo {howpublished} {{NIST Atomic
  Spectra Database (ver. 5.8), [Online]. Available:
  {\tt{https://physics.nist.gov/asd}} [10/30/2020]. National Institute of
  Standards and Technology, Gaithersburg, MD.}} (\bibinfo {year}
  {2020})\BibitemShut {NoStop}%
\bibitem [{\citenamefont {Safronova}(2022)}]{safronovapriv}%
  \BibitemOpen
  \bibfield  {author} {\bibinfo {author} {\bibfnamefont {M.~S.}\ \bibnamefont
  {Safronova}},\ }\href@noop {} {\bibinfo {title} {private communication}}
  (\bibinfo {year} {2022})\BibitemShut {NoStop}%
\end{thebibliography}
%

\end{document}